\documentclass[review]{elsarticle}

\journal{Physics Reports}

\pdfoutput=1

\usepackage[bookmarks=false,hyperfootnotes=false,colorlinks]{hyperref}
\usepackage{graphicx}
\usepackage{amsmath}
\usepackage{amssymb}
\usepackage[caption=false]{subfig}
\usepackage{xspace}
\usepackage{slashed}
\usepackage{verbatim}
\usepackage{bbm}
\usepackage{amsfonts}
\usepackage{import}
\usepackage{booktabs}
\usepackage[export]{adjustbox}
\usepackage{multirow}
\usepackage[svgnames]{xcolor}

\AtBeginDocument{%
  \hypersetup{
    citecolor=blue,
    linkcolor=blue,   
    urlcolor=blue
    }
  }

\def\as{\alpha_s}
\newcommand{\asbar}{{\bar \alpha}_s}
\newcommand{\zcut}{z_\text{cut}}
\newcommand{\Rsub}{R_\text{sub}}

\newcommand{\sanepruning}{\text{Y-pruning}}

\begin{document}


\title{Jet Substructure at the Large Hadron Collider:\\ A Review of Recent Advances in Theory and Machine Learning}
 
\author{Andrew J. Larkoski}
\ead{larkoski@reed.edu}
\address{Physics Department, Reed College, Portland, OR 97202, USA}

\author{Ian Moult}
\ead{ianmoult@lbl.gov}
\address{Berkeley Center for Theoretical Physics, University of California, Berkeley, CA 94720, USA}
\address{Theoretical Physics Group, Lawrence Berkeley National Laboratory, Berkeley, CA 94720, USA}

 \author{Benjamin Nachman}
 \ead{benjamin.philip.nachman@cern.ch}
 \address{Physics Division, Lawrence Berkeley National Laboratory, Berkeley, CA 94720, USA}

\date{\today{}}

\begin{abstract}
\noindent Jet substructure has emerged to play a central role at the Large Hadron Collider (LHC), where it has provided numerous innovative new ways to search for new physics and to probe the Standard Model in extreme regions of phase space. In this article we provide a comprehensive review of state of the art theoretical and machine learning developments in jet substructure. This article is meant both as a pedagogical introduction, covering the key physical principles underlying the calculation of jet substructure observables, the development of new observables, and cutting edge machine learning techniques for jet substructure, as well as a comprehensive reference for experts.  We hope that it will prove a useful introduction to the exciting and rapidly developing field of jet substructure at the LHC.
%
%
%
\end{abstract}

\maketitle 

\tableofcontents{}

\newcommand{\Dobs}[2]{D_{#1}^{(#2)}} 

\DeclareRobustCommand{\Sec}[1]{Sec.~\ref{#1}}
\DeclareRobustCommand{\Secs}[2]{Secs.~\ref{#1} and \ref{#2}}
\DeclareRobustCommand{\App}[1]{App.~\ref{#1}}
\DeclareRobustCommand{\Tab}[1]{Table~\ref{#1}}
\DeclareRobustCommand{\Tabs}[2]{Tables~\ref{#1} and \ref{#2}}
\DeclareRobustCommand{\Fig}[1]{Fig.~\ref{#1}}
\DeclareRobustCommand{\Figs}[2]{Figs.~\ref{#1} and \ref{#2}}
\DeclareRobustCommand{\Eq}[1]{Eq.~(\ref{#1})}
\DeclareRobustCommand{\Eqs}[2]{Eqs.~(\ref{#1}) and (\ref{#2})}
\DeclareRobustCommand{\Ref}[1]{Ref.~\cite{#1}}
\DeclareRobustCommand{\Refs}[1]{Refs.~\cite{#1}}

\newcommand{\runone}{Run 1\xspace}
\newcommand{\runtwo}{Run 2\xspace}
\newcommand{\ptD}{\ensuremath{p_{T}D}\xspace}
\newcommand{\fcut}{\ensuremath{f_{\mathrm{cut}}}\xspace}
\newcommand{\rsub}{\ensuremath{R_{\mathrm{sub}}}\xspace}

\newcommand{\antikt}       {anti-\ensuremath{k_{t}}\xspace}
\newcommand{\antiktfour}    {anti-\ensuremath{k_{t}}, \ensuremath{R=0.4}\xspace}
\newcommand{\antiktten}   {anti-\ensuremath{k_{t}}, \ensuremath{R=1.0}\xspace}
\newcommand{\antikteight}   {anti-\ensuremath{k_{t}}, \ensuremath{R=1.0}\xspace}

\newcommand{\de}{\ensuremath{^\circ}}
\newcommand{\ten}[1]{\ensuremath{\times \text{10}^\text{#1}}}
\newcommand{\unit}[1]{\ensuremath{\text{\,#1}}\xspace}
\newcommand{\mum}{\ensuremath{\,\mu\text{m}}\xspace}
\newcommand{\micron}{\ensuremath{\,\mu\text{m}}\xspace}
\newcommand{\cm}{\ensuremath{\,\text{cm}}\xspace}
\newcommand{\mm}{\ensuremath{\,\text{mm}}\xspace}
\newcommand{\mus}{\ensuremath{\,\mu\text{s}}\xspace}
\newcommand{\keV}{\ensuremath{\,\text{ke\hspace{-.08em}V}}\xspace}
\newcommand{\MeV}{\ensuremath{\,\text{Me\hspace{-.08em}V}}\xspace}
\newcommand{\MeVns}{\ensuremath{\text{Me\hspace{-.08em}V}}\xspace} 
\newcommand{\GeV}{\ensuremath{\,\text{Ge\hspace{-.08em}V}}\xspace}
\newcommand{\GeVns}{\ensuremath{\text{Ge\hspace{-.08em}V}}\xspace} 
\newcommand{\gev}{\GeV}
\newcommand{\TeV}{\ensuremath{\,\text{Te\hspace{-.08em}V}}\xspace}
\newcommand{\TeVns}{\ensuremath{\text{Te\hspace{-.08em}V}}\xspace} 
\newcommand{\PeV}{\ensuremath{\,\text{Pe\hspace{-.08em}V}}\xspace}
\newcommand{\keVc}{\ensuremath{{\,\text{ke\hspace{-.08em}V\hspace{-0.16em}/\hspace{-0.08em}}c}}\xspace}
\newcommand{\MeVc}{\ensuremath{{\,\text{Me\hspace{-.08em}V\hspace{-0.16em}/\hspace{-0.08em}}c}}\xspace}
\newcommand{\GeVc}{\ensuremath{{\,\text{Ge\hspace{-.08em}V\hspace{-0.16em}/\hspace{-0.08em}}c}}\xspace}
\newcommand{\GeVcns}{\ensuremath{{\text{Ge\hspace{-.08em}V\hspace{-0.16em}/\hspace{-0.08em}}c}}\xspace} 
\newcommand{\TeVc}{\ensuremath{{\,\text{Te\hspace{-.08em}V\hspace{-0.16em}/\hspace{-0.08em}}c}}\xspace}
\newcommand{\keVcc}{\ensuremath{{\,\text{ke\hspace{-.08em}V\hspace{-0.16em}/\hspace{-0.08em}}c^\text{2}}}\xspace}
\newcommand{\MeVcc}{\ensuremath{{\,\text{Me\hspace{-.08em}V\hspace{-0.16em}/\hspace{-0.08em}}c^\text{2}}}\xspace}
\newcommand{\GeVcc}{\ensuremath{{\,\text{Ge\hspace{-.08em}V\hspace{-0.16em}/\hspace{-0.08em}}c^\text{2}}}\xspace}
\newcommand{\GeVccns}{\ensuremath{{\text{Ge\hspace{-.08em}V\hspace{-0.16em}/\hspace{-0.08em}}c^\text{2}}}\xspace} 
\newcommand{\TeVcc}{\ensuremath{{\,\text{Te\hspace{-.08em}V\hspace{-0.16em}/\hspace{-0.08em}}c^\text{2}}}\xspace}

\newcommand{\pbinv} {\mbox{\ensuremath{\,\text{pb}^\text{$-$1}}}\xspace}
\newcommand{\fbinv} {\mbox{\ensuremath{\,\text{fb}^\text{$-$1}}}\xspace}
\newcommand{\nbinv} {\mbox{\ensuremath{\,\text{nb}^\text{$-$1}}}\xspace}
\newcommand{\mubinv} {\ensuremath{\,\mu\mathrm{b}^{-1}}\xspace}
\newcommand{\percms}{\ensuremath{\,\text{cm}^\text{$-$2}\,\text{s}^\text{$-$1}}\xspace}
\newcommand{\lumi}{\ensuremath{\mathcal{L}}\xspace}
\newcommand{\Lumi}{\ensuremath{\mathcal{L}}\xspace}


\newcommand{\PT}{\ensuremath{p_{\mathrm{T}}}\xspace}
\newcommand{\pt}{\ensuremath{p_{\mathrm{T}}}\xspace}
\newcommand{\kt}{\ensuremath{k_{\mathrm{T}}}\xspace}
\newcommand{\ET}{\ensuremath{E_{\mathrm{T}}}\xspace}
\newcommand{\HT}{\ensuremath{H_{\mathrm{T}}}\xspace}
\newcommand{\et}{\ensuremath{E_{\mathrm{T}}}\xspace}
\newcommand{\Em}{\ensuremath{E\hspace{-0.6em}/}\xspace}
\newcommand{\Pm}{\ensuremath{p\hspace{-0.5em}/}\xspace}
\newcommand{\PTm}{\ensuremath{{p}_\mathrm{T}\hspace{-1.02em}/\kern 0.5em}\xspace}
\newcommand{\PTslash}{\PTm}
\newcommand{\ETm}{\ensuremath{E_{\mathrm{T}}^{\text{miss}}}\xspace}
\newcommand{\MET}{\ETm}
\newcommand{\ETmiss}{\ETm}
\newcommand{\ETslash}{\ensuremath{E_{\mathrm{T}}\hspace{-1.1em}/\kern0.45em}\xspace}
\newcommand{\VEtmiss}{\ensuremath{{\vec E}_{\mathrm{T}}^{\text{miss}}}\xspace}
\newcommand{\ptvec}{\ensuremath{{\vec p}_{\mathrm{T}}}\xspace}
\newcommand{\ptvecmiss}{\ensuremath{{\vec p}_{\mathrm{T}}^{\kern1pt\text{miss}}}\xspace}
\newcommand{\toptau}{\ensuremath{\tau_{32}}\xspace}
\newcommand{\Vtau}{\ensuremath{\tau_{21}}\xspace}

\newcommand{\Mjet}{\ensuremath{M_{\mathrm{jet}}}\xspace}

\newcommand{\ttbar}{\ensuremath{t\overline{t}}\xspace} 
\newcommand{\bbbar}{\ensuremath{b\overline{b}}\xspace} 
\newcommand{\ttH}{\ensuremath{t\overline{t}H}\xspace} 
\newcommand{\W}{\ensuremath{W}\xspace}
\newcommand{\Z}{\ensuremath{Z}\xspace}
\newcommand{\Hig}{\ensuremath{H}\xspace}

\section{Introduction}

The Large Hadron Collider (LHC) is currently the center of attention in particle physics, providing a unique opportunity to probe the dynamics of the Standard Model (SM) at the TeV scale, and to search for new physics. One of the major new developments which has come to play a central role at the LHC is jet substructure. Jets are collimated sprays of particles resulting from quarks and gluons produced at high energy; jet substructure is a set of tools to exploit information from the radiation pattern inside these jets. For example, jet substructure can be used to identify boosted hadronically decaying electroweak bosons and top quarks. Jet substructure techniques have provided innovative advances in probing the SM, in addition to improving the sensitivity for new physics searches.
The surge of interest in jet substructure at the LHC has been driven by the extended energy reach, which has inspired new theoretical ideas and reconstruction techniques to probe this previously unexplored and exciting regime.

The renewed theoretical interest in jet structure has resulted in a renaissance for the theoretical understanding of jets. The resulting rapid progress in this field has resulted in precise predictions for a wide variety of observables. This analytical understanding has also led to observable engineering with new tailored techniques that are already deployed in the big experiments; a healthy symbiosis of experimental and theoretical ideas has helped propel this process forward \cite{Adams:2015hiv,Altheimer:2013yza,Altheimer:2012mn,Abdesselam:2010pt}. Theoretical developments in jet substructure have also had a broader impact on QCD both in the vacuum as well as in medium.

Due to the central role that jet substructure is now playing at the LHC, with its mature and sophisticated set of theoretical and experimental techniques, it is time to provide a comprehensive review of jet substructure. The goal of this review is to be
\begin{enumerate}
\item a state-of-the-art reference for those looking for an overview of the field;
\item a primer on both theoretical and machine learning techniques for newcomers; and
\item an outline of the challenges going forward, and the work that has yet to be done in the field.
\end{enumerate}

Due to the wide range of topics that are covered, we must be somewhat selective in our presentation. We have therefore taken the approach of emphasizing representative examples, and underlying physical principles rather than details. The specifics of calculational techniques for jet substructure, for example, are beyond the scope of this review; however, we have attempted to provide a comprehensive source of references, which we believe will in itself be a useful resource. There are also a number of more specialized and earlier reviews on different aspects of jet substructure, to which we refer the interested reader \cite{Salam:2009jx,Shelton:2013an,Plehn:2011tg,Cacciari:2015jwa,Adams:2015hiv,Altheimer:2013yza,Altheimer:2012mn,Abdesselam:2010pt,marzani:2019}. For the reader interested in more specialized techniques, these can be supplemented by reviews and texts on QCD relevant for jet substructure calculations \cite{Sterman:1995fz,iain_notes,Schwartz:2013pla,Ellis:1991qj,Collins:2011zzd,Becher:2014oda,Luisoni:2015xha,Campbell:2017hsr}, as well as several standard texts and reviews on deep learning \cite{Goodfellow-et-al-2016,1404.7828,DNNnaturereview} (for particle physics~\cite{Guest:2018yhq,Radovic:2018dip}), and the standard machine learning packages \cite{scikit-learn,keras,tensorflow,2016arXiv160502688short,jia2014caffe,pytorch,CNTK} (for particle physics~\cite{Hocker:2007ht}).

This publication consists of two main sections: \Sec{sec:theory} covers theory developments, and \Sec{sec:machine_learning} covers applications of machine learning to jet substructure. 
%
%
In the theory portion of the review, we begin in \Sec{sec:calcs} with a review of the theoretical aspects of jet substructure calculations. The goal here is not to describe technical details of calculations, but rather to emphasize the different physics relevant to jet substructure calculations. This also lays the foundations enabling the reader to evaluate the quality of different calculations, and appreciate difficulties in extending these calculations to more complicated observables. Then, in \Sec{sec:moreobs} we survey the wide range of jet substructure calculations which have been performed using these techniques. Instead of providing an exhaustive listing, we have chosen to discuss representative observables in more detail, while providing references to other calculations. In \Sec{sec:moreloops} we discuss prospects for improving the precision of jet substructure calculations, and highlight those observables which are most amenable to high precision calculations. In \Sec{sec:newfronts} we discuss new frontiers in jet substructure, attempting to highlight the broad range of connections to other areas of physics, both theoretical and experimental, where we hope ideas from jet substructure will prove fruitful. Finally, we conclude with a wish list of goals, which we hope will drive theory progress in the field in the years to come.


In \Sec{sec:machine_learning} we review applications of machine learning (ML) to jet substructure, which is a topic of significant current interest, both from the theoretical and experimental communities. To a large extent, jet physics is driving the use of modern ML tools in high energy physics due to the complex and rich structure inside jets. We have attempted to make this section self contained, introducing both technical aspects of machine learning, as well as highlighting their applications to problems of current interest in jet substructure. As with the other sections in this review, our goal is not to provide a comprehensive review, but simply a broad overview of this exciting and rapidly developing field. 


\section{Theory Developments}
\label{sec:theory}

The past several years have seen significant theoretical efforts focused around designing and predicting observables measured on jets.  The guiding principles for constructing a jet substructure observable is that the observable should be 1) sensitive to the physics you want to probe, and 2) be calculable from first principles in Quantum Chromodynamics (QCD).  That an observable should be sensitive to the physics you want to probe is vital for the ability to draw certain conclusions from the measurement of the observable. The property of calculability is more subtle, and is largely shaped by our limited theoretical tools for addressing QCD beyond perturbation theory. A sufficient condition to ensure calculability within the perturbation theory of QCD is infrared and collinear (IRC) safety, which has therefore played a central role as an organizing principle for jet observables.


Heuristically, IRC safety is typically stated in the following way~\cite{Ellis:1991qj}:
\begin{quote}
An observable is infrared and collinear safe if it is insensitive to infinitesimally soft or exactly collinear emissions.
\end{quote}
Because QCD is a gauge theory with massless particles, the Feynman diagrammatic perturbation theory in the strong coupling constant $\alpha_s$ is degenerate, which is manifested as divergences in the soft (low energy) or collinear limits of particles.  However, as guaranteed by the Kinoshita-Lee-Nauenberg theorem \cite{Kinoshita:1962ur,Lee:1964is}, when the soft and collinear regions of phase space are inclusively summed over, the divergences exactly cancel between the real and virtual contributions to the cross section at each perturbative order.  The property of IRC safety ensures that the phase space restrictions that the measured value of an observable imposes do not disrupt this cancellation.

The importance of infrared and collinear safety for jet physics was first recognized in the jet algorithm of Sterman and Weinberg \cite{Sterman:1977wj}.  Shortly after, Clavelli argued that the jet mass is an infrared and collinear safe observable and produced what could be called the first jet substructure theory prediction computed from first principles perturbative QCD \cite{Clavelli:1979md}.  Building on their work on studying the thrust observable \cite{Farhi:1977sg}, Catani, Trentadue, Turnock, and Webber (CTTW) produced the first resummed jet substructure observable, calculating the heavy jet mass in $e^+e^-\to$ hadrons events to next-to-leading logarithmic accuracy \cite{Catani:1991bd,Catani:1992ua}.  These calculations initiated the theory of perturbatively calculable jet observables, whose accuracy could be systematically improved order-by-order in perturbation theory, laying the foundations for modern jet substructure calculations.



Especially motivated by the high collision energy, the exceptional resolution of the experiments at the LHC, and the introduction of fast \cite{Cacciari:2005hq} and experimentally well-behaved IRC safe jet finding algorithms \cite{Cacciari:2008gp}, the theory of jet substructure has become a mature, lively field, enabling applications well beyond those envisioned in the seminal works of \Refs{Seymour:1993mx,Butterworth:2002tt,Butterworth:2008iy}.  No longer are jet observables and calculations restricted to the mass, but include a whole menagerie of observables that are sensitive to multiple hard prongs in the jet, or to coherent soft emission, or to correlations between radiation inside and outside the jet.  The construction of these observables is still shaped by the requirement of IRC safety, but more and more observables have been introduced that lack this property.  Through the development of new techniques, classes of these IRC unsafe observables have broadened the definition of calculable within perturbative QCD, while retaining predictive power.

In this section, theory calculations directly of jet substructure observables or for observables that have importance for jet substructure are reviewed.  Section~\ref{sec:calcs} starts by describing the different aspects of a calculation for an IRC safe jet observable and the most important physics that goes into a calculation.  The use of fully-exclusive event generators and the definition of the the accuracy of a calculation is also reviewed.  Section~\ref{sec:moreobs} provides an overview of the status of theory predictions for jet substructure observables and the physics they are meant to capture.  Some of the observables reviewed here have been measured on jets in ATLAS or CMS, and appropriate references are provided for those observables.  The efforts devoted to high-precision calculations in jet substructure are reviewed in Section~\ref{sec:moreloops}.  Because the observables are typically dominated by soft or collinear radiation, an important aspect of achieving high-precision in jet physics is resummation to higher logarithmic accuracy.  Section~\ref{sec:newfronts} summarizes the frontiers of jet substructure calculations, highlights several areas where the field can make progress, and also reviews how the field of jet substructure has influenced QCD theory more broadly. It concludes with a set of recommendations for data-theory comparison, which we believe will provide a firm foundation from which to build a strong future program.


\subsection{Aspects of Jet Substructure Calculations}
\label{sec:calcs}

In this section we will describe aspects of jet substructure calculations, focusing on the case of IRC safe observables. We also review the standard theoretical approaches which have proved most powerful. While there has been recent interest in the calculation of more general classes of observables, these are beyond the scope of this brief review. However, many of the principles involved in these more sophisticated calculations are similar to those discussed here.

We take as an example a single measurement, the jet mass, $m_J$, and consider the case that $m_J \ll p_{TJ}$, the jet's transverse momentum. A calculation of the jet mass that can be directly compared with experimental measurements must incorporate the following three ingredients:
\begin{itemize}
\item Resummation of logarithmically enhanced contributions $\alpha_s \log^2\left( \frac{p_{TJ}}{m_J}\right)$.
\item Fixed-order perturbative corrections in $\alpha_s$.
\item Non-perturbative corrections from hadronization and other effects.
\end{itemize}
The goal of this section will be to describe the physics underlying each of these contributions, and the accuracy to which different theoretical tools are able to describe them. This will allow us to compare in detail the theoretical sophistication of different jet substructure calculations, with examples in the forthcoming sections illustrating each of these features. It will also allow us to describe the difficulties inherent in the calculation of more complicated observables. 

\begin{figure}[t]
\begin{center}
\includegraphics[width=0.95\columnwidth]{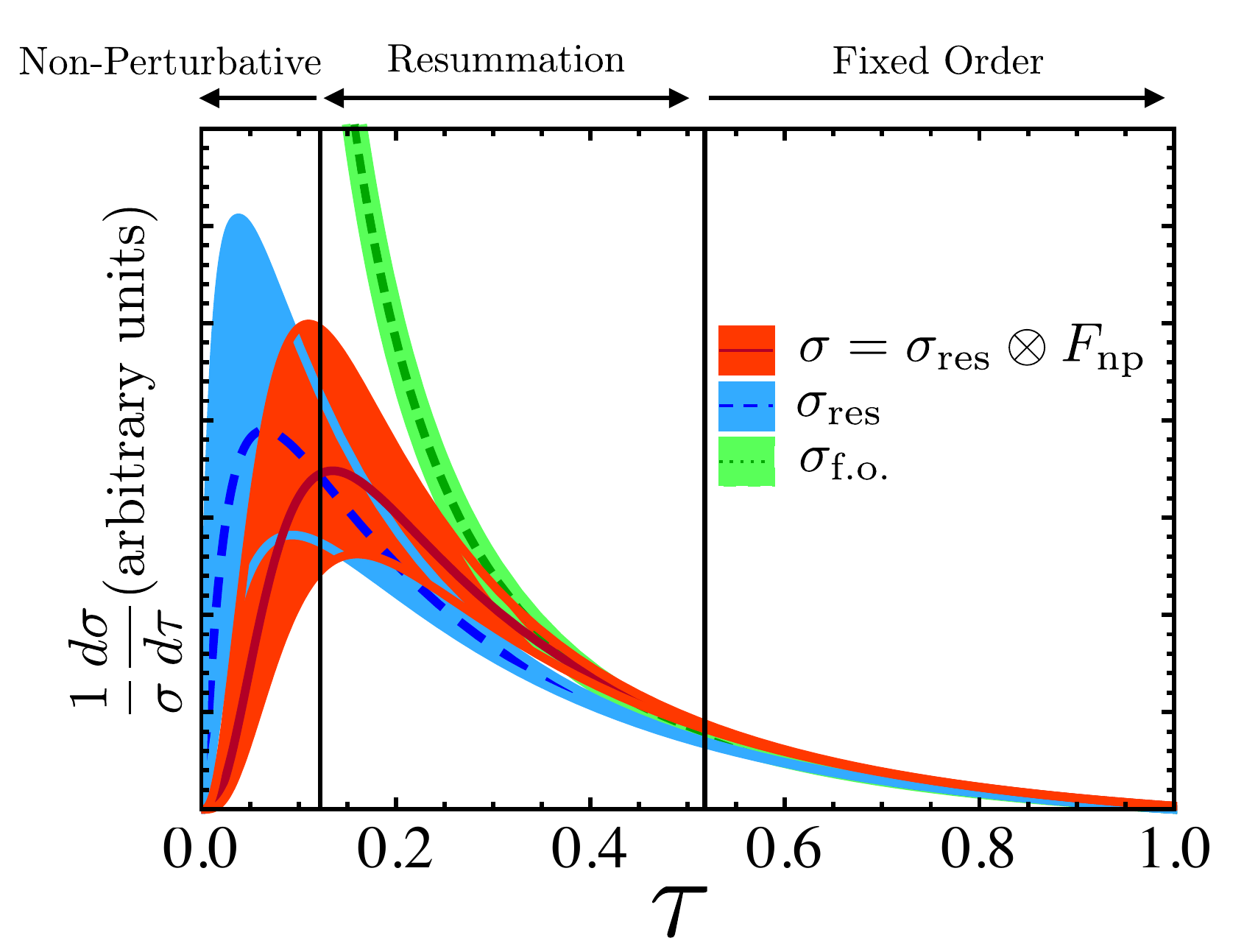} 
\end{center}
\vspace{-0.4cm}
\caption{A typical jet substructure calculation, emphasizing the regions where different contributions dominate the physical description of the observable.  Here, $\sigma_\text{f.o.}$ is the fixed-order prediction for the cross section, $\sigma_\text{res}$ includes resummation of large logarithms and $\sigma$ is the complete theory prediction including non-perturbative effects through a shape function $F_\text{np}$.} 
\label{fig:structure_calculation}
\end{figure}

A schematic illustration of these different ingredients for an observable $\tau$, which can be taken to be the jet mass, is illustrated in Fig.~\ref{fig:structure_calculation}. The green curve shows the prediction of the observable in fixed order QCD, which diverges as $\tau\to0$ due to the fact that logarithmically enhanced terms are not resummed. The resummation of the logarithmically enhanced terms causes the distribution to go to zero as $\tau\to0$, as shown in the blue curve. Finally, the inclusion of non-perturbative contributions shifts the distribution at small values of $\tau$, where the observable is sensitive to fluctuations at the scale $\Lambda_{\text{QCD}}$.  This can be implemented in a calculation with a non-perturbative shape function, $F_\text{np}$.  Shaded bands are representative of theoretical uncertainties.  We have used a general observable $\tau$ to emphasize that while we will focus on $m_J$ as a concrete example in this section, the behavior of Fig.~\ref{fig:structure_calculation} is generic for a wide range of observables.

\subsubsection{Resummation}

We begin with a discussion of the perturbative aspects of a calculation. Since we have restricted our focus to IRC safe observables, a perturbative expansion in the strong coupling constant, $\alpha_s$, gives finite results order-by-order in perturbation theory. Relative to the Born process, calculations are referred to as leading order (LO), next-to-leading-order (NLO), and so forth. However, due to the multi-scale nature of jet substructure problems (in this case both $m_J$ and $p_{TJ}$ enter as relevant scales), the perturbative expansion involves not only constants, but functions of the ratio $p_{TJ}/m_J$. In particular, due to the soft and collinear divergences of QCD, large logarithms $\log(p_{TJ}/m_J)$ appear at each order in perturbation theory.  To simplify the discussion, we work with the cumulative cross section
\begin{align}
R(m_J)=\int\limits_0^{m_J} \frac{d\sigma}{d m_J'}\, dm_J'\,.
\end{align}
We choose to express the cumulative cross section as
\begin{align}\label{eq:xsecexpress}
R(m_J)=C(\alpha_s) \Sigma(m_J, \alpha_s) + D(m_J, \alpha_s)\,,
\end{align}
where the different functions on the right side have different properties.
Here,
\begin{align}
C(\alpha_s)=1+\sum\limits_{n=1}^\infty \left( \frac{\alpha_s}{2\pi}\right)^n C_n\,,
\end{align}
where the $C_n$ are constants, and
\begin{align}\label{eq:cum}
&\ln \Sigma(m_J, \alpha_s)=\sum\limits_{n=1}^\infty \sum\limits_{m=1}^{n+1} \left( \frac{\alpha_s}{2\pi}\right)^n G_{nm} \ln^m \frac{m_J}{p_{TJ}}\\
&=\left( \frac{\alpha_s}{2\pi}\right) \left( G_{12} \log^2  \frac{p_{TJ}}{m_J}  + G_{11} \log  \frac{p_{TJ}}{m_J} \right) \nonumber \\
&+\left( \frac{\alpha_s}{2\pi}\right)^2 \left( G_{23} \log^3  \frac{p_{TJ}}{m_J}  + G_{22} \log^2  \frac{p_{TJ}}{m_J} + G_{21} \log  \frac{p_{TJ}}{m_J} \right) \nonumber\\
&+\cdots \nonumber\,,
\end{align}
referred to as the radiator, contains the logarithmically enhanced terms.  The $G_{nm}$ are constants, independent of $m_J$ and $p_{TJ}$.  We see that increasingly high powers of the logarithm appear at each order in perturbation theory. Furthermore, as $m_J \to 0$, this result diverges at any fixed order in perturbative theory.  These logarithmically enhanced terms are intimately related to the soft and collinear behavior of QCD, since in the limit $m_J\to0$, only soft particles, or particles collinear to the jet direction, can contribute.  The $D(m_J, \alpha_s)$ term is referred to as the power corrections, as all terms in it scale like a positive power of $m_J$, and so vanish in the $m_J\to 0$ limit.

For typical jets at the LHC, we have $m_J \sim \mathcal{O}(10)$ GeV, while $p_{TJ}\sim  \mathcal{O}(500)$ GeV.  These values motivate the scaling $\log(p_{TJ}/m_J)\sim 1/\alpha_s$, for which the traditional fixed-order expansion in $\alpha_s$ is invalidated. Each term in a vertical column of \Eq{eq:cum} has the same scaling in $\alpha_s$, and terms which are higher-order in $\alpha_s$ in a given column are not suppressed. To obtain a reliable prediction, one must resum all terms in a given column, using an understanding of the all-orders structure of QCD. Here one uses a counting of leading logarithm (LL), next-to-leading logarithm (NLL), and so forth to distinguish this expansion from the standard fixed-order expansion. Unlike the fixed-order expansion, there is some ambiguity in the precise organization of the resummation. For a detailed review of different countings, see Ref.~\cite{Almeida:2014uva}. Here we will use the conventions of CTTW \cite{Catani:1992ua}, where we define the order using the cumulative cross section in \Eq{eq:cum}.

With the scaling $\log(p_{TJ}/m_J)\sim 1/\alpha_s$, we have that the resummed expansions include all terms at the following orders
\begin{align}
\text{LL}&: \mathcal{O}(\alpha_s^{-1})\nonumber\\
\text{NLL}&: \mathcal{O}(1)\nonumber\\
\text{NNLL}&: \mathcal{O}(\alpha_s)\nonumber\\
\text{NNNLL}&: \mathcal{O}(\alpha_s^2) 
\end{align}
Therefore, a LL calculation includes only the most singular terms, namely all terms that scale as $\mathcal{O}(\alpha_s^{-1})$. This gives rise to the familiar Sudakov form factor \cite{Sudakov:1954sw}
\begin{align}\label{eq:sudakov}
\Sigma(m_J)=\exp\left[\left( \frac{\alpha_s}{2\pi}\right) G_{12} \log^2  \frac{p_{TJ}}{m_J}+\cdots\right]\,,
\end{align}
which implements the physical behavior that $\Sigma(m_J)\to 0$ as $m_J\to 0$. The characteristic Sudakov peak is illustrated by the blue curve in Fig.~\ref{fig:structure_calculation}.  LL calculations are useful for understanding qualitative aspects of jet substructure observables. However, since they include only terms scaling as $\mathcal{O}(\alpha_s^{-1})$, they miss $\mathcal{O}(1)$ corrections, and they therefore typically do not provide a quantitative description of the distribution. 

NLL calculations include all terms that scale like $\mathcal{O}(1)$, and therefore this is the first order at which corrections are $\mathcal{O}(\alpha_s)$, and hence suppressed by the strong coupling constant. Such calculations should therefore begin to describe quantitative features of the distribution.  For calculations beyond NLL one obtains a reliable estimate of perturbative uncertainties, and an understanding of the perturbative convergence. This is particularly important for comparison with experimental measurements, and therefore motivates higher-order resummations, which will be discussed in \Sec{sec:moreloops}.

A variety of different approaches exist for performing resummation, each of which has its own advantages and disadvantages. These include explicit calculation of the radiator function \cite{Banfi:2004yd,Banfi:2014sua},
the use of generating functionals \cite{Konishi:1979cb,Dokshitzer:1982fh,Dokshitzer:1982ia,Dokshitzer:1982xr}, factorization theorems \cite{Collins:1985ue,Collins:1988ig,Collins:1989gx}, and effective field theory techniques \cite{Bauer:2000ew,Bauer:2000yr,Bauer:2001yt,Bauer:2001ct,Bauer:2002nz}.  The explicit calculation of the radiator can be performed for generic observables in $e^+e^-\to $ hadrons events to NNLL \cite{Banfi:2004yd,Banfi:2014sua}. It has the advantage that it can be applied to generic observables \cite{Banfi:2016zlc,Banfi:2018mcq}, and only requires the soft and collinear factorization of matrix elements in QCD.

All-orders factorization theorems were pioneered in the work of Refs.~\cite{Collins:1985ue,Collins:1988ig,Collins:1989gx}. Using the soft-collinear effective theory (SCET) \cite{Bauer:2000ew,Bauer:2000yr,Bauer:2001yt,Bauer:2001ct,Bauer:2002nz,Rothstein:2016bsq}, which provides a particularly powerful framework for proving factorization, factorization theorems have been proven for a number of processes of interest for jet substructure, such as mass measurements for top quark jets \cite{Fleming:2007xt, Fleming:2007qr} and the thrust event shape \cite{Schwartz:2007ib,Becher:2008cf}. There has recently been significant effort in extending all-orders factorization theorems to more differential jet substructure observables \cite{Bauer:2011uc,Pietrulewicz:2016nwo,Larkoski:2015kga,Larkoski:2015zka,Larkoski:2014tva,Procura:2014cba}.

For high-order resummation, the most powerful techniques rely on factorization theorems and resummation through renormalization group evolution. This is a vast topic to which we cannot do justice in this brief review. A number of excellent reviews exist on these topics \cite{Sterman:1995fz,iain_notes,Becher:2014oda,Luisoni:2015xha}.  In cases for which factorization can be demonstrated, higher logarithms can be systematically resummed by computing anomalous dimensions of field-theoretic objects to higher perturbative orders. This has allowed the highest-order resummation for $e^+e^-$ event shapes to NNNLL \cite{Becher:2008cf,Abbate:2010xh,Chien:2010kc,Hoang:2014wka}, as well as for the highest precision jet substructure calculations to NNLL \cite{Frye:2016okc,Frye:2016aiz}, to be discussed in \Sec{sec:moreloops}.

\begin{figure}[t]
\begin{center}
\includegraphics[width=0.95\columnwidth]{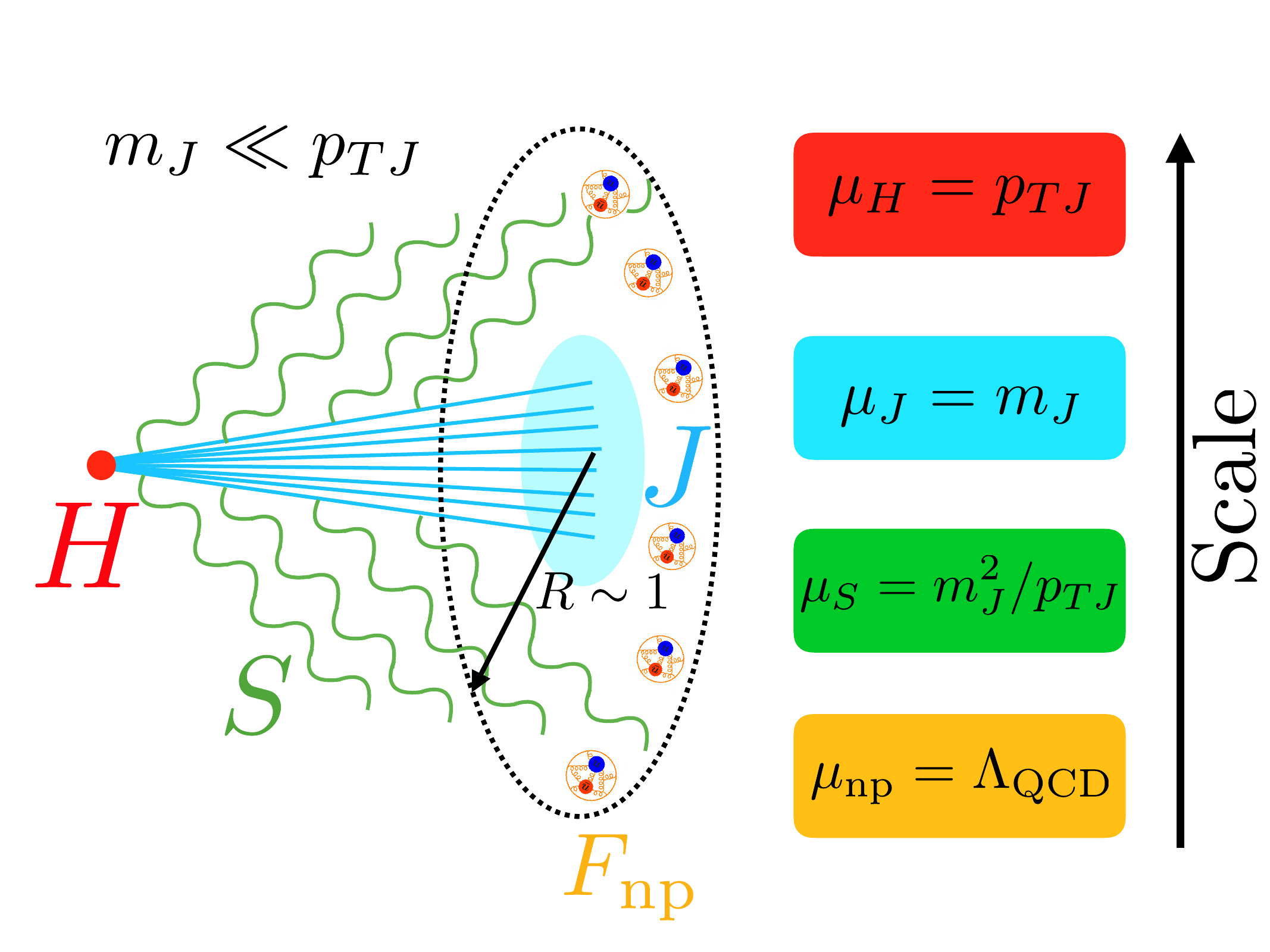} 
\end{center}
\vspace{-0.4cm}
\caption{The different scales relevant for the description of a mass measurement on a jet. Sudakov double logarithms, $\log^2 (p_{TJ}/m_J)$, are resummed by renormalization group evolution between the different scales. 
} 
\label{fig:mass_schematic}
\end{figure}

Taking dijets in $e^+e^-$ collisions as a simple example (to avoid complications deriving from the colored initial state), for the differential cross section of the left and right hemisphere jet masses in the limit where $m_L,m_R \ll Q$ we have the schematic factorized expression
\begin{align}\label{eq:schematic_fact}
\frac{d\sigma }{dm_L \, dm_R}=&\sigma_0 H(Q^2;\mu) \cdot J(m_L;\mu) \otimes J(m_R;\mu) \otimes S(m_L,m_R;\mu)\,.
\end{align}
Here, $\sigma_0$ is the electroweak production cross section, and $H(Q^2)$ is the hard function that incorporates virtual corrections to $e^+e^-\to q\bar q$ at center of mass energy $Q$. These two components are independent of the measurement made on the jet. The radiation within the jets is described by the functions in the second line: $J(m_L)$ is the jet function that describes collinear radiation that contributes to the left hemisphere mass (and similarly for $J(m_R)$), and $S(m_L,m_R)$ is the soft function that describes the contribution of soft radiation (both perturbative and non-perturbative) to the hemisphere masses. A schematic picture of a single jet showing the radiation described by each of the functions is shown in \Fig{fig:mass_schematic}. For simplicity, in \Eq{eq:schematic_fact} we have not included an explicit function describing non-perturbative corrections. The QCD confinement scale $\Lambda_{\text{QCD}}\sim 1$ GeV is included in \Fig{fig:mass_schematic} to remind the reader that non-perturbative effects associated with hadronization can play an important role, and we will discuss them in \Sec{sec:non_pert}. The $\otimes$ symbol denotes convolution over the various contributions to a hemisphere mass.  Each of these functions can be given field theoretic definitions in terms of matrix elements of operators in the effective theory.   

Sudakov double logarithms are resummed by the renormalization group evolution of the different functions
\begin{align}\label{eq:RG_linear}
\mu \frac{d}{d\mu} F(m;\mu)=\int dm' \gamma_F(m-m';\mu) F(m';\mu)\,,
\end{align}
where here $F$ denotes any of the functions appearing in \Eq{eq:schematic_fact}.
The boundary conditions of the renormalization group evolution can be chosen so that large logarithms are not present in their perturbative expansion. All logarithms are then generated by the renormalization group evolution. For the case of the jet mass, these scales are indicated in \Fig{eq:schematic_fact}. Variation of these scales provides an estimate of the perturbative uncertainty, giving rise to the uncertainty bands illustrated in \Fig{fig:structure_calculation}. To achieve a complete resummation of all logarithms, each function appearing in the factorization theorem of \Eq{eq:schematic_fact} can only depend on  a single scale.

Factorization theorems of this structure do not necessarily exist for more differential jet substructure observables. A large amount of the interest in observables discussed in this article is on the resummation of large logarithms that do not naturally fit into such a factorization theorem, or that require a refactorization of the functions appearing in \Eq{eq:schematic_fact} to ensure that each function depends only on a single scale. An example of this that will be discussed further in \Sec{sec:ngls} is the case when $m_L \ll m_R$, or $m_R \ll m_L$, so that the soft function depends on two disparate scales, and the factorization of \Eq{eq:schematic_fact} no longer resums all large logarithms.

\subsubsection{Fixed-Order Corrections}

While resummation plays an essential role in jet substructure calculations, it does not provide a complete description of the perturbative cross section. Indeed, while resummation is appropriate when $\log(p_{TJ}/m_J)\sim 1/\alpha_s$, for a range of masses when $\log(p_{TJ}/m_J)\sim 1$ it is no longer appropriate. In particular, it includes a tower of terms which are no longer enhanced in the resummed cross section, but does not include terms that do not involve logarithms. One must therefore merge a resummed calculation with a standard fixed order calculation to provide an accurate description of the distribution throughout the entire phase space. This is particularly important to get correct endpoint behavior of distributions, as well as to decrease perturbative uncertainties in the transition region between resummation and fixed order.  These terms that are important at higher masses appear in the $D(m_J, \alpha_s)$ term of Eq.~\ref{eq:xsecexpress}.

 Unlike resummation, where the large logarithms at higher orders can be predicted using the universal infrared structure of QCD, fixed-order corrections at higher orders in $\alpha_s$ are not universal, and are typically difficult to compute, particularly when a measurement is made on the final state radiation. In particular, with jet measurements, the phase space integrals often cannot be computed analytically, and therefore must be performed using a fixed-order Monte Carlo program.  A variety of fixed-order programs exist, both for hadron colliders, such as MCFM8~\cite{Campbell:1999ah, Campbell:2010ff, Campbell:2015qma, Boughezal:2016wmq}, NLOJet++ \cite{Nagy:1997yn,Nagy:1998bb,Nagy:2001xb,Nagy:2001fj,Nagy:2003tz}, and NNLOJet \cite{Ridder:2015dxa,Ridder:2016rzm}, as well as for $e^+e^-$ collisions such as EVENT2 \cite{Catani:1996vz,Catani:1996jh}, EERAD3 \cite{Gehrmann-DeRidder:2007nzq}, and CoLoRFulNNLO \cite{DelDuca:2016csb,DelDuca:2016ily,Tulipant:2017ybb,Kardos:2018kth}.  Fixed-order contributions to several traditional jet substructure observables have been studied in Refs.~\cite{Field:2012rw,Larkoski:2015uaa}.


Higher-loop fixed-order calculations relevant for jet substructure are particularly intensive due to the fact that one is typically interested in jet substructure observables that are only first non-zero with several emissions within the jet. NLO amplitudes for hadron colliders with two partons within  a jet are available; for example, $W/Z/H+2$ jets \cite{Bern:1996ka,Bern:1997sc,Berger:2006sh, Badger:2007si, Glover:2008ffa, Dixon:2009uk, Badger:2009hw, Badger:2009vh}. To extend this to NNLO, to match the precision available for $e^+e^-$ observables \cite{Garland:2002ak,GehrmannDeRidder:2007hr,GehrmannDeRidder:2007jk,Gehrmann-DeRidder:2007nzq,Weinzierl:2008iv,DelDuca:2016ily}, will require the NNLO calculation of five-point scattering amplitudes. Significant recent progress is being made on the fixed-order calculations of relevant and related amplitudes \cite{Badger:2013yda,Gehrmann:2015bfy,Dunbar:2016aux,Badger:2017jhb,Abreu:2017hqn,Chawdhry:2018awn,Abreu:2018jgq,Badger:2018enw,Abreu:2018zmy,Abreu:2018aqd}. While jet substructure studies have mostly focused on the resummation of logarithmically enhanced terms, as the precision increases and  as higher-order calculations become available, fixed-order corrections will play a more important role.

\subsubsection{Non-Perturbative Corrections}\label{sec:non_pert}

Finally, we have so far been discussing perturbative calculations, namely calculations at the level of quarks and gluons. A complete calculation which can be compared with an experimental measurement must also take into account non-perturbative effects, such as hadronization. While we have focused on IRC safe observables, for which non-perturbative effects are not required to give a finite result, this does not imply that such effects give a numerically small contribution.  Indeed one can show by considering a single low-energy emission at the scale $\Lambda_\text{QCD}$ off of the hard core, for $m_J^2 \lesssim p_{T J}\Lambda_{\text{QCD}}$, non-perturbative effects dominate, and must be incorporated. For sufficiently inclusive, additive observables, they can be included using a shape function  \cite{Korchemsky:1999kt,Korchemsky:2000kp,Bosch:2004th,Hoang:2007vb,Ligeti:2008ac}, $F_{\text{np}}$, which is convolved with the perturbative distribution
\begin{align}
\frac{d \sigma_{\text{had}}}{dm_J} =\int d\epsilon  \frac{d \sigma_{\text{pert} } (m_J^2-p_{T J}\epsilon)}{dm_J}  F_{\text{np}}(\epsilon)\,.
\end{align}
The shape function is a non-perturbative object, and as such is not currently calculable from first principles.  Calculations in lattice QCD, a common tool to perform first principles non-perturbative calculations, are difficult due to the fact that the shape function is described by a matrix element of lightlike Wilson lines, and the lattice is formulated with Euclidean time.
Because of this situation, models for the shape function must be used.  A common model is \cite{Stewart:2014nna} 
\begin{align}\label{eq:shape_func}
F_{\text{np}}(\epsilon) =\frac{4\epsilon}{\Omega^2} e^{-2\epsilon/\Omega}\,.
\end{align}
Here, $\Omega$ is a mass scale on the order of $\Lambda_\text{QCD}$.  This integrates to unity and has first moment equal to $\Omega$.

The shape function can be expanded in moments, with higher moments suppressed by powers of $p_{TJ} \Lambda_{\text{QCD}}/m_J^2$. The leading contribution is a shift of the distribution and is determined by a universal non-perturbative parameter \cite{Akhoury:1995sp,Dokshitzer:1995zt} multiplied by a calculable, observable dependent number \cite{Dokshitzer:1995zt,Lee:2006fn,Lee:2007jr}. Hadron mass effects can break this universality, but can also be included \cite{Salam:2001bd,Mateu:2012nk}. An important benefit of all-orders factorization theorems is the ability to give definitions to non-perturbative contributions as matrix elements of field-theoretic operators. While these matrix elements cannot be evaluated analytically, this allows one to prove relations between non-perturbative contributions to different observables \cite{Lee:2006fn,Lee:2007jr,Mateu:2012nk}, and to understand their dependence on jet kinematics \cite{Stewart:2014nna}.

While non-perturbative effects have not received much attention in the jet substructure literature, they are in fact a limiting factor in improving the precision of jet substructure calculations. This is particularly important since many jet substructure discriminants are used experimentally at high purity; in other words, when cutting on small values of the observable, where non-perturbative effects play an important role. One interesting approach to reduce non-perturbative contributions is the use of jet grooming techniques, which will be discussed later.

\subsubsection{Extension to General Jet Observables}

Although we have focused in this section on the case of jet mass, the three features of a calculation that we have highlighted are much more general. Indeed, although many of the observables we will discuss in this review have a more involved structure than the jet mass, their complexity typically arises from added complications to one of the three components of the calculation considered above. 

Most importantly, in jet substructure, one often focuses on multiple measurements made on jets, or combinations of grooming procedures and measurements. In these cases, it is primarily the logarithmic structure of the observable that is complicated due to the presence of additional scales imposed by the measurements. Multiple possible distinct ratios of scales can then appear in the arguments of logarithms at each perturbative order. While more sophisticated techniques are required to perform the resummation, the principle is identical to the single-scale case. In this case, effective field theory techniques with degrees of freedom living at each of the scales, and the use of renormalization group evolution equations between the different scales has proven particularly powerful. In the presence of multiple scales it is important to carefully specify to which order different logarithms are resummed. This will play an important role in our discussion of non-global logarithms, grooming, and jet radius logarithms.

We have also in this section not discussed observables which are not IRC safe, which deserve a brief mention. Observables which are not IRC safe suffer from infrared and collinear divergences do not necessarily give a finite result at each order in perturbation theory. (Here we refer to a generic point in phase space. IRC safe observables may exhibit singularities at the boundaries of the phase space, as is the case for the jet mass, or at isolated points in the interior of the phase space \cite{Catani:1997xc}.)  Different examples of such behavior are known. These include the standard case, where a non-perturbative function is required to absorb the infrared or collinear singularities. For example for identified hadron production a fragmentation function is required to absorb the collinear singularities \cite{Collins:1981uw}. In this case, predictivity is maintained due to the universality of the collinear limit, allowing the fragmentation functions to be measured in data and applied to other processes (see, e.g., \cite{deFlorian:2007aj}).  However even in such cases, the same bullets discussed above still apply, and the resummation of large logarithmic contributions can still be achieved through renormalization group evolution, similar to the case of the perturbative factorization theorems we discussed. A more exotic scenario which arises in a variety of jet substructure observables, and will be discussed in more detail below, is Sudakov safety \cite{Larkoski:2013paa}. In this case, the observable can exhibit non-analytic behavior in $\alpha_s$, for example, $\sqrt{\alpha_s}$, so that it cannot be computed in a perturbative expansion in $\alpha_s$, but no non-perturbative functions are required to achieve a finite result. The study of Sudakov safe observables has, to this point, relied on the use of a formulation in terms of standard IRC safe observables to which the theoretical techniques discussed above do apply \cite{Larkoski:2015lea}.

\subsubsection{Parton Showers}

While we have focused primarily on analytic techniques in this section, and will continue to do so throughout this review, parton shower generators also play an extremely important role at the LHC, particularly for jet substructure. Much like analytic calculations, parton shower generators are themselves a large topic relying on a number of specialized techniques, for which many excellent reviews are available \cite{Buckley:2011ms,Skands:2011pf,Skands:2012ts,Seymour:2013ega,Gieseke:2013eva,Hoche:2014rga}. A large number of parton shower generators exist, each of which has its own advantages and disadvantages, and emphasizes different physics aspects. Examples include Pythia \cite{Sjostrand:2006za,Sjostrand:2007gs,Sjostrand:2014zea}, a $p_T$-ordered dipole shower; Vincia \cite{Giele:2007di,Giele:2011cb,Fischer:2016vfv}, Sherpa \cite{Gleisberg:2003xi,Gleisberg:2008ta}, Ariadne \cite{Pettersson:1988zu,Lonnblad:1992tz}, and DIRE \cite{Hoche:2015sya}, dipole-antenna showers; Herwig++/Herwig7 \cite{Marchesini:1991ch,Corcella:2000bw,Corcella:2002jc,Gieseke:2003rz,Bahr:2008pv,Bellm:2015jjp}, an angular-ordered dipole shower; Deductor \cite{Nagy:2007ty,Nagy:2008eq,Nagy:2008ns,Nagy:2012bt,Nagy:2014mqa,Nagy:2015hwa,Nagy:2017dxh}, which is based on a quantum density matrix; Geneva \cite{Alioli:2012fc,Alioli:2015toa} which is based on reweighting a parton shower to the results of an analytic resummation, and Whizard, which specializes in providing higher multiplicity matrix elements in the hard scattering, which are then showered \cite{Kilian:2007gr}. Instead of providing a detailed description, which is beyond the scope of this review, here we will review the extent to which parton shower generators capture the physics important for jet substructure, emphasizing again the three elements discussed above, namely the resummation of logarithmically enhanced contributions, fixed-order corrections in $\alpha_s$, and non-perturbative corrections.

\begin{figure}[t]
\begin{center}
\includegraphics[width=0.95\columnwidth]{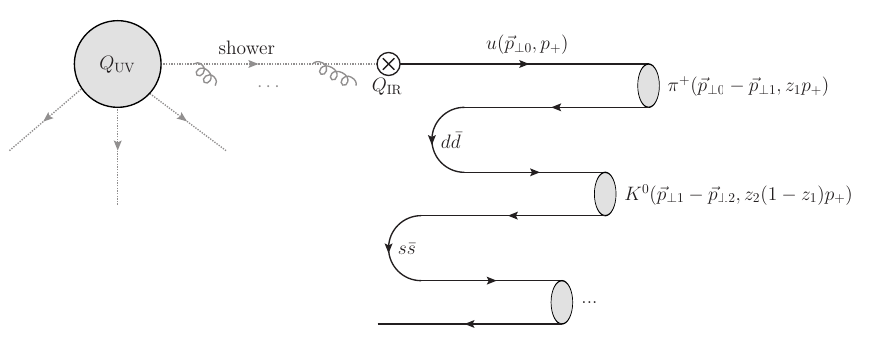} 
\end{center}
\vspace{-0.4cm}
\caption{A schematic of the factorization employed in a parton shower program into a hard scattering, denoted $Q_{\text{UV}}$, a shower describing perturbative soft and collinear emissions, and an exclusive hadronization model, describing the conversion into hadrons. Taken from  Ref.~\cite{Skands:2012ts}  } 
\label{fig:MC_fact}
\end{figure}

The goal of an event generator, an important component of which is the parton shower, is to provide a complete description of a general hard scattering process, to which a general measurement can been applied. This is in contrast to many analytic techniques, such as the factorization theorems discussed above, which are specific to a particular observable. A parton shower relies on the factorization of a process into a hard scattering, a perturbative shower, describing soft and collinear emissions, and a non-perturbative hadronization process into final-state hadrons. This is shown schematically in \Fig{fig:MC_fact}. This factorization is similar in spirit to that described in \Eq{eq:schematic_fact}, although it differs in that it does not factorize the shower into soft and collinear contributions separately. Furthermore, parton shower generators have the advantage that they generate fully exclusive events consisting of hadrons, allowing in principle arbitrary questions to be asked about the final state, well beyond what is possible in an analytic calculation. However, this generality comes at the loss of theoretical precision as compared to dedicated calculations, as well as an increased reliance on models. There is therefore a fruitful interplay between parton shower and analytic calculations for different jet substructure observables.

\begin{figure}[t]
\begin{center}
\includegraphics[width=0.85\columnwidth]{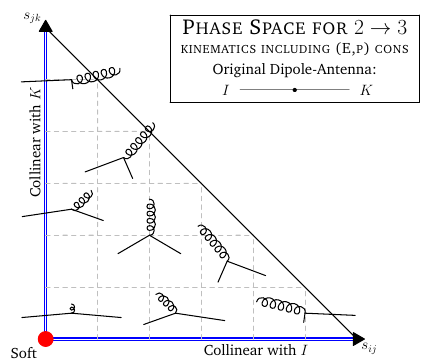} 
\end{center}
\vspace{-0.4cm}
\caption{The configuration of the final state partons in a parton shower based on a Markovian implementation of $2\to 3$ splitting. Approximations for the splitting kernel are made throughout the entire phase space, based on the soft and collinear limits.  Taken from Ref.~\cite{Giele:2011cb} } 
\label{fig:MC_split}
\end{figure}

The perturbative parton shower is based on a Markov chain implementation of parton splitting, using an approximation of the matrix element in the soft and collinear limits. This is typically implemented either via $1\to2$ splittings based on the QCD splitting functions augmented by soft coherence \cite{Catani:1990rr}, or using $2\to 3$ splittings.  The complete phase space for $2\to 3$ parton splitting is shown schematically in \Fig{fig:MC_split}, highlighting the different soft and collinear regions of phase space.  Either of these implementations reproduces the LL Sudakov factor of \Eq{eq:sudakov} for a generic observable. Through an appropriate choice of scheme for the strong coupling constant, this can be extended to NLL for many observables \cite{Catani:1990rr}. Parton showers also implement a number of corrections beyond a strict LL analytic calculation, for example, they implement exact momentum conservation, which is not included in an analytic calculation. A formalism for NNLL resummation based on $2\to 4$ splittings has been presented in Ref.~\cite{Li:2016yez}, which will allow for an improved accuracy in the description of the perturbative shower.  Other work toward NNLL resummation was presented in Ref.~\cite{Hoche:2017iem,Hoche:2017hno}, where the NLO splitting functions were implemented into a parton shower framework.

Higher fixed-order corrections to the hard scattering can be achieved through matching to higher-order perturbative calculations, which include both virtual and real corrections to the Born process. As in the case of analytic resummation, such corrections are essential to provide a complete description of the phase space, particularly away from regions that are dominated by soft and collinear emissions.  This is now well-established at NLO, and a variety of different techniques exist and are implemented into standard programs \cite{Catani:2001cc,Lonnblad:2001iq,Frixione:2002ik,Mrenna:2003if,Frixione:2003ei,Lavesson:2005xu,Mangano:2006rw,Frixione:2007vw,Frixione:2008ym}. There has also been significant recent progress in matching specific processes at NNLO \cite{Alioli:2012fc,Hamilton:2013fea,Hoeche:2014aia,Karlberg:2014qua,Hoche:2014dla,Hamilton:2015nsa}.

A major difference between analytic calculations and parton showers is in the description of hadronization. As discussed in \Sec{sec:non_pert} in analytic calculations of jet shape observables, due to the inclusive nature of the observables considered, hadronization can be captured by a shape function which smears the observable by $\mathcal{O}(\Lambda_{\text{QCD}})$. On the other hand, parton showers implement a fully exclusive model of hadronization, populating the final state phase space with hadrons. This hadronization process is described by sophisticated models, typically either the string \cite{Andersson:1983ia,Andersson:1998tv} or cluster model \cite{Webber:1983if,Marchesini:1987cf}, the parameters of which are tuned to data. See for example \Refs{Buckley:2011ms,Skands:2011pf,Skands:2012ts} for a more detailed discussion. For a jet shape-type observable, such as the jet mass, such models will be well-reproduced by the simple shape functions described earlier. However, a complete hadronization model allows for much more detailed questions, for example details of the flavor content of a jet, or the distribution of radiation between jets \cite{Bartel:1981kh,Althoff:1985wt,Aihara:1986hz,Sheldon:1986gy,Akrawy:1991ag}, to be modelled. The extent to which parton showers provide an accurate description of data often depends on the extent to which they have been tuned. We will highlight in \Sec{sec:help_others} particular measurements in jet substructure that may help improve parton shower descriptions of jets.

\subsection{Status of Jet Substructure Calculations}
\label{sec:moreobs}

Using the techniques discussed in the previous section, there has been extensive work to make first-principles predictions for distributions of jet substructure observables.  Because jet substructure observables are typically more exclusive than $e^+e^-$ collision event shapes, the phase space restrictions imposed by the observables can be quite complicated.  Even the simplest calculations provide significant insight into the structure and dominant physics of observables.

In this section, we survey some of the calculations for jet substructure that have been completed in recent years.  As this is a huge list, we restrict detailed discussion to calculations of observables that have been measured; either at the LHC or previous collider experiments.  A discussion of calculations of the (appropriately defined) jet mass will be presented in Section~\ref{sec:moreloops}.  Broadly, jet substructure observables can be classified by the jet topology to which they are sensitive.  This section will for the most part respect this organization.  We begin in Section \ref{sec:1prong} discussing observables that are sensitive to one-prong structure in jets; that is, are sensitive to radiation off of a single hard core in the jet.  In Section \ref{sec:2prong}, we discuss the calculation of jet observables that are sensitive to two-prong structure in the jet.  All observables in both of these classes that we discuss are constructed from IRC safe observables.  However, there are many very useful jet observables that are not IRC safe, some of which have been both calculated and measured.  We review these observables in Section \ref{sec:newstruct}.

\subsubsection{Calculations for One-Prong Jets}\label{sec:1prong}

For an observable to be sensitive to radiation off of a single hard core in the jet it is convenient to choose the observable to vanish if the jet has only one constituent, so that non-zero values of the observable probe the structure of radiation.  The simplest case of such observables are those that are non-zero with a single emission from the hard core. This greatly restricts the form of possible IRC safe observables, and almost all such observables can be broadly categorized as an angularity or an energy correlation function.  While the term wasn't used until recently, angularities are among the oldest IRC safe observables.  Thrust \cite{Farhi:1977sg} and broadening \cite{Rakow:1981qn,Ellis:1986ig} are both angularity-type observables.  Angularities in their modern form were developed in Refs.~\cite{Berger:2003iw,Almeida:2008yp,Ellis:2010rwa} and can be defined as
\begin{equation}
\tau^{(\alpha)} = \frac{1}{E_J}\sum_{i\in J} E_i \theta_{i\hat J}^\alpha\,.
\end{equation}
Here, this definition is appropriate for jets in $e^+e^-$ collisions where $E_J$ is the jet energy, $E_i$ is the energy of particle $i$ in the jet, and $\theta_{i\hat J}$ is the angle between particle $i$ and an appropriate axis of the jet.  For jets produced in $pp$ collisions, energies are changed to transverse momenta, and angles are changed to (pseudo)rapidity-azimuth distances.  The angular exponent $\alpha$ controls the sensitivity of the observable to collinear radiation; smaller $\alpha$ is more sensitive to collinear emissions.  For IRC safety, $\alpha>0$.  For small radius jets with the jet axis defined as the four-vector momentum sum of constituents (the momentum axis), the angularity with $\alpha = 2$ coincides with the jet thrust, and $\alpha = 1$ is the jet broadening.

Energy correlation function observables were introduced in Refs.~\cite{Banfi:2004yd,Jankowiak:2011qa,Larkoski:2013eya} as an alternative to angularities.  The two-point energy correlation function can be defined as
\begin{equation}
e_2^{(\alpha)} = \frac{1}{E_J^2}\sum_{i<j\in J} E_i E_j \theta_{ij}^\alpha\,,
\end{equation}
again, written in a form appropriate for jets in $e^+e^-$ collisions.  The sum runs over all distinct pairs of particles in the jet, and unlike the angularities, the factor $\theta_{ij}$ is the angle between particles $i$ and $j$ and does not reference an axis.  This has the advantage that the energy correlation functions by definition are insensitive to soft recoil effects, or ``recoil-free''.  We will discuss recoil sensitivity in the following.  Note that both the angularities and energy correlation functions are 0 if the jet has one constituent: either that constituent lies along the jet axis or there are no pairs of distinct particles in the jet.

\begin{figure}[t]
\begin{center}
\includegraphics[width=0.95\columnwidth]{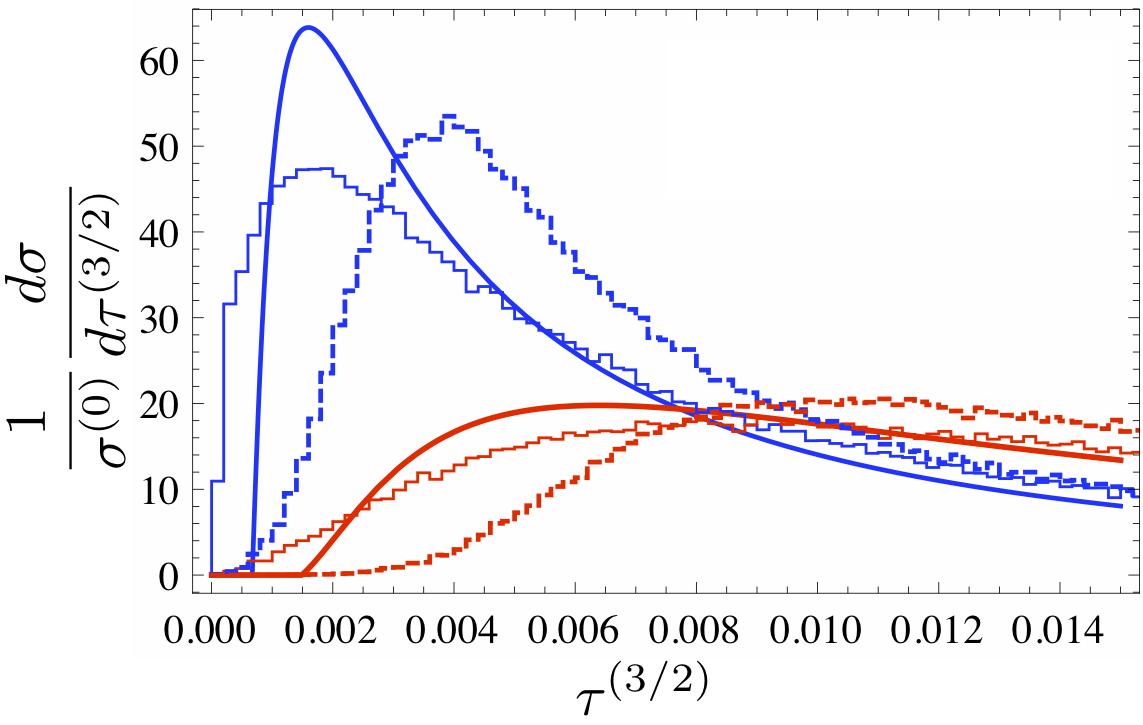} 
\end{center}
\vspace{-0.4cm}
\caption{Differential cross section of the angularity $\tau^{(3/2)}$ for both quark (blue) and gluon (red) jets produced in $e^+e^-$ collisions.  NLL analytic predictions are solid, and Pythia predictions are the histograms with parton level (solid) and hadron level (dashed).  Adapted from Ref.~\cite{Ellis:2010rwa}.} 
\label{fig:scetshapes}
\end{figure}

Angularities measured on individual jets produced in $e^+e^-$ collisions were first calculated to NLL accuracy in Ref.~\cite{Ellis:2010rwa}.  The authors developed a factorization formula for distinct angularity measurements on widely-separated jets, for various values of the angular exponent.  In Fig.~\ref{fig:scetshapes}, we show a plot from that paper in which they calculated the distribution of the angularity with $\alpha = 3/2$.  Their analytic calculation is solid, and simulation from Pythia is denoted by the histograms.  The solid histograms are parton level, while the dashed histograms include effects of hadronization.  Predictions for both quark (blue) and gluon (red) jets are plotted.

A limitation of the analysis of Ref.~\cite{Ellis:2010rwa} is that they ignored recoil and so could not calculate observables like broadening.  A recoil-sensitive observable is one for which the displacement of hard collinear radiation from the jet axis contributes to the observable at leading power in the soft limit.  The sensitivity to recoil for the angularities can be understood simply.  Consider a jet with two particles separated by an angle $\theta$, with one of the particles having an energy fraction $z$.  The angle of this particle to the momentum axis in the collinear limit is
\begin{equation}
\theta_{z\hat J}=(1-z)\theta\,.
\end{equation}
Then, the value of the angularity for this jet is
\begin{equation}
\tau^{(\alpha)} = z\left((1-z) \theta\right)^\alpha + (1-z)\left(z \theta\right)^\alpha\,.
\end{equation}
In the soft limit $z\to 0$, this simplifies to
\begin{equation}
\left.\tau^{(\alpha)}\right|_{z\to 0} = (z+z^\alpha)\theta^\alpha\,.
\end{equation}
Note that the $z^\alpha$ term comes from the contribution of the displacement of the harder particle from the jet axis.  When $\alpha > 1$, we can ignore this contribution in the $z\to 0$ limit, while that term dominates for $\alpha \leq 1$.  To calculate the angularities for $\alpha \leq 1$ requires correctly accounting for this recoil effect, which greatly complicates precision calculations.  Recently, groups have calculated the broadening observable in $e^+e^-$ collisions and fully included the effects of recoil \cite{Dokshitzer:1998kz,Chiu:2012ir,Becher:2012qc,Becher:2011pf}.

However, just including the effects of recoil may be undesirable, depending on your application.  For example, Ref.~\cite{Larkoski:2013eya} demonstrated that, from calculations to NLL accuracy, recoil-free observables like the energy correlation functions are more powerful quark versus gluon jet discriminants than their recoil-sensitive counterparts.  To have an observable that is maximally sensitive to the radiation off of the hard core in a jet, one wants to consider observables that are insensitive to recoil.  One option is to use the energy correlation functions.  Repeating the exercise above for a jet with two constituents for the energy correlation functions, one finds that there is a single contribution to the observable in the soft limit:
\begin{equation}
\left.e_2^{(\alpha)}\right|_{z\to 0} = z\theta^\alpha\,,
\end{equation}
with no recoil term.

Another option for eliminating recoil is to change the definition of the jet axis.  The momentum axis is one IRC safe choice for the jet axis, but there are others.  The momentum axis coincides with the axis which minimizes the value of thrust measured about it; thus it is sometimes referred to as the thrust axis.  One can correspondingly define a ``broadening axis'': the axis that minimizes the value of jet broadening calculated about it.  For a jet with two constituents, one can show that the broadening axis coincides with the direction of the hardest particle in the jet.  Therefore, recoil is eliminated because the harder particle is not displaced from the jet axis.  For jets with many particles, the broadening axis lies along the direction of the dominant energy flow.

\begin{figure}[t]
\begin{center}
\includegraphics[width=0.95\columnwidth]{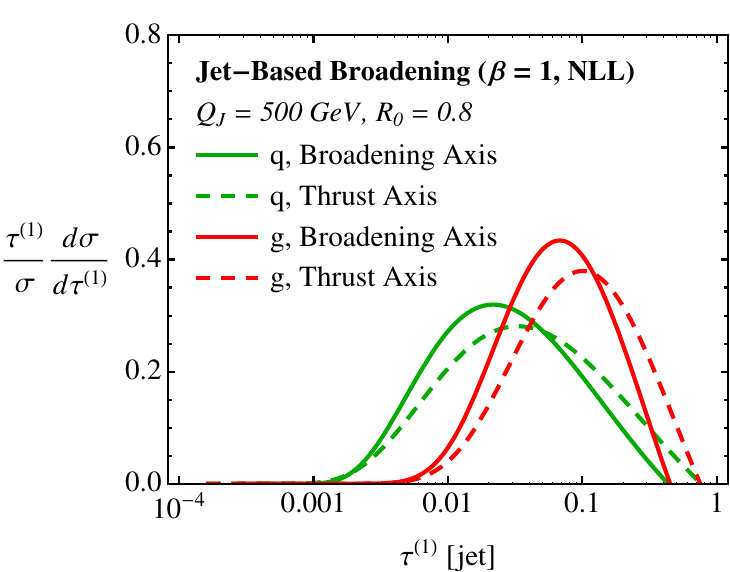} 
\end{center}
\vspace{-0.4cm}
\caption{Distribution of the jet broadening measured about the broadening (solid) and thrust (dashed) axes of the jet calculated at NLL accuracy.  The jets are either initiated by a quark (green) or a gluon (red).  Taken from Ref.~\cite{Larkoski:2014uqa}.} 
\label{fig:broadaxes}
\end{figure}

The jet broadening axis was introduced in Ref.~\cite{Larkoski:2014uqa}, and is similar to the spherocity axis \cite{Georgi:1977sf} for $e^+e^-$ events.  Ref.~\cite{Larkoski:2014uqa} demonstrated that resummation of angularities about the broadening axis was straightforward and presented results at NLL accuracy.  The calculation of the jet broadening ($\tau^{(1)}$) measured about both the thrust and broadening axes is shown in Fig.~\ref{fig:broadaxes}, for both quark and gluon jets.  As expected, the distribution for the jet broadening about the broadening axis lies at smaller values than about the thrust axis.  While the definition of the broadening axis is quite unwieldy, there is a jet algorithm recombination scheme that has been developed that directly clusters jets about their hard core \cite{salam_wta,Larkoski:2014uqa,Bertolini:2013iqa}.  When two particles are clustered, the winner-take-all (WTA) recombination scheme assigns their clustered direction to be along the harder of the two particles.  This is in contrast to standard $E$-scheme recombination \cite{Blazey:2000qt}, which assigns the clustered particle a direction determined by the vector sum of the particles' momenta.  Jet axes in the WTA scheme are formally identical to the broadening axis to leading power in the soft limit.

Calculations of angularities or angularity-like observables has recently been extended in several ways.  Generalized angularities that have both angular and energy weighting exponents were defined in Ref.~\cite{Larkoski:2014pca}.  While these observables are no longer in general IRC safe, their distribution can still be calculated at NLL accuracy in terms of a single non-perturbative moment.  Ref.~\cite{Hornig:2016ahz} extended definitions of angularities from $e^+e^-$ collisions to dijet events in $pp$ collisions.  Refs.~\cite{Larkoski:2014tva,Procura:2014cba,Procura:2018zpn} calculated the distributions of jets on which multiple angularities have been measured.  Measuring multiple angularities constrains the radiation in different ways, depending on the relative size of the angularities.

While it does not yield a probability distribution of an ensemble of jets, the jet shape observable \cite{Ellis:1990ek} is useful for understanding average radiation patterns from jets.  The jet shape is defined as the fraction of the jet's energy that is contained within a cone of radius $r$ about the jet axis.  Therefore, for every jet, the jet shape is a curve as a function of $r$ that vanishes at $r=0$ and equals 1 when $r=R$, the jet radius.  Calculations of the jet shape \cite{Seymour:1997kj,Li:2011hy,Larkoski:2012eh,Chien:2014nsa,Isaacson:2015fra,Hornig:2016ahz,Kang:2017mda} have focused on the curve averaged over an ensemble of jets, evaluated at fixed-order, or including some amount of resummation.  The first complete calculation resummed to NLL accuracy was only recently presented in \Ref{Cal:2019hjc}.  This averaged jet shape curve manifests several properties expected of jets; for example, the collinear singularity, the effective strengthening of the collinear singularity due to the running of $\alpha_s$, and that the radiation in gluon jets is, on average, at wider angles than for quark jets.  We will discuss the necessity of resummation of $r$ in the jet shape in Section \ref{sec:logr}.

\subsubsection{Calculations for Two-Prong Jets}\label{sec:2prong}

Observables that are sensitive to radiation off of multiple hard prongs in the jet can be defined as the natural generalizations of the angularities or energy correlation functions.  The generalization of thrust to multiple hard directions was started with triplicity and related observables \cite{Brandt:1978zm,Wu:1979iz,Nachtmann:1982nj} long ago, and much more recently with the observable called $N$-jettiness \cite{Stewart:2010tn}.  In $e^+e^-$ collisions, $N$-jettiness identifies $N$ axes and calculates the local jet thrust about each axis, and then sums all contributions.  If $N$-jettiness is small, then radiation in the event is localized about at most $N$ hard directions.  A generalization of this to subjets within jets was introduced in Refs.~\cite{Kim:2010uj,Thaler:2010tr,Thaler:2011gf} called $N$-subjettiness.  For a jet produced in $e^+e^-$ collisions, $N$-subjettiness is defined as
\begin{equation}
\tau_N^{(\alpha)} = \frac{1}{E_J}\sum_{i\in J}E_i \min\{
\theta_{i,1}^\alpha,\theta_{i,2}^\alpha,\dotsc,\theta_{i,N}^\alpha
\}\,.
\end{equation}
Here, the sum runs over all particles in the jet, and the angle is calculated from the axis that is closest to the particle $i$.  For $N=1$, $N$-subjettiness coincides with jet angularities, and the location of the subjet axes can be defined in any IRC safe way.

The generalization of the energy correlation functions to sensitivity to more prongs is less constrained, so here, we will present only the original definitions.  (Definitions of more general classes of energy correlation functions can be found in Refs.~\cite{Moult:2016cvt,Komiske:2017aww}, and applications to higher point substructure have been considered in \cite{Larkoski:2014zma}.)  For sensitivity to radiation off of two hard cores in a jet produced in $e^+e^-$ collisions, we define the three-point energy correlation function as \cite{Larkoski:2013eya}
\begin{equation}
e_3^{(\alpha)} = \frac{1}{E_J^3}\sum_{i<j<k\in J} E_iE_j E_k \theta_{ij}^\alpha \theta_{jk}^\alpha \theta_{ik}^\alpha\,.
\end{equation}
$N$-point energy correlation functions are defined similarly, and include products of the energies of $N$ distinct particles and ${N \choose 2}$ pairwise angles.  Both $N$-subjettiness and the energy correlation functions are IRC safe, for $\alpha > 0$.  Forms of these observables appropriate for jets in $pp$ collisions are found by the substitutions to $(p_T,\eta,\phi)$ coordinates.

Typically, the higher point subjettiness or the energy correlation functions are not useful directly in this form, however.  To discriminate jets with one hard prong versus two hard prongs of radiation, for example, requires comparing the value of different subjettinesses or energy correlation functions.  If, say, $\tau_1^{(\alpha)}$ is large while $\tau_2^{(\alpha)}$ is small, then radiation in the jet is localized about two hard directions; i.e., the jet is two-pronged.  This is efficiently encoded in the ratio of appropriate observables.  The optimal ratios for discrimination can be found by studying the scaling of various contributions of radiation to the jet in the soft and collinear limits.  The optimal ratio of $N$-subjettinesses is
\begin{equation}
\tau_{2,1}^{(\alpha)} = \frac{\tau_2^{(\alpha)}}{\tau_1^{(\alpha)}}\,,
\end{equation}
while for energy correlation functions, the optimal ratio is called $D_2^{(\alpha)}$ \cite{Larkoski:2014gra}:
\begin{equation}
D_2^{(\alpha)} = \frac{e_3^{(\alpha)}}{(e_2^{(\alpha)})^3}\,.
\end{equation}
Interestingly, while the individual $N$-subjettiness or energy correlation function observables are IRC safe, $\tau_{2,1}^{(\alpha)}$ and $D_2^{(\alpha)}$ are not IRC safe generically \cite{Soyez:2012hv}.  They are, however, Sudakov safe \cite{Larkoski:2013paa} and therefore still calculable in perturbation theory, a property we will discuss in the following section.  Here, we will consider these observables on jets for which a cut on the jet mass is made, which renders them IRC safe.

Ref.~\cite{Larkoski:2015kga} presented the first NLL accurate calculations of two-prong jet observables with the calculation of $D_2^{(\alpha)}$ for jets produced in $e^+e^-$ collisions.  Their calculation proceeded by factorization of the cross section into factors that described individual components of the two-prong jet radiation.  Because the jet mass and $D_2^{(\alpha)}$ were both measured on the jet, this defined multiple configurations of radiation within the jet, each with its own factorization formula.  One of these regions was first studied in Ref.~\cite{Bauer:2011uc}, and consists of a jet with two hard, relatively collinear prongs, surrounded by soft radiation. Another region which consists of one hard prong and one soft, wide angle prong was introduced in Ref.~\cite{Larkoski:2015zka} to resum non-global logarithms, which will be discussed in Section \ref{sec:moreloops}.

\begin{figure}[t]
\begin{center}
\includegraphics[width=0.95\columnwidth]{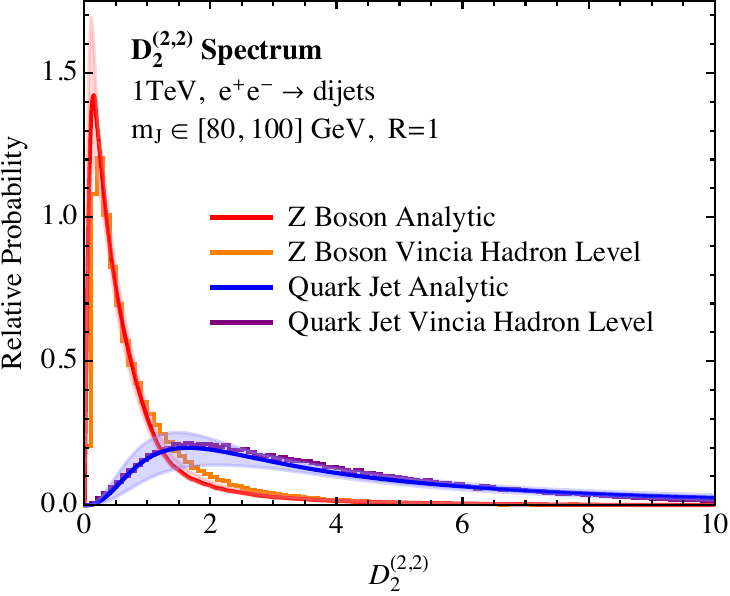} 
\end{center}
\vspace{-0.4cm}
\caption{NLL distributions of $D_2^{(2)}$ measured on quark jets (blue) and hadronically-decaying $Z$ bosons (red) in $e^+e^-$ collisions, compared to histograms from Vincia.  Shaded bands represent theoretical uncertainties.  Taken from Ref.~\cite{Larkoski:2015kga}.} 
\label{fig:D2}
\end{figure}

A plot of the $D_2^{(2)}$ distribution with a mass cut around the $Z$ boson mass from Ref.~\cite{Larkoski:2015kga} is shown in Fig.~\ref{fig:D2}.  Here, distributions measured on both quark jets (dominantly one-prong) and boosted, hadronically decaying $Z$ bosons (dominantly two-prong) produced in $e^+e^-$ collisions are shown.  Also plotted are the distributions as simulated in the Vincia parton shower at hadron level.  The factorization formula for the cross section enables a non-perturbative correction model to be easily included in the calculation.  $D_2^{(\alpha)}$ has been studied in data at the LHC during Run 2 \cite{atlasr2d2,Aad:2015rpa,ATLAS:2016wlr}. 

\begin{figure}[t]
\begin{center}
\includegraphics[width=0.95\columnwidth]{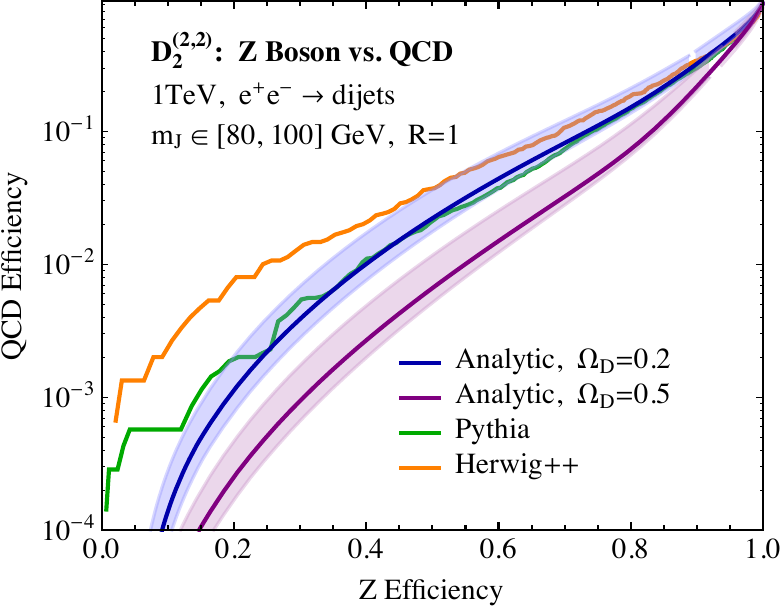} 
\end{center}
\vspace{-0.4cm}
\caption{The effect of varying non-perturbative parameters on the discrimination power of the $D_2^{(\alpha)}$ observable. This is reflected in large variations in the predicted efficiencies both in parton shower and analytic calculations. Taken from Ref.~\cite{Larkoski:2015kga}.} 
\label{fig:D2_shape}
\end{figure}

Demanding that the jet have a two-prong structure as defined by a measured value of $D_2^{(\alpha)}$ or $\tau_{1,2}^{(\alpha)}$ introduces new scale hierarchies into the jet.  For a jet of a given $p_T$, there is only a finite amount of phase space available for perturbative radiation to set the jet mass to be well above $\Lambda_\text{QCD}$.  Requiring that there are two well-defined, perturbative prongs in the jet further enforces that the scale of the prongs lies between the jet mass and $\Lambda_\text{QCD}$.  The perturbative phase space volume for two-prong structure in the jet is therefore significantly smaller than that for a perturbative mass.  Therefore, we expect that two-prong observables are much more sensitive to hadronization corrections than one-prong observables.  A demonstration of this sensitivity is illustrated in Fig.~\ref{fig:D2_shape}.  Here, a sliding cut in the $D_2^{(2)}$ observable is made and all jets to the left of the cut are kept, which sweeps out a signal versus background efficiency curve.  Analytic predictions with two different values of the non-perturbative scale are compared to the output of the Pythia and Herwig++ parton showers.  Both the large difference between the different analytic predictions and Pythia and Herwig++ is evidence that measuring and tuning non-perturbative parameters is vital for a good description of these observables that are sensitive to multiple prongs. 

The first resummed calculations of the $N$-subjettiness ratio $\tau_{2,1}^{(\alpha)}$ were presented in Refs.~\cite{Salam:2016yht,Dasgupta:2015lxh}.  Though the calculations were limited to LL accuracy, they are sufficient to inform differences between $N$-subjettiness and the energy correlation functions.  Unlike the energy correlation functions, $N$-subjettiness depends sensitively on how the axes are defined, and different axes can improve or degrade discrimination power.  However, resummation of $\tau_{2,1}^{(\alpha)}$ is most important when $\tau_2^{(\alpha)}\ll \tau_1^{(\alpha)}$, where the subjet axes are unambiguous and identical for any IRC safe definition.

The first calculation of groomed jet mass observables on two-prong jets was performed in Ref.~\cite{Dasgupta:2015yua}.  Ref.~\cite{Salam:2016yht} presented the first calculation of observables sensitive to two-prong structure on groomed jets.  Jet grooming is the systematic removal of soft, wide-angle radiation in the jet that likely came from contamination radiation, uncorrelated with the hard, final state radiation.  We will discuss the utility of jet groomers more in Section \ref{sec:jetmass}, but typically grooming is done to make the measurement and calculation more robust.  The approach of Ref.~\cite{Salam:2016yht} was to use jet grooming to tag two-prong jet substructure and discriminate from one-prong jets.  This observation led to the new definition of ratios of $N$-subjettiness observables with different amounts of grooming, referred to as dichroic ratios.

The jet groomer used in Ref.~\cite{Salam:2016yht} was the modified mass drop tagger (mMDT) groomer, introduced in Ref.~\cite{Dasgupta:2013ihk}.  The mMDT grooming algorithm first re-clusters the jet with the Cambridge/Aachen jet algorithm \cite{Dokshitzer:1997in,Wobisch:1998wt}, which orders emissions by their relative angle.  Beginning at the largest angle emissions, mMDT steps through the Cambridge/Aachen branchings and removes those branches that fail the requirement
\begin{equation}\label{eq:mmdtreq}
\frac{\min[p_{Ti},p_{Tj}]}{p_{Ti}+p_{Tj}}> \zcut
\end{equation}
Here, $i$ and $j$ are the two branches at a branching, $p_{Ti}$ is the $p_T$ of branch $i$, and $\zcut$ is the parameter of mMDT grooming.  Typically $\zcut \sim 0.1$.  The removal of branches stops when Eq.~\ref{eq:mmdtreq} is satisfied.  More details about jet grooming and other groomers will be discussed in Section \ref{sec:jetmass}.

\begin{figure}[t!]
\begin{center}
\includegraphics[width=0.95\columnwidth]{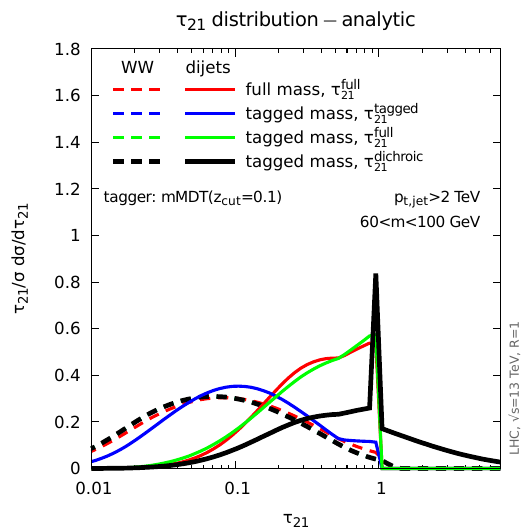} 
\end{center}
\vspace{-0.4cm}
\caption{LL distributions of $\tau_{1,2}^{(2)}$ measured on QCD jets (solid) and hadronically-decaying $W$ bosons (dashed), with various levels of jet grooming.  Taken from Ref.~\cite{Salam:2016yht}.} 
\label{fig:tau12}
\end{figure}

Plots of the LL predictions from Ref.~\cite{Salam:2016yht} of these $N$-subjettiness ratios are shown in Fig.~\ref{fig:tau12}.  Distributions for boosted, hadronically-decaying $W$-bosons are shown in dashed, while jets from QCD are solid.  The different colors represent $N$-subjettiness ratios with different amounts of grooming with mMDT. The black curve, for example, corresponds to making a cut on the mMDT jet mass around the $W$ mass, measuring $\tau_1^{(2)}$ on the mMDT jet, and $\tau_2^{(2)}$ on the full, ungroomed jet.  Such a configuration significantly reduces the overlap between one- and two-prong jets, improving discrimination power.  While groomed and ungroomed $N$-subjettiness calculations have since been extended \cite{Napoletano:2018ohv} and the observables have been measured at the LHC \cite{ATLAS:2012am,Aad:2013gja,Aad:2015rpa}, these calculations will have to be extended beyond LL accuracy for comparison with data.

\begin{figure}[t]
\begin{center}
\includegraphics[width=0.95\columnwidth]{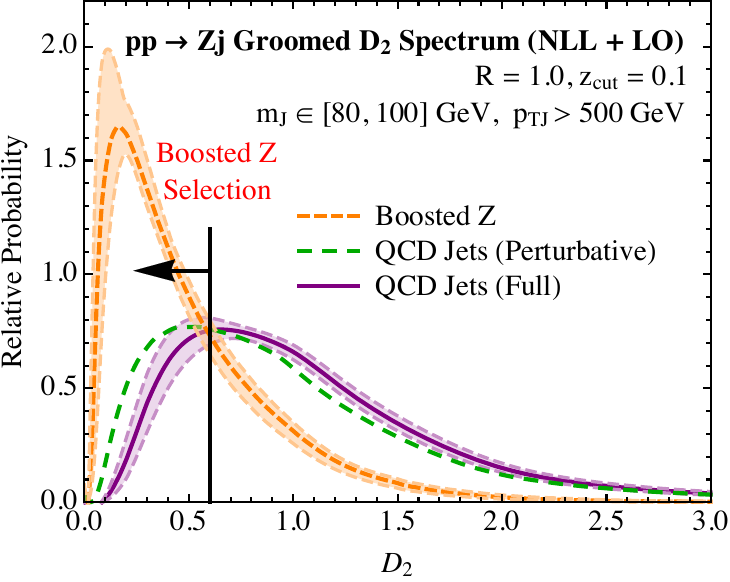} 
\end{center}
\vspace{-0.4cm}
\caption{NLL distributions of groomed $D_2^{(2)}$ in $pp\to Zj$. Shaded bands represent theoretical uncertainties.  Taken from Ref.~\cite{Larkoski:2017iuy}.} 
\label{fig:D2_pp}
\end{figure}

$D_2$ with mMDT grooming was first calculated on jets produced in $pp$ collisions in \Ref{Larkoski:2017iuy}, which included a careful treatment of both perturbative and non-perturbative contributions to the distribution. A plot of the distribution for both signal ($Z$) and QCD jets in $pp\to Zj$ events is shown in \Fig{fig:D2_pp}. It is also interesting to compare this with the distributions in Fig.~\ref{fig:D2}. The grooming procedure has a large effect on the distribution for QCD jets, but a much smaller effect on the distribution for $Z$ jets, as expected.

\begin{figure}[t]
\begin{center}
\includegraphics[width=0.95\columnwidth]{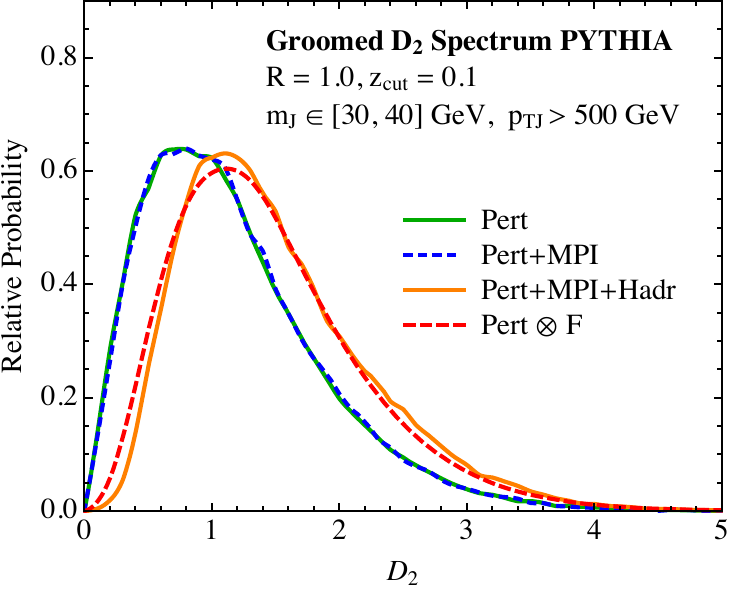} 
\end{center}
\vspace{-0.4cm}
\caption{Non-perturbative effects for groomed $D_2$. Contributions from the underlying event are completely negligible, and non-perturbative effects can be incorporated using a shape function, $F$, extracted from $e^+e^-$.  Taken from Ref.~\cite{Larkoski:2017iuy}.} 
\label{fig:np_shift}
\end{figure}

Of particular importance is the fact that non-perturbative effects are under theoretical control due to the grooming procedure. In particular, contributions from the underlying event are completely negligible, and non-perturbative effects from hadronization can be incorporated using a universal non-perturbative parameter, which is independent of whether the jet was initiated by a quark or gluon, is independent of the jet mass, and can be extracted from jets in $e^+e^-$ collisions. In \Fig{fig:np_shift} these properties are verified in parton shower Monte Carlo, and the final non-perturbative distribution is predicted using the shape function of \Eq{eq:shape_func} with a non-perturbative parameter extracted from simulated jets in $e^+e^-$ at the $Z$-pole. We hope that we will have a direct comparison between theoretical calculations and experimental measurements, as well as more detailed studied of non-perturbative effects for two-prong observables, in the near future.

There have been a few other two-prong jet calculations in the literature.  Refs.~\cite{Dasgupta:2015yua,Dasgupta:2016ktv} calculated jet mass distributions combined with jet taggers and groomers for hadronically-decaying $W$ bosons at LL accuracy.  By boosting the calculation of thrust, Ref.~\cite{Feige:2012vc} calculated the $\tau_{2,1}^{(2)}$ distribution for hadronically-decaying boosted $Z$ bosons at NNNLL accuracy.  There have also been studies of fixed-order corrections to two-prong jet substructure observables.  The LO and NLO distributions of $\tau_{2,1}^{(\alpha)}$ were calculated in Ref.~\cite{Larkoski:2015uaa} using NLOJet++ for jets in $e^+e^-$ collisions, for various $\alpha$ and several different axis choices.  This study demonstrated that distributions and discrimination power of $N$-subjettiness is highly sensitive to the choice of axes.

\subsubsection{Calculations for Three-Prong Jets}\label{sec:3prong}

Theoretical calculations for two-prong jet substructure are sufficiently involved that little progress has been made on the study of jets on which higher-point structure is resolved.  In particular, the resolution of three-prongs in a jet is relevant for distinguishing hadronically-decaying top quarks from QCD jets.  The first calculations for three-prong jets  was presented in \Ref{Field:2012rw} which calculated the jet planar flow \cite{Almeida:2008yp,Almeida:2008tp,Thaler:2008ju}, a measure of the aplanarity of the jet constituents, to LO.  Note that a jet with two constituents is necessarily planar (the momenta of the two particles lies in a plane), so this observable is first non-zero for jets with three constituents.  The first resummed calculations of three-prong jet observables was presented in \Ref{Dasgupta:2018emf}.  There, the authors studied the widely-used CMS \cite{CMS:2009lxa,CMS:2014fya}, Johns Hopkins \cite{Kaplan:2008ie}, and $Y$-splitter \cite{Butterworth:2002tt,Brooijmans:2008zza} top taggers.  A significant development in this work was the calculation of a new, IRC safe, top tagging variant of the CMS top tagger, and the incorporation of grooming into top tagging observables.  Nevertheless, progress is just being made in understanding these intricate jets, so there are still significant developments to be made.

\subsubsection{Calculations for New Structures and Probes}\label{sec:newstruct}

In this section, we review calculations of other structures or probes of jets, beyond those sensitive to pronged substructure.  Most of the focus of this section will be on IRC unsafe observables.  In particular, we will consider the charged hadron multiplicity, charged-particle jet distributions, and the collinear splitting function.  The observables chosen to be  discussed in detail are motivated by their utility and the existence of experimental measurements.  We will also briefly discuss fragmentation functions and recent efforts to define and calculate observables that count the number of IRC safe-defined subjets in a jet.

The charged particle multiplicity in a jet has a long theoretical history in QCD \cite{Bolzoni:2012ii,Dremin:1999ji,Capella:1999ms,Dremin:1994bj,Dremin:1993vq,Catani:1992tm,Catani:1991pm,Malaza:1985jd,Gaffney:1984yd,Malaza:1984vv,Mueller:1983cq,Konishi:1978yx,Brodsky:1976mg} and has been measured at nearly all collider experiments for the past several decades.  Particle multiplicity is not IRC safe, because an infinitesimally soft or exactly collinear splitting changes the multiplicity, inhibiting real and virtual divergences from canceling.  Nevertheless, the running of the multiplicity distribution in jet energy can be perturbatively calculated; more precisely, moments of the multiplicity distribution are perturbatively renormalized.  The mean charged particle multiplicity can be calculated for any jet energy given the input of one non-perturbative quantity: the mean multiplicity at one value of the jet energy.  Because of its long history, the charged particle multiplicity is one of the most precisely calculated jet observables, predicted at NNLL matched to NNNLO in Ref.~\cite{Bolzoni:2012ii}.

\begin{figure}[t]
\begin{center}
\includegraphics[width=0.95\columnwidth]{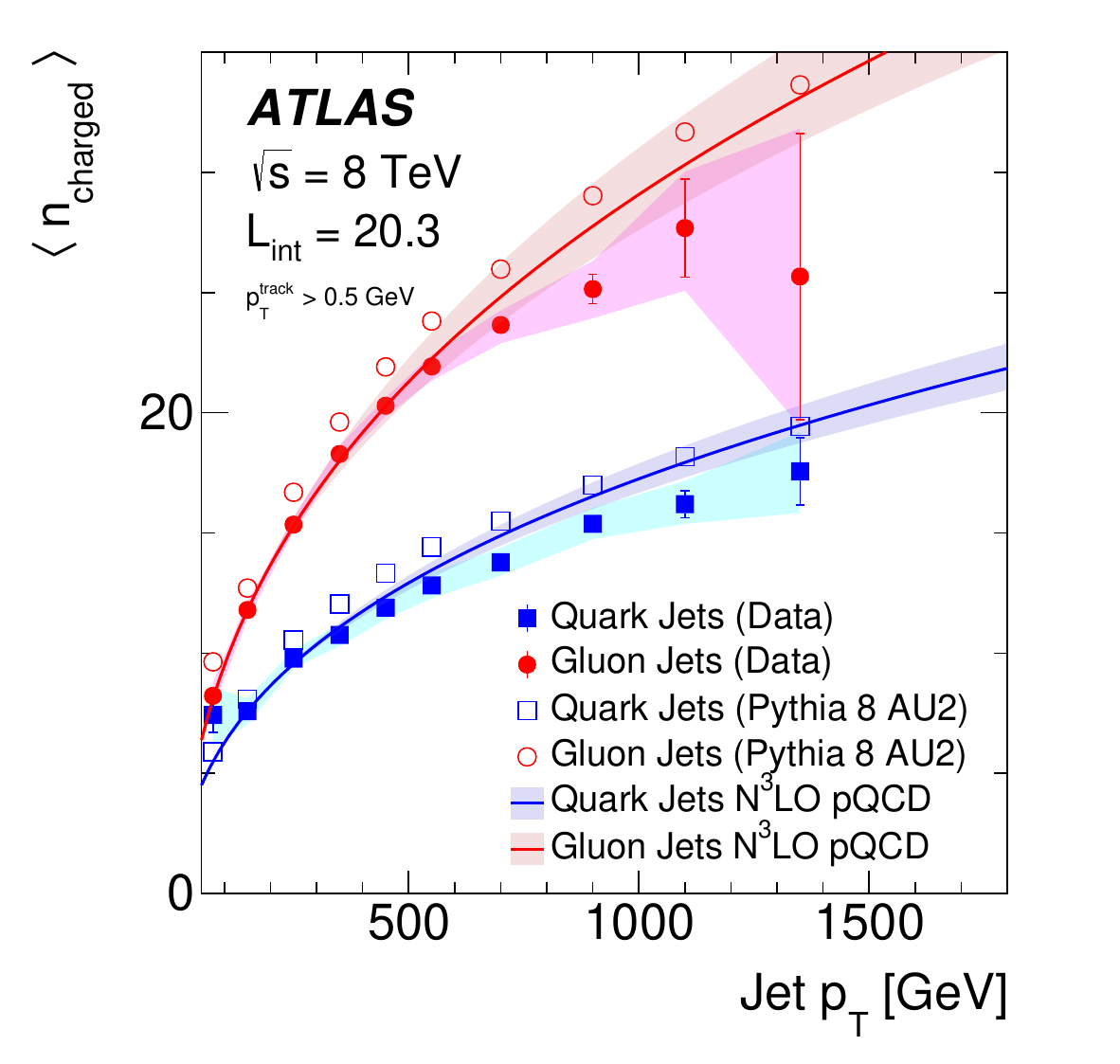} 
\end{center}
\vspace{-0.4cm}
\caption{Plot comparing the NNNLO prediction of Refs.~\cite{Capella:1999ms,Dremin:1999ji} (solid line) of quark (lower) and gluon (upper) jet mean charged particle multiplicities as a function of jet $p_T$ to the ATLAS measurement.  Taken from Ref.~\cite{Aad:2016oit}.
} 
\label{fig:partmult}
\end{figure}

The mean charged particle multiplicity has been measured at the LHC \cite{Aad:2011sc,Aad:2011gn,Chatrchyan:2012mec,Aad:2016oit}.  
%
In Fig.~\ref{fig:partmult} we show a plot from Ref.~\cite{Aad:2016oit} which compares a measurement of the mean quark and gluon multiplicity measured at ATLAS to the predictions from Refs.~\cite{Capella:1999ms,Dremin:1999ji}.  Within ATLAS, individual quark and gluon multiplicities were identified by separating the sample into low- and high-rapidity regions.  Quark and gluon jets have a different rapidity dependence and so can be separated if the rapidity dependence is known.  The ratio of the mean charged particle multiplicity between quark and gluon jets is seen to approach the expected asymptotic value of $C_A/C_F = 9/4$, though uncertainties are relatively large in the highest momentum bins.
Particle multiplicity is among the most powerful observables for discrimination of quark and gluon jets \cite{Gallicchio:2011xq,Gallicchio:2012ez,Larkoski:2014pca}, and further theoretical studies could provide understanding as to why it works so well and provide insight into the construction of new observables.

Another fundamental jet observable is its electric charge.  The jet charge was introduced in Ref.~\cite{Field:1977fa} as an energy weighted sum of the electric charges of the hadrons in the jet:
\begin{equation}
Q_{\kappa} = \sum_{i\in J} \left(
\frac{E_i}{E_J}
\right)^{\kappa} q_i\,,
\end{equation}
where $q_i$ is the electric charge of particle $i$.  The exponent $\kappa$ enables control over the sensitivity of the jet charge observable to soft particles in the jet, with $\kappa = 0$ the unweighted sum over particle charges and $\kappa = \infty$ the charge of the hardest particle in the jet.  The jet charge is not IRC safe, as the splitting of a soft gluon into quarks can change the charge of the jet.  Furthermore, the perturbative degrees of freedom have fractional charges, while measured hadrons have integer charges.  


The jet charge has been studied recently theoretically \cite{Krohn:2012fg,Waalewijn:2012sv} and measured at the LHC \cite{Aad:2015cua,Sirunyan:2017tyr}.  It is one of the more powerful probes for identifying the initiating quark flavor of a jet and discriminating the hadronic decays of $W$ and $Z$ bosons from one another~\cite{Krohn:2012fg,Aad:2015eax}.  As with multiplicity, only the evolution with energy of the jet charge can be calculated perturbatively; the jet charge distribution at a given energy is required non-perturbative input.  Additionally, the parameter $\kappa$ must be greater than 0 to ensure that the jet charge is infrared (soft) safe.  Then, charged parton evolution can be described by Altarelli-Parisi evolution of jet charge fragmentation functions.  These generalized fragmentation functions were defined in Refs.~\cite{Waalewijn:2012sv} and used to predict moments of the jet charge distribution, as a function of jet energy.

\begin{figure}[t]
\begin{center}
\includegraphics[width=0.95\columnwidth]{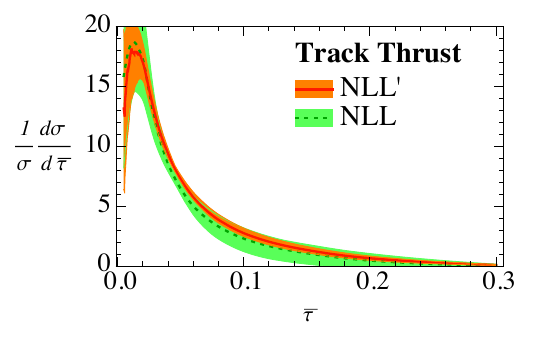} 
\end{center}
\vspace{-0.4cm}
\caption{NLL and NLL$'$ calculations of the charged-track thrust distribution for jets in $e^+e^-$ collisions.  Theoretical uncertainties are represented by the shaded bands. Taken from Ref.~\cite{Chang:2013iba}.} 
\label{fig:track_thrust}
\end{figure}

Working exclusively with charged particles has experimental advantages.  The angular resolution of charged particles is substantially better than the resolution of the calorimetry.  This enables the collision origin of charged particles to be uniquely identified, reducing effects of contamination from secondary proton collisions per bunch crossing.  Therefore, measuring more standard jet observables, like thrust, exclusively on charged particles can be experimentally beneficial.  This was studied in Refs.~\cite{Waalewijn:2012sv,Chang:2013rca,Chang:2013iba} which defined track functions which are fragmentation functions that follow charged particle production.  Ref.~\cite{Chang:2013iba} calculated the charged-track thrust observable on a jet, and a plot from that paper is shown in Fig.~\ref{fig:track_thrust}.  Their calculations include evolution of the track functions to NLL and NLL$'$ accuracy.  NLL$'$ accuracy includes the logarithms resummed at NLL, but also the pure ${\cal O}(\alpha_s)$ contribution (that contributes to the total cross section corresponding to the $C(\alpha_s)$ term in Eq.~\ref{eq:xsecexpress}).  This formally only contributes at NNLL accuracy, but by including it, theoretical uncertainties can be significantly reduced, as illustrated in Fig.~\ref{fig:track_thrust}.

As mentioned above, collinear parton evolution is governed by the Altarelli-Parisi splitting functions \cite{Gribov:1972ri,Dokshitzer:1977sg,Altarelli:1977zs}, which themselves cannot be directly measured in an IRC safe way.  While measuring the longitudinal momentum fraction eliminates soft singularities, collinear singularities are exposed.  Additionally, in a jet with many emissions of many particles, it is not immediately clear how to define the splitting that you want to measure.  The collinear splitting functions are a sensitive probe of fundamental interactions of partons and collective phenomena, and so a theoretical framework to predict and measure them is desirable.

Both of the issues discussed above have resolutions.  To identify a well-defined splitting of partons in the jet, we can exploit the mMDT groomer.  In its algorithm, mMDT orders particles in the jet by their relative angle, and removes those wide angle emissions that fail the hardness criteria.  The branching that passes the criteria can be defined to be the splitting of interest.   We then define the momentum sharing factor $z_g$ as the smallest momentum fraction in the branching that passes:
\begin{equation}
z_g=\frac{\min[p_{Ti},p_{Tj}]}{p_{Ti}+p_{Tj}}>\zcut\,,
\end{equation}
where $i$ and $j$ are the particles in the branching.  Note that because this passed the mMDT criteria and it is the softer emission, $\zcut<z_g<1/2$.  Here, $\zcut$ is the mMDT groomer parameter, and typically $\zcut$ is chosen to be about $0.1$.

To solve the collinear unsafety issue, there are two ways forward that have been identified.  First, we can measure another quantity that regulates the collinear divergence; for example, the jet mass.  For a non-zero groomed jet mass, the angle between the emissions that pass mMDT grooming must be non-zero, so there is no collinear singularity.  The region where the mass is small is suppressed by a Sudakov factor, and so when integrated over, the Sudakov factor exponentially suppresses the region where the particles in the branching become collinear.  This eliminates the collinear singularity, though at the expense of requiring the inclusion of all-orders effects in the form of a Sudakov factor.  Observables that are IRC unsafe but can be made calculable in perturbation theory in this manner are called Sudakov safe \cite{Larkoski:2013paa}.  Other examples of Sudakov safe observables include the two-prong ratio observables $\tau_{2,1}^{(\alpha)}$ and $D_2^{(\alpha)}$ when no mass cut on the jet is imposed.

\begin{figure}[t]
\begin{center}
\includegraphics[width=0.95\columnwidth]{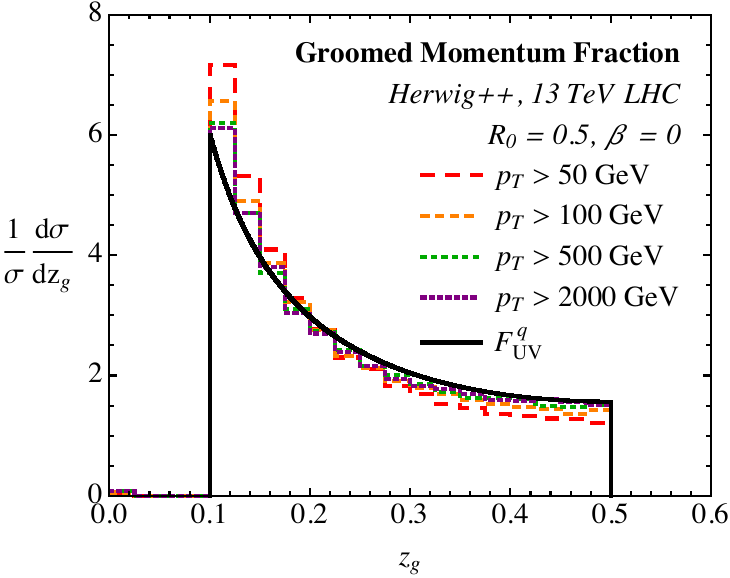} 
\end{center}
\vspace{-0.4cm}
\caption{Distribution of the momentum sharing $z_g$ observable in the ultraviolet limit $F_\text{UV}^q$ for the first splitting that passes the mMDT jet groomer, compared to simulation from Herwig++ for jets with a range of transverse momenta. Taken from Ref.~\cite{Larkoski:2015lea}.} 
\label{fig:zg_dist}
\end{figure}

For the momentum sharing observable $z_g$, Ref.~\cite{Larkoski:2015lea} introduced another method to calculate its distribution perturbatively.  Because the observable $z_g$ only has collinear singularities, its distribution can also be described by a generalized fragmentation function that describes the value of $z_g$ for a jet with one constituent (when there is no splitting).  Using the fragmentation function, one can calculate its perturbative evolution by demanding that the cross section is independent of the scale at which the fragmentation function is evaluated.  This method predicts that the $z_g$ flows to an ultraviolet fixed point, where its distribution is precisely given by the Altarelli-Parisi splitting functions.  A plot of this prediction is illustrated in Fig.~\ref{fig:zg_dist}, comparing the ultraviolet prediction against simulation from Herwig++ of jets with various transverse momenta.  As the transverse momenta of the jets increases, the Herwig++ simulation is observed to approach the expected ultraviolet fixed-point distribution.  This momentum sharing observable has been measured at the LHC \cite{CMS-PAS-HIN-16-006}, and we will discuss other applications in Section \ref{sec:newfronts}.

Recently, there have been efforts to define, in an IRC safe way, the number of subjets in a jet. These extend early studies counting jets in $e^+e^-$ events \cite{Catani:1991pm}.   Refs.~\cite{Bhattacherjee:2015psa,Sakaki:2018opq} defined an associated jet rate observable that counts the number of nearby jets that are produced in association with the jet of interested.  The authors developed a non-linear evolution equation and calculated this rate to LL accuracy.  This observable is useful for quark versus gluon jet discrimination because gluons will have more nearby jets than quarks due to their larger color charge.  Extending the work in Ref.~\cite{Bertolini:2013iqa} that introduced closed form expressions that closely reproduced jets found with traditional algorithms, Ref.~\cite{Bertolini:2015pka} calculated the fractional jet multiplicity.  The fractional jet multiplicity is defined as a weighted sum over the energies of particles in the event that effectively assign a probability that an emission is associated with a particular jet.  For well-separated emissions, this observable just counts the number of emissions and returns an integer.  For more ambiguous configurations of particles, the fractional jet multiplicity can be non-integral, and its distance from an integer is a measure of how well-defined the jets are.  A related idea was introduced and studied in \Ref{Neill:2018uqw}.  As the infrared region of phase space is approached in parton evolution, an effective measure of entropy of a jet increases because the volume of allowed phase space for emissions grows.  \Ref{Neill:2018uqw} calculated the entropy of a jet as a function of the parton shower ``time'' under which the multiplicity of jet increases.  Ref.~\cite{Frye:2017yrw} introduced a method to count subjets in a jet based on an iterated implementation of the soft drop jet groomer.  To LL accuracy, the distribution of this observable is Poissionian, with the mean values proportional to the color Casimir factors of the jet flavor.  Additionally, this counting observable can be shown to be equivalent to the likelihood ratio for quark versus gluon jet discrimination, and so is formally the best observable for this application to LL accuracy.

Finally, we discuss the fragmentation of partons into hadrons.  There is a long history of theoretical studies of fragmentation \cite{Georgi:1977mg,Ellis:1978ty,Collins:1981uw}, and it can be thought of as one of the first jet substructure observables, though it is not IRC safe.  Because QCD is a confining gauge theory, fragmentation is a fundamental issue in all collision experiments, and is, as mentioned in earlier sections, the ultimate limitation to any precision theoretical study.  The fragmentation process is sensitive to the medium in which it occurs, and can therefore be a powerful probe into the properties of the heavy ion medium versus the vacuum.   There has been significant recent theoretical effort to understand fragmentation of light partons within identified jets \cite{Procura:2009vm,Procura:2011aq,Chien:2015ctp,Arleo:2013tya,Dai:2016hzf,Kaufmann:2015hma,Elder:2017bkd,Makris:2017arq,Neill:2018wtk}, heavy quark fragmentation \cite{Kang:2017yde,Fickinger:2016rfd,Bodwin:2015iua,Fleming:2012wy,Bain:2017wvk,Ilten:2017rbd,Dai:2018ywt,lee:2019,Makris:2018npl}, and fragmentation in heavy ion collisions \cite{Kang:2016ofv,Kang:2017frl,Cacciari:2012mu}.

\begin{figure}[t]
\begin{center}
\includegraphics[width=0.95\columnwidth]{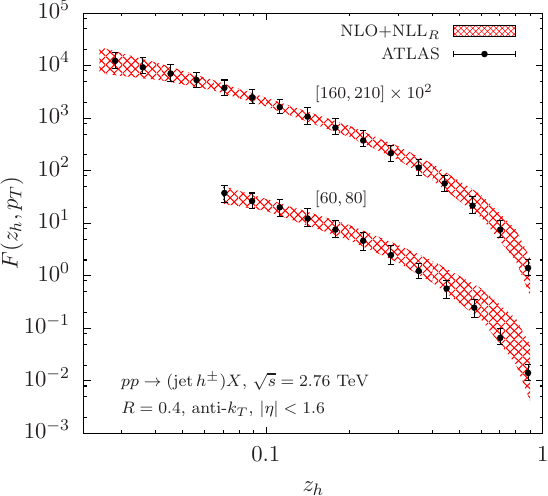} 
\end{center}
\vspace{-0.4cm}
\caption{Jet fragmentation function $F(z_h,p_T)$ calculated at NLO with NLL resummation of jet radius logarithms as a function of the hadron momentum fraction $z_h$.  The theory predictions with uncertainties are red hashed, while the data from ATLAS \cite{ATLAS:2015mla} are the black dots. Taken from Ref.~\cite{Kang:2016ehg}.} 
\label{fig:zh_dist}
\end{figure}

Recently, there has been work toward defining exclusive and semi-inclusive jet substructure in which the momentum of an individual hadron is identified in a jet \cite{Baumgart:2014upa,Bain:2016rrv,Bain:2016clc,Kang:2016ehg,Neill:2016vbi,Kang:2018qra,Liu:2018ktv}.  This has important applications to understanding hadro-production in high transverse momentum jets at the LHC.  Because some or all of the momentum components of a single hadron are measured, the corresponding observables are not IRC safe.  However, they can be formulated in the fragmentation function language,  and their perturbative evolution can be calculated.  Ref.~\cite{Kang:2016ehg} defined a semi-inclusive fragmenting jet function in which the longitudinal component of the momentum of a hadron is reconstructed.  They calculated the distribution of the hadron's momentum to NLO matched to NLL resummation of the jet radius, which enabled comparison to measurements that had been made at the LHC \cite{Aad:2011sc,Chatrchyan:2012gw,ATLAS:2015mla}.  A plot of the calculation of the jet fragmentation function from Ref.~\cite{Kang:2016ehg} compared to data from ATLAS is shown in \Fig{fig:zh_dist}.

\subsection{Toward Higher-Order Resummation}
\label{sec:moreloops}

Most of the observables discussed in the previous section were calculated to the lowest non-trivial order (either LL or LO), focusing on the understanding of jet structure from theoretical calculations.  Especially for observables that are sensitive to multiple hard prongs in the jet, even LL calculations can be significantly challenging due to the complicated phase space regions that must be considered.  Therefore, the push to higher theoretical precision begins with simpler observables, and uses what is learned there to extend to more complicated observables.

In this section, we will discuss efforts to push the calculation of jet substructure observables to high precision.  We start in Section \ref{sec:jetmass} with calculations of the jet mass.  In the regime where the jet mass is small compared to the energy or transverse momentum of the jet, the jet mass can suffer from non-global logarithms (NGLs), which are the leading manifestation of the correlation of in-jet and out-of-jet scales.  Despite their notorious difficulty to compute and control, recently there has been progress on understanding NGLs, which we review in Section \ref{sec:ngls}.  With the exceptional granularity of the LHC detectors and high-luminosity environment, it is often advantageous to find jets with radii that are sufficiently small so as to necessitate the resummation of logarithms of the jet radius, which will be reviewed in Section \ref{sec:logr}.

\subsubsection{Jet Mass} \label{sec:jetmass}

The simplest, interesting, IRC safe jet observable is its mass.  Starting with the analysis of Ref.~\cite{Catani:1991bd,Catani:1992ua}, the jet mass has been widely studied in $e^+e^-$ collisions.  Closely related to the jet mass is thrust which, in the soft and/or collinear limits, is equivalent to the sum of the masses of the two hemispheres in $e^+e^-\to$ hadrons events.  The state-of-the-art calculations for both thrust and heavy jet mass in $e^+e^-$ collisions are accurate to NNNLL+NNLO \cite{Becher:2008cf,Abbate:2010xh,Chien:2010kc}.

\begin{figure}
\begin{center}
\includegraphics[width=0.95\columnwidth]{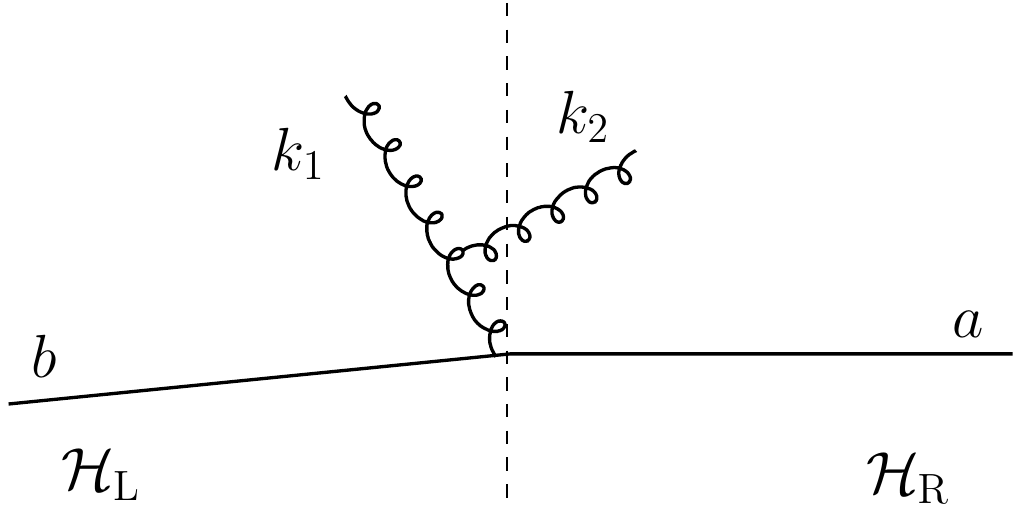} 
\end{center}
\vspace{-0.4cm}
\caption{Illustration of the leading emission configuration in $e^+e^-\to$ dijets events that produce NGLs.  Here, the right hemisphere has a mass less than that of the left hemisphere.  Taken from Ref.~\cite{Dasgupta:2001sh}.
} 
\label{fig:ngl_viz}
\end{figure}

To calculate the mass of jets produced in $pp$ collisions at the LHC involves many complications beyond those encountered in $e^+e^-$ collisions.  First, unlike the case in $e^+e^-$ collisions, just measuring the largest mass jet in the event does not control all possible large logarithms present in the cross section.  This was first observed in a study of the lightest hemisphere jet mass in $e^+e^-$ collisions \cite{Dasgupta:2001sh}.  For the light hemisphere mass, it is possible that a re-emission from the heavy hemisphere sets the light hemisphere mass.  This is illustrated in Fig.~\ref{fig:ngl_viz}.  The leading manifestation of this is called a non-global logarithm, which exists because the light hemisphere mass measurement does not globally constrain radiation in the event.  It is also exclusively a non-Abelian phenomena that would not exist if QCD were an Abelian gauge group.  Ref.~\cite{Dasgupta:2001sh} developed a Monte Carlo to resum these non-global logarithms and calculated the leading NGL for the light hemisphere mass $m_L$ to be
\begin{equation}\label{eq:ngl_2}
S_2=-C_F C_A\frac{\pi^2}{3}\left(
\frac{\alpha_s}{2\pi} \log\frac{Q^2}{m_L^2}
\right)^2\,.
\end{equation}
Here, $Q$ is the center-of-mass collision energy.  Note that this is proportional to the adjoint Casimir $C_A$, manifestly demonstrating its non-Abelian nature.  In general, NGLs will exist in the jet mass calculations in $pp$ collisions and must be accounted for somehow for a precision prediction.  

The global event environment in $pp$ collisions is much more active than in $e^+e^-$ collisions.  Because protons are not point particles, in every hard proton collision there are multiple secondary scatterings of partons within the protons referred to as the underlying event \cite{Affolder:2001xt}.  To first approximation, this underlying event is low-energy radiation distributed approximately uniformly in rapidity, although no field-theoretic definition exists yet. The extent to which the underlying event contributes to different observables is a subtle question, which despite significant progress is not completely understood \cite{Collins:1988ig,Gaunt:2014ska,Zeng:2015iba,Rothstein:2016bsq}. Including the effects of underlying event in a calculation therefore requires a model.

While more of an experimental issue, the high luminosity of the LHC also means that there are a significant number of secondary proton collisions per bunch crossing.  The pile-up radiation from these secondary proton collisions can deposit substantial amounts of energy in the event, greatly biasing potential jet measurements.  However, unlike underlying event, pile-up is truly uncorrelated with the hard scattering, and as such can in principle be completely removed from the event.  This is important for comparisons of data with precision calculations.

\begin{figure}[t]
\begin{center}
\includegraphics[width=0.95\columnwidth]{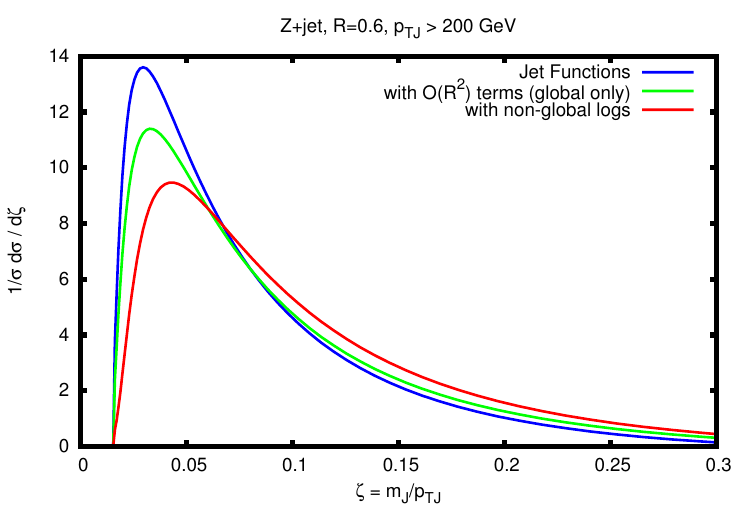} 
\end{center}
\vspace{-0.4cm}
\caption{Plot comparing the distribution of the ratio of jet mass to jet $p_T$ calculated to NLL accuracy in $pp\to Z+j$ events with (red) and without (green) the inclusion of non-global logarithms.  Taken from Ref.~\cite{Dasgupta:2012hg}.} 
\label{fig:sd_mass_ngl}
\end{figure}

Early precision calculations of the jet mass in $pp$ collisions \cite{Chien:2012ur,Dasgupta:2012hg,Jouttenus:2013hs} addressed the effects of NGLs and underlying event in different ways.  Ref.~\cite{Dasgupta:2012hg} computed the jet mass distribution in $pp\to Z+j$ events and included the leading effects of NGLs into their calculation directly.  Fig.~\ref{fig:sd_mass_ngl} illustrates their prediction for the jet mass distribution at NLL accuracy with and without the inclusion of NGLs.  The effect of NGLs is especially large near the peak of the distribution, where the ratio of the jet $p_T$ to the jet mass is the largest.  Additionally, no method for suppression of effects from underlying event was included.  The approach of Ref.~\cite{Jouttenus:2013hs}, by contrast, was to make a cut on the global observable $N$-jettiness \cite{Stewart:2010tn} in addition to calculating the jet mass in $pp\to H+j$ at NNLL accuracy.  The cut on $N$-jettiness suppresses effects of underlying event and greatly reduces NGLs, at the cost of restricting radiation throughout the event.

\begin{table*}[t]
  \centering
  \begin{tabular}{lccccc}\hline
             & highest logs  & transition(s) & Sudakov peak & NGLs & NP:
             $m^2 \lesssim$\\
\hline
    plain mass & $\as^n L^{2n\phantom{-1}}$ & ---  & $L\simeq
    1/\sqrt{\asbar}$ & yes & $\Lambda_\text{NP}\, p_t \,R$\\
\hline 
    trimming & $\as^n L^{2n\phantom{-1}}$ & $\zcut$, $ r^2 \zcut$ 
    &    $L\simeq 1/\sqrt{\asbar} - 2\ln r$ & yes & $\Lambda_\text{NP}\, p_t \,\Rsub$\\
    pruning & $\as^n L^{2n\phantom{-1}}$ & $\zcut$, $\zcut^2$    &
    $L\simeq 2.3/\sqrt{\asbar}$ & yes &  $\Lambda_\text{NP}\, p_t \,R$\\
    MDT     & $\as^n L^{2n-1}$ & $\zcut$, $\frac14\zcut^2$, $\zcut^3$
    & --- & yes& 
    $\Lambda_\text{NP}\, p_t \,R$
    \\
\hline 
    $\sanepruning$ & $\as^n L^{2n-1}$ & $\zcut$    & (Sudakov tail) &
    yes &  $\Lambda_\text{NP}\, p_t \,R$\\
    mMDT    & $\as^n L^{n\phantom{2-1}}$  & $\zcut$               &
    --- & no & $\Lambda_\text{NP}^2/\zcut$\\
\hline
  \end{tabular}
  \caption{
    %
    Table summarizing the results from Ref.~\cite{Dasgupta:2013ihk}.  The largest logarithms for plain mass, trimmed, pruned, MDT, and mMDT groomed masses are listed, where $L=\log p_{TJ}^2/m_J^2$.  NGLs only do not exist for the mMDT mass distribution and non-perturbative corrections are suppressed by two powers of the QCD scale $\Lambda_\text{NP}$.  Adapted from Ref.~\cite{Dasgupta:2013ihk}.
  }
  \label{tab:summary}
\end{table*}

Instead of suppressing radiation globally throughout the event, jet grooming techniques have been introduced that directly reduce the effects of underlying event, pile-up, and other contamination radiation within jets.
Jet groomers are algorithms that remove radiation from an identified jet that is likely not from the final state.  Among the early jet groomers that have seen wide use experimentally are filtering \cite{Butterworth:2008iy}, trimming \cite{Krohn:2009th}, and pruning \cite{Ellis:2009me}.  While the details of the algorithms differ, their broad function is the same.  Contamination radiation, from underlying event, initial-state radiation, or pile-up, is most likely relatively low $p_T$ with respect to the $p_T$ of the jet and relatively uniformly distributed over the jet area.  This is in contrast to final state radiation, which is collinear with the jet axis.  Jet groomers systematically identify soft, wide-angle radiation in a jet and remove it if it fails a threshold criteria.

The first theory calculations of the effects of jet groomers on the jet mass were presented in Refs.~\cite{Dasgupta:2013ihk,Dasgupta:2013via}.  These authors calculated the jet mass distribution on jets that had been groomed with trimming, pruning, and the mass drop tagger (MDT, developed in Ref.~\cite{Butterworth:2008iy}) to NLL accuracy.  In their analysis, they also estimated the contribution to the jet mass distribution from NGLs for each of the groomers studied.  Non-global radiation is expected to have properties similar to underlying event or initial state radiation, and be relatively low energy and approximately uniformly distributed over the jet.  Therefore, it may be expected that jet groomers would also remove NGLs from the jet mass distribution, and render the prediction more robust.

However, this is not what was found.  Each of the groomers studied retained NGLs in the mass distributions.  With an explicit calculation, the authors were able to identify how to construct a jet groomer that eliminated NGLs and defined the new algorithm called the modified mass drop tagger (mMDT).  The summary of their results is presented in Tab.~\ref{tab:summary}.  Of the groomers they studied, only the mMDT groomer successfully removed the effects of NGLs.  Shortly after this work, Ref.~\cite{Larkoski:2014wba} introduced the soft drop grooming algorithm as a generalization of the mMDT groomer.  The soft drop groomer proceeds similarly to mMDT described in Section \ref{sec:2prong}.  It starts by reordering emissions in the jet according to their relative angle using the Cambridge/Aachen jet algorithm \cite{Dokshitzer:1997in,Wobisch:1998wt}.  Starting with emissions at the widest angle, the procedure steps through the Cambridge/Aachen branching history removing those branches that fail the requirement
\begin{equation}\label{eq:sdreq}
\frac{\min[p_{Ti},p_{Tj}]}{p_{Ti}+p_{Tj}}> \zcut \left(
\frac{R_{ij}}{R}
\right)^\beta
\end{equation}
Here, $p_{Ti}$ is the $p_T$ of branch $i$, $R_{ij}$ is the angle between branches $i$ and $j$, $R$ is the jet radius, and $\zcut$ and $\beta$ are parameters of the soft drop grooming algorithm. The value $\beta = 0$ coincides with mMDT, and typically $\zcut \sim 0.1$.  The procedure terminates when a branching satisfies Eq.~\ref{eq:sdreq}.  Like mMDT, Ref.~\cite{Larkoski:2014wba} demonstrated that soft drop eliminates NGLs from the groomed jet mass distribution.

\begin{figure}
\begin{center}
\includegraphics[width=0.95\columnwidth]{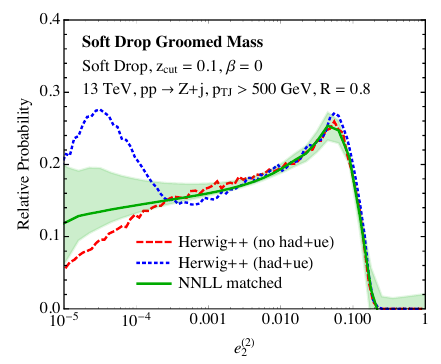} 
\end{center}
\vspace{-0.4cm}
\caption{The mMDT (soft drop with $\beta = 0$) jet mass distribution at NNLL accuracy matched to NLO fixed-order in $pp\to Z+j$ events compared to parton and hadron level Herwig++.  Here, $e_2^{(2)}=m_J^2/p_{T}^2$.  Taken from Ref.~\cite{Frye:2016aiz}.
} 
\label{fig:sd_mass}
\end{figure}

By identifying jet groomers that remove NGLs, precision resummed predictions for jet substructure observables  could then be made.  Using the factorization and resummation framework of SCET, Refs.~\cite{Frye:2016okc,Frye:2016aiz} demonstrated that the mMDT or soft drop groomed jet mass could be calculated to formally any logarithmic accuracy, with no NGLs of the jet mass present in the distribution.  This was demonstrated with predictions for the soft drop groomed jet mass in $pp\to Z+j$ events performed to NNLL accuracy, matched to NLO fixed-order predictions from MCFM \cite{Campbell:2002tg,Campbell:2003hd}.  Fig.~\ref{fig:sd_mass} illustrates the prediction of the mMDT groomed jet mass compared with the output of Herwig++.  Theoretical uncertainties are illustrated by the green band, and there is good agreement between the parton shower and the theory predictions.  The disagreement at small values of the mass is due to non-perturbative contributions to the mass from the effects of hadronization.

\Ref{Marzani:2017mva,Marzani:2017kqd} presented calculations that complement those in \Refs{Frye:2016okc,Frye:2016aiz}.  While only performing the resummation to NLL accuracy, \Ref{Marzani:2017mva,Marzani:2017kqd} included effects from the finite value of the soft drop parameter $\zcut$.  These predictions have also seen significant interest from experiment, and  already, both ATLAS and CMS have measured mMDT/soft drop groomed jet mass \cite{Aaboud:2017qwh,Sirunyan:2018xdh}.  As they are unfolded measurements, they were directly compared to the theoretical predictions from both \Refs{Frye:2016okc,Frye:2016aiz} and \Ref{Marzani:2017mva,Marzani:2017kqd} with agreement within uncertainties. Detailed comparisons can be found in the experimental references.  Further advancements in precision groomed jet mass calculations were presented in \Refs{Kang:2018jwa,Kang:2018vgn}.  These references introduced an additional factorization of the cross section calculations from \Refs{Frye:2016okc,Frye:2016aiz} that enabled resummation of the jet radius and grooming parameter $z_\text{cut}$.  Additionally, \Ref{Kang:2018vgn} incorporated the ingredients for NNLL resummation of the groomed mass, as well as NNLL resummation of general jet angularities.  This requires two-loop anomalous dimensions for soft functions with various angular dependencies, which were presented in \Ref{Bell:2018vaa,Bell:2018gce}.  While not completed early enough to be included in the first experimental studies, these references compared their calculations to data, and found good agreement.

As theoretical and experimental techniques improve in the future, these comparisons will enable quantitative and detailed studies of jet properties.  Further precision jet substructure predictions will utilize mMDT, soft drop, or related grooming techniques to eliminate the effects of NGLs and other contamination.  Some promising directions will be discussed in Sec.~\ref{sec:newfronts}.


\subsubsection{Non-Global Logarithms} \label{sec:ngls}

NGLs are not currently under sufficient theoretical control, and therefore are ideally removed for precision data-theory comparison. However, for a variety of observables sensitive to soft physics, which are important as probes of color flow, hadronization, and underlying event, NGLs are unavoidable, and are therefore important to understand theoretically. Furthermore, NGLs manifest fascinating emergent phenomena of the non-Abelian nature of QCD. There has therefore been significant theoretical work in attempting to understand their structure.  Shortly after their identification, Dasgupta and Salam showed in Monte Carlo simulation that radiation was suppressed near the boundary of jets due to non-global effects \cite{Dasgupta:2002bw}.  This so-called buffer region reduces the dependence of jet observables on the explicit shape of the jet boundary.  Shortly after this, Banfi, Marchesini, and Smye (BMS) developed an evolution equation \cite{Banfi:2002hw} that describes NGLs in the leading number of colors $N_c$ and leading-logarithmic approximations.  Since then, there has been substantial work devoted to understanding the nature of NGLs at fixed-order and their all-orders description within the BMS evolution equation \cite{Dasgupta:2002dc,Appleby:2002ke,Weigert:2003mm,Rubin:2010fc,Banfi:2010pa,Kelley:2011tj,Hornig:2011iu,Hornig:2011tg,Kelley:2011aa,Kelley:2012kj,Hatta:2013iba,Khelifa-Kerfa:2015mma,Hagiwara:2015bia}.  
An observable which is directly sensitive to non-global correlations has also been proposed and calculated \cite{Larkoski:2015npa}.
NGLs also have connections to factorization-violating effects in jet veto cross sections \cite{Forshaw:2006fk,Forshaw:2008cq,Forshaw:2009fz,DuranDelgado:2011tp,Martinez:2018ffw}, and this direction remains an active area of research.

The BMS equation for NGLs in hemisphere mass measurements in $e^+e^-\to $ hadrons events can be expressed as
\begin{equation}\label{eq:BMS}
\partial_L g_{n\bar n} = \int_\text{heavy} \frac{d\Omega_j}{4\pi} W_{n\bar n}^j [U_{n\bar nj}g_{nj}g_{j\bar n}-g_{n\bar n}]\,.
\end{equation}
Here, $L$ is the NGL multiplied by color and coupling factors:
\begin{equation}
L=\frac{N_c \alpha_s}{\pi} \log\frac{m_H}{m_L}\,,
\end{equation}
and $m_H$ ($m_L$) is the heavy (light) hemisphere mass.  $g_{n\bar n}$ is the all-orders expression for the leading-color and LL NGLs from the dipole formed from the light-like directions $n$ and $\bar n$ in the light and heavy hemisphere, respectively.  The angular integral extends over light-like vectors $n_j$ that lie in the heavy hemisphere, and $W_{n\bar n}^j$ is the matrix element for eikonal emission:
\begin{equation}
W_{n\bar n}^j = \frac{n\cdot\bar n}{(n\cdot n_j)(n_j\cdot \bar n)}\,.
\end{equation}
The factor $U_{n\bar nj}$ represents the resummed virtual contributions
\begin{equation}
U_{n\bar nj} = \exp\left[
L\int_\text{light}\frac{d\Omega_q}{4\pi}\left(
W_{n\bar n}^q-W_{nj}^q-W_{j\bar n}^q
\right)
\right]\,,
\end{equation}
where the integral is over the light hemisphere, and is responsible for the buffer region near the hemisphere boundary.  As a non-linear integro-differential equation, the BMS equation has not been solved exactly.  With the boundary condition $g_{n\bar n}(L=0)=1$, the term generated at order $\alpha_s^2$ in the BMS equation agrees exactly with Eq.~\ref{eq:ngl_2}.   The non-linearity of the BMS equation should be contrasted with the linear renormalization (Callan-Symanzik) evolution in Eq.~\ref{eq:RG_linear}.  At the very least, the non-linearity suggests that one needs to think about logarithmic resummation of NGLs in a very different way than for standard, global logarithms.
 
\begin{figure}
\begin{center}
\includegraphics[width=0.95\columnwidth]{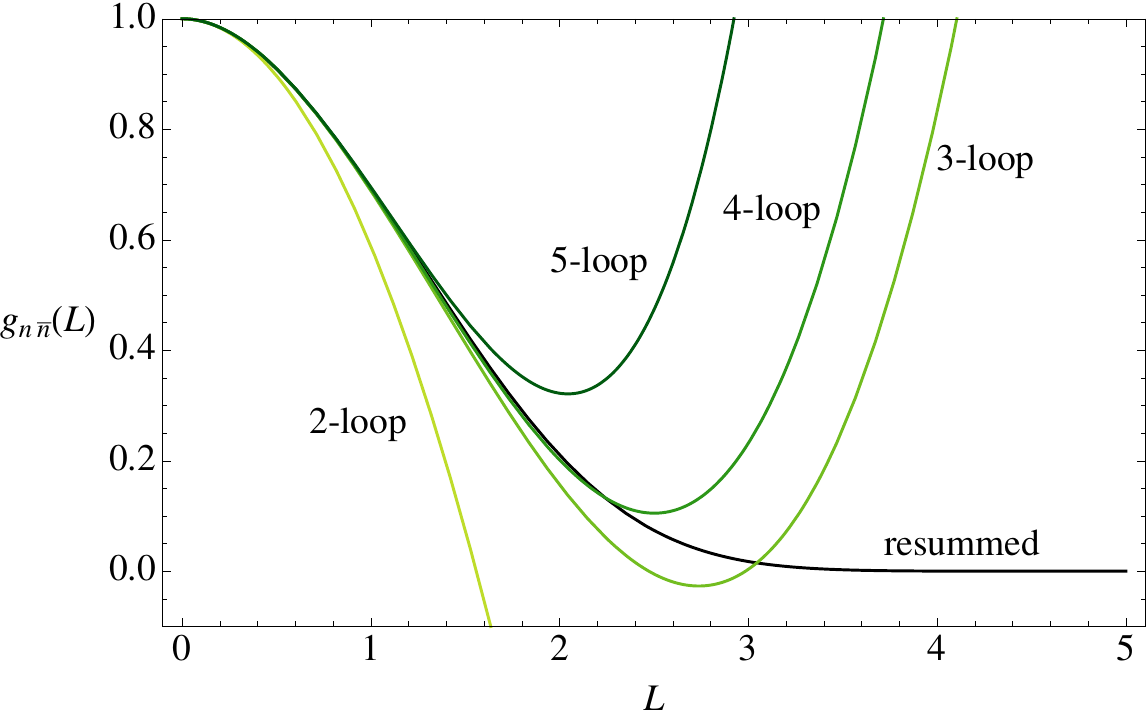} 
\end{center}
\vspace{-0.4cm}
\caption{Non-global logarithms of the ratio of the heavy to the light hemisphere masses $m_H/m_L$ in $e^+e^-\to$ hadrons collisions calculated through 5-loop order compared the numerical evaluation of the BMS equation.   Taken from Ref.~\cite{Schwartz:2014wha}.
} 
\label{fig:ngl_5loop}
\end{figure}

Recently, there have been efforts by several groups to develop systematic all-orders descriptions of NGLs.  \Ref{Schwartz:2014wha} provided the first explicit high-orders calculations of NGLs, by expanding the BMS equation to five-loop order.  To calculate the corresponding NGLs to this order, the authors exploited techniques of the modern amplitudes program, expressing the result in terms of generalized polylogarithm functions \cite{2001math3059G,2002math8144G,2011arXiv1105.2076G}.  Their work provided the first concrete evidence that the fixed-order expansion of NGLs has a finite radius of convergence.  This is in contrast to global logarithms, which can be resummed to all orders into the form of Eq.~\ref{eq:cum}.  Their results for the light hemisphere mass in $e^+e^-\to$ hadrons collisions up through 5 loops is shown in Fig.~\ref{fig:ngl_5loop}.  That the 5 loop result seems to deviate from the numerical solution of the BMS equation before the 4 loop result suggests that the radius of convergence of NGLs is about 1.  Now, there exist 12-loop results for NGLs in hemisphere mass distributions that provide further support for this observation \cite{SimonBMS}, as well as arguments based on the analytic structure of the BMS equation \cite{Larkoski:2016zzc}.

\Ref{Caron-Huot:2015bja} developed an extension to the BMS equation to NLL accuracy by exploiting the relationship of the BMS equation to high-energy small-$x$ parton evolution \cite{Weigert:2003mm,Hatta:2008st}.  This relationship is exact for conformal theories, and suggests a general form for NGL evolution to all orders and including all color effects.  Shortly after \Ref{Caron-Huot:2015bja}, \Ref{Larkoski:2015zka} developed a systematic approximation approach to NGLs called the dressed gluon expansion.  The dressed gluons are defined by all-orders factorization theorems for the production of multiple subjets in the heavy hemisphere and resum all NGLs down to an unresolved infrared scale.  This work was extended in Refs.~\cite{Neill:2015nya,Larkoski:2016zzc,Neill:2016stq,Neill:2018yet} and Ref.~\cite{Larkoski:2016zzc} proved that in the leading color and LL approximations, the dressed gluon expansion corresponds to an iterative solution of the BMS equation.  Refs.~\cite{Becher:2015hka,Becher:2016mmh,Becher:2016omr} put forward a different approach also based on SCET. Rather than introducing an infrared resolution scale, they wrote factorization formulae directly for non-global observables. In these formulae, the relevant low-energy physics is encoded in Wilson lines along the directions of the energetic particles and the solution of the renormalization group equations satisfied by these multi-Wilson-line operators resum all logarithmically enhanced contributions, including NGLs. 
In particular, they have applied their framework to various definitions of jet mass and out-of-jet energy \cite{Becher:2016omr,Balsiger:2018ezi,Balsiger:2019tne} and also to the narrow broadening event shape \cite{Becher:2017nof}, which suffers from both NGLs and rapidity logarithms.


As these works demonstrate, there is still a lot to understand about NGLs, and their calculation and effect on collider physics will remain and active area of research into the future.

\subsubsection{Resumming Logarithms of the Jet Radius} \label{sec:logr}

Motivated both by the fine granularity of the ATLAS and CMS detectors as well as the high luminosity collision environment, it can be advantageous to use a small jet radius.  The effects of contamination on a jet scale like the area of the jet, so contamination can be reduced by finding jets with a small radius $R$.  However, by restricting radiation in the jet to be collimated within an angle $R\ll 1$, this produces potentially large logarithms of the jet radius, $\log 1/R$, that appear in the cross section.  For precision calculations and controlled perturbation theory of jet rates with small radii, these logarithms in general need to be resummed.  The resummation and calculation of jet radius logarithms has been an active area of research \cite{Seymour:1997kj,Gerwick:2012fw,Banfi:2012jm,Becher:2013xia,Stewart:2013faa,Catani:2013oma,Alioli:2013hba,Dasgupta:2014yra,Chien:2015cka,Dasgupta:2016bnd,Kolodrubetz:2016dzb,Idilbi:2016hoa,Dai:2016hzf,Kang:2017mda,Cal:2019hjc}.

\begin{figure}
\begin{center}
\includegraphics[width=0.95\columnwidth]{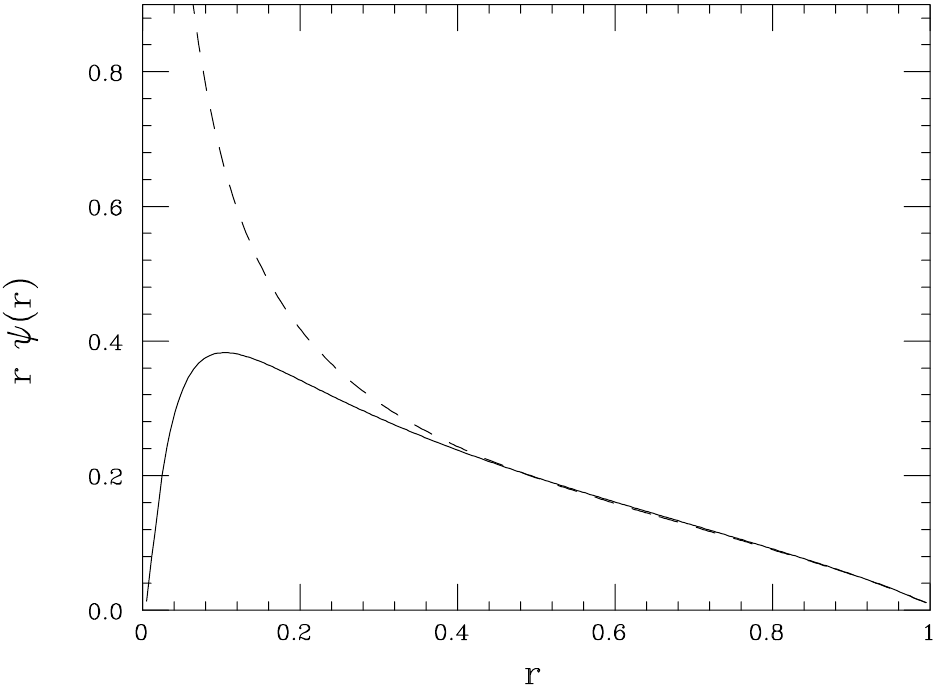} 
\end{center}
\vspace{-0.4cm}
\caption{
Comparison of the differential jet shape for 50 GeV $k_T$ jets at LO (dashed) compared to LL resummed (solid) as a function of the subjet cone radius $r$.  Taken from Ref.~\cite{Seymour:1997kj}.
} 
\label{fig:shape_resum}
\end{figure}

Standard jet radii used to find jets at the LHC are about $R=0.4-0.6$, which is not small enough to produce large logarithms.  However, depending on the application, effective jet radius logarithms can be large.  One of the first studies of resummation of jet radii was presented in Ref.~\cite{Seymour:1997kj} in the calculation of the jet shape \cite{Ellis:1990ek}.  The jet shape $\Psi(r)$ is the fraction of the radius $R$ jet's energy (or $p_T$) that is localized within a cone of radius $r$ about the jet axis.  As $r$ becomes small, radiation in the jet is highly restricted, resulting in large logarithms.  Fig.~\ref{fig:shape_resum} compares the differential jet shape $\psi(r)$ (the derivative of $\Psi(r)$) for $k_T$ jets \cite{Catani:1993hr} calculated at LO to including LL resummation of the logarithms of $r$.  Large deviations between the fixed order and resummed calculations are observed below $r \simeq 0.2$, demonstrating the necessity of resummation in this region.

\begin{figure}
\begin{center}
\includegraphics[width=0.95\columnwidth]{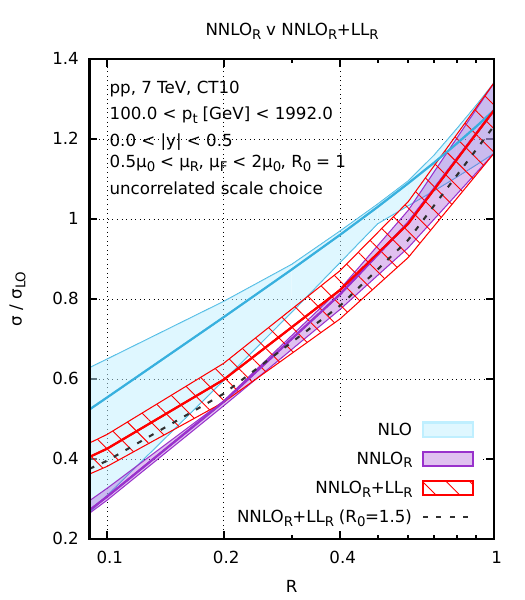} 
\end{center}
\vspace{-0.4cm}
\caption{
Comparison of the jet cross section as a function of radius $R$ in $pp\to jj$ events calculated at NLO, approximate NNLO (NNLO$_\text{R}$), and matched to LL resummation.  Taken from Ref.~\cite{Dasgupta:2016bnd}.
} 
\label{fig:logr_resum}
\end{figure}

Recently, there have been a few groups that have developed methods to resum logarithms of the jet radius systematically, and improve jet cross section predictions.  \Ref{Dasgupta:2014yra} developed a fragmentation function formalism for resumming the leading logarithms in the rate for production of small-radius jets.  The fragmentation function tracks the rate for new jets with small radius to be produced, as the resolution scale flows from high to low energies.  Unlike the tower of logarithms discussed in Sec.~\ref{sec:calcs}, LL accuracy for logarithms of the jet radius (denoted as LL$_\text{R}$) consists of the tower of terms that have a form
\begin{equation}
\text{LL}_\text{R} = \sum_n c_n \left(\alpha_s \log \frac{1}{R^2}\right)^n\,,
\end{equation}
where the $c_n$ are numerical coefficients.  This method was applied to calculate the jet rate in $pp\to$ dijets events at the LHC in \Ref{Dasgupta:2016bnd}.   Fig.~\ref{fig:logr_resum} shows a result from that study, comparing the jet rate  as a function of the radius $R$ at NLO, approximate NNLO (NNLO$_\text{R}$), and NNLO$_\text{R}$ matched to LL$_\text{R}$ resummation.  Similar to that observed in Fig.~\ref{fig:shape_resum}, resummation of the jet radius becomes important below about $R \lesssim 0.2$.  

Resummation of logarithms of the jet radius have also been approached using effective field theory techniques in Ref.~\cite{Chien:2015cka} (Related methods were used in Refs.~\cite{Becher:2015hka,Becher:2016mmh,Becher:2016omr}.).  These authors demonstrated factorization of modes that are sensitive to the boundary of the jet, and therefore sensitive to the jet radius.  Taking the results of the calculation of the two-loop soft function with an out-of-jet energy veto in $e^+e^-$ collisions from Ref.~\cite{vonManteuffel:2013vja}, the authors of Ref.~\cite{Chien:2015cka} explicitly demonstrated that their factorization theorem generates the correct logarithms of the jet radius at NNLO. However, they stressed that they would need to include multi-Wilson line operators to also account for all logarithms of the out-of-jet energy. Fitting the missing NNLO logarithm numerically, the found agreement with the analytic result of Ref.~\cite{Becher:2015hka}.  While these factorization methods have still only been applied and calculated in the context of $e^+e^-$ collisions, they are a promising direction for achieving NLL resummation of jet radius logarithms and beyond.

\subsection{New Frontiers}
\label{sec:newfronts}

While jet substructure is now a mature field, it continues to find interesting new applications, both experimentally and theoretically. In particular, with recent theory advances, we are now at a stage where we can begin to think about making precision measurements of Standard Model parameters with jet substructure. In this section we highlight several new frontiers in the field of jet substructure, many of which build upon the theory developments discussed in earlier sections of this review. We begin by highlighting several new applications of jet substructure to measurements at the LHC, focusing on ways in which techniques from jet substructure can be used to improve the theoretical robustness of these measurements. We then discuss how ideas developed in the context of jet substructure can have an impact in QCD theory more broadly. The role of Open Data in the future of jet substructure is then briefly discussed. Finally, we conclude this section with a wish list, discussing areas where theoretical progress is needed, and data-theory comparisons could soon be performed.

\subsubsection{New Applications of Jet Substructure}

Jet substructure has been used in a wide variety of applications at the LHC, from searches to measurements. In this section we highlight several new applications of ideas from jet substructure with an emphasis on how these techniques can be used to improve our theoretical understanding or the accuracy of our description of a particular physical system. We have chosen to focus on precision mass measurements using grooming, jet substructure as a probe of the heavy ion medium, applications of jet substructure to understanding heavy flavor, and the study of $\alpha_s$ in resummation-sensitive observables.\\

\noindent {\small{\sf Precision Mass Measurements With Grooming}}\\

In this section, we discuss the prospects for precision mass measurements with grooming. 
This is particularly interesting and most promising for the top quark. Masses are not physical quantities, but are instead scheme dependent parameters of the Standard Model Lagrangian. Due to its non-vanishing color charge, the top quark mass is particularly sensitive to QCD effects, and must be carefully defined. In standard approaches to top quark mass measurements parton shower simulation is heavily used to generate fully exclusive events, and it is difficult to provide a precise theoretical definition of the mass that is being extracted. 

In the case that the top quark decays inclusively into a jet whose mass is measured, a rigorous factorization theorem can be proven for the case of $e^+e^-\to$ dijets \cite{Fleming:2007xt,Fleming:2007qr}.  This enables one to express the final distribution in terms of Lagrangian parameters,  like the top quark mass, in a well-defined renormalization scheme.  An appropriate scheme can be chosen to remove renormalon ambiguities, increasing the theoretical precision \cite{Hoang:2008yj}. This factorization theorem includes all effects, both perturbative and non-perturbative. The mass obtained in a particular theoretically-motivated scheme can then be converted to the mass parameter of a parton shower simulation program using a conversion based on a comparison of predictions for a physical observable \cite{Hoang:2008xm,Butenschoen:2016lpz}. 

It is interesting to consider extending this approach to proton colliders, and considering the decay of a boosted top quark into a large $R$ jet. In attempting to do so, one finds that underlying event shifts the peak of the jet mass distribution by about 1 to 2 GeV. In other words, the observable is sensitive to underlying event at the level of the desired accuracy on the top quark mass. This is problematic due to the lack of theoretical understanding of the underlying event. Inspired by the development of grooming tools from jet substructure, one can ask whether groomers can be used to minimize effects from underlying event, allowing for a precise and robust top quark mass measurement at the LHC from jet substructure.

\begin{figure}[t]
\begin{center}
\includegraphics[width=0.95\columnwidth]{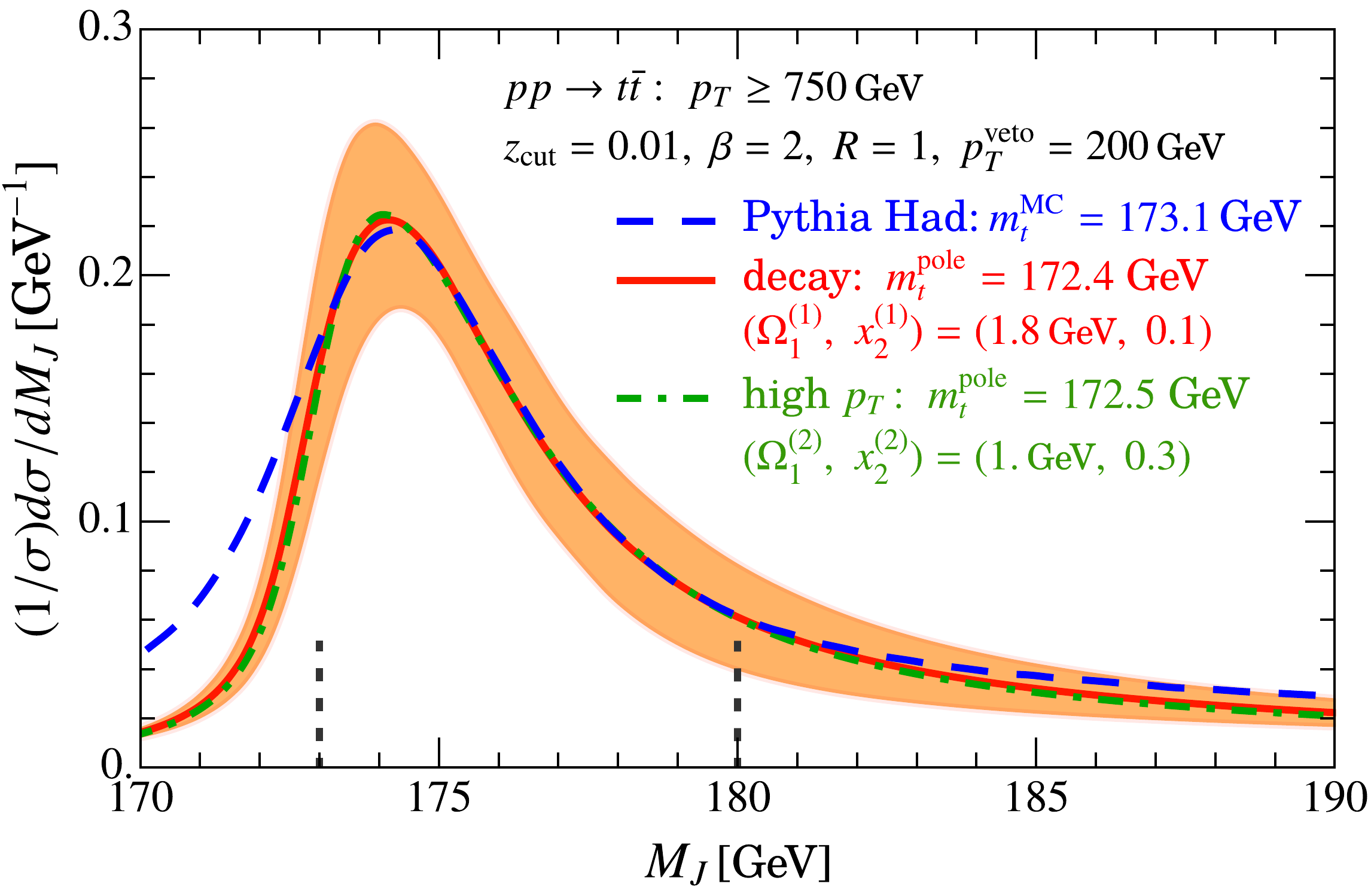} 
\end{center}
\vspace{-0.4cm}
\caption{
Comparison of factorization theorem predictions and Pythia of boosted top quark mass distribution near the peak region in $pp\to t\bar t$ events at the LHC.  The vertical dashed lines correspond to the fitting region for top quark mass extraction.  Taken from \Ref{Hoang:2017kmk}.
} 
\label{fig:top_mass}
\end{figure}

The top quark mass measured from jets on which soft drop grooming is performed has recently been studied theoretically in \Ref{Hoang:2017kmk}.  Extending Ref.~\cite{Frye:2016aiz}, those authors derived active-parton factorization theorems to describe the effects of soft drop on a boosted top quark.  The parameters of the soft drop groomer can be tuned appropriately so that contamination from underlying event is mitigated, but emissions at the scale of the width of the top quark are unaffected.  Depending on which emission in the boosted top quark jet passed the soft drop criteria, \Eq{eq:sdreq}, \Ref{Hoang:2017kmk} identified two factorization theorems, referred to as ``decay'' or ``high $p_T$''.  A strength of these factorization theorems is that because soft drop removes low-energy radiation in jets, the top quark jets decouple from the rest of the event.  This enables formal extraction of the top quark mass jet-by-jet, and not process-by-process, which significantly simplifies analysis.  \Fig{fig:top_mass} compares the boosted top quark jet mass distributions as predicted in the NLL factorization theorems of \Ref{Hoang:2017kmk} to Pythia parton shower simulation.  To account for hadronization effects, the factorization theorems are augmented with non-perturbative model functions, described by two parameters.  It should be possible to improve the precision of the factorization theorems to NNLL accuracy in the near future to produce precision predictions. While this is still in the early stages, and its feasibility must be studied experimentally, it shows a promising application of techniques from jet substructure to the measurement of fundamental parameters in the Standard Model.\\


\noindent {\small{\sf Jet Substructure as a Probe of the QCD Medium}}\\

The suppression of hadron and jet cross sections observed at RHIC \cite{Adcox:2001jp,Adler:2002xw,Adcox:2004mh,Arsene:2004fa,Back:2004je,Adams:2005dq} and the LHC \cite{Aad:2010bu,Aamodt:2010jd,Chatrchyan:2011sx,CMS:2012aa,Aad:2012vca,Aad:2014bxa,Adam:2015ewa} indicates a modification of jet evolution in the Quark Gluon Plasma (QGP). This is consistent with theoretical calculations, which indicate that an energetic parton propagating in a strongly-coupled medium will lose a significant fraction of its energy due to interactions with the medium prior to hadronization \cite{Gyulassy:1993hr,Baier:1996sk,Zakharov:1997uu,Gyulassy:2000er,Wiedemann:2000za,Arnold:2002ja}. The interaction of colored partons with the medium, and the corresponding modification of the QCD shower, therefore provides a powerful probe of the nature of the QGP. Recently there has been interest in using observables designed for jet substructure to probe jet evolution in the QGP, with the hope of better understanding the medium \cite{Salgado:2003rv,Armesto:2003jh,Jacobs:2005pk,Polosa:2006hb,Vitev:2008rz,Renk:2009hv,Chien:2015hda,Chien:2016led,He:2015pra,Cao:2016gvr,Mehtar-Tani:2016aco,KunnawalkamElayavalli:2017hxo,Milhano:2017nzm,Casalderrey-Solana:2016jvj,Casalderrey-Solana:2015vaa,Tachibana:2017syd,Chang:2017gkt,Li:2017wwc,Sievert:2018imd,Kang:2018wrs}. Such measurements typically probe specific aspects of the shower, due to their more differential nature, and therefore may allow for different effects to be disentangled.


As a particular example we highlight the observable $z_g$, introduced in \Sec{sec:newstruct}, which provides (in theory) a direct measure of the splitting function. Some models postulate that the vacuum splitting functions are modified in the QGP due to interactions with the medium. These medium-modifed splitting functions have been analytically calculated \cite{Ovanesyan:2011kn} using an extension of SCET \cite{Ovanesyan:2011xy,Kang:2014xsa} with a model of the medium as a background field. Using these medium-modified splitting functions combined with a model of the medium, calculations of the $z_g$ observable were performed in \Refs{Chien:2016led,Mehtar-Tani:2016aco}, and compared to LHC data \cite{CMS-PAS-HIN-16-006}. However, a better understanding of the systematics and unfolding is required to ensure that the data and theory comparison can be meaningfully interpreted. Nevertheless, due to the direct connection of the $z_g$ observable with the splitting function, it seems promising for understanding the nature of the QGP, and the modifications to splitting functions due to the presence of the medium.

A variety of parton shower programs are now available which take into account interactions with the medium \cite{Zapp:2013vla,Armesto:2009fj,Renk:2008pp,Schenke:2009gb,Zapp:2008gi,Kordell:2017hmi}.  More measurements of $z_g$ have been performed in several experiments \cite{Kauder:2017cvz,Kauder:2017mhg,Caffarri:2017bmh,Sirunyan:2017bsd} and recent parton shower studies have incorporated $z_g$ in analysis \cite{KunnawalkamElayavalli:2017hxo,Milhano:2017nzm,Chien:2018dfn}, indicating that this provides a promising probe of the medium. While particular tunes are able to reproduce certain measurements, a completely coherent picture has not yet emerged. It is hoped that the measurement and theoretical study of a range of jet substructure observables, each of which are designed to probe the shower in a particular way, as well as an understanding of the correlations between observables, may hold the key to obtaining a consistent theory of the QGP.  For a review of the broader applications of jet substructure to heavy ion physics see \Ref{Andrews:2018jcm}.\\

\noindent {\small{\sf Heavy Flavor, Fragmentation and Jet Substructure}}\\

One of the standard probes of QCD is identified hadron production, where we have classic factorization proofs, allowing for the theoretical description in terms of perturbatively calculable coefficients and non-perturbative fragmentation functions \cite{Georgi:1977mg,Ellis:1978ty,Collins:1981uw}. Recent theoretical advances for studying jets have allowed this to be extended to the measurement of identified hadrons within jets on which some other property of a jet, for example its mass, is measured. The required theoretical formalism was first presented in \cite{Procura:2009vm}, where a factorization in terms of ``fragmenting jet functions", which extend the notion of fragmentation functions to fragmentation within a jet, was derived. Fragmentation within a jet allows a study of both perturbative and non-perturbative aspects of jets, and is of interest both in vacuum, and in the medium.

There has recently been significant theoretical work extending these ideas to study a wide variety of fragmentation processes within jets. This includes more detailed studies and extension of ``standard" fragmentation in jets \cite{Kaufmann:2015hma,Dai:2016hzf,Arleo:2013tya,Chien:2015ctp,Procura:2011aq,Neill:2016vbi}, heavy flavor fragmentation in jets \cite{Bain:2016clc,Fleming:2012wy,Baumgart:2014upa,Bodwin:2015iua,Fickinger:2016rfd,Kang:2017yde}, fragmentation in heavy ion \cite{Cacciari:2012mu,Kang:2016ofv,Kang:2017frl}, as well as new jet substructure observables to study heavy flavor fragmentation \cite{Ilten:2017rbd}. These have provided significant new tools to study fragmentation within jets, and allow for direct comparison with theoretical calculations and extraction of non-perturbative functions.

\begin{figure}
\begin{center}
\includegraphics[width=0.95\columnwidth]{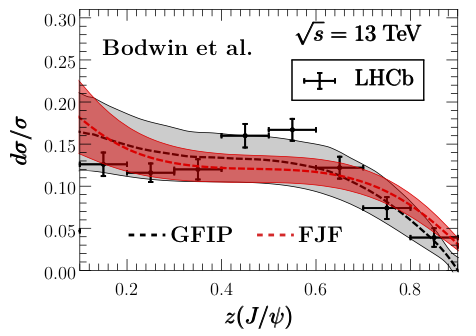} 
\end{center}
\vspace{-0.4cm}
\caption{Theoretical predictions of the $z(J/\psi)$ distribution compared with measurements from LHCb. The different theory predictions are described in the text. Taken from Ref.~\cite{Bain:2017wvk}.
} 
\label{fig:JPSI}
\end{figure}

Due to the wealth of recent calculations and measurements, here we highlight a particular theoretical calculation of $J/\psi$ production in a jet \cite{Bain:2016clc,Bain:2017wvk}, which provides a probe of the mechanism of $J/\psi$ production and decay. In \Fig{fig:JPSI} we show a calculation of $z(J/\psi)$, which denotes the energy fraction of the $J/\psi$ within the jet \cite{Bain:2016clc,Bain:2017wvk}. Here we show predictions using fragmenting jet functions (FJFs), as well gluon fragmentation improved Pythia (GFIP), which implements a FJF inspired picture into Pythia. The required non-perturbative matrix elements use the fits of Ref.~\cite{Bodwin:2014gia}.  The analytic calculations are compared with recent measurements from LHCb, showing good agreement. This particular example also highlights the important role that jet substructure can play at LHCb, where it has been less commonly applied, but where we hope that newly developed jet substructure techniques can have significant impact. 

\vspace{0.2cm}

\noindent {\small{\sf $\alpha_s$ in Resummation Sensitive Observables}}\\

Event shape observables measured at LEP have played an important role in measurements of the strong coupling constant, $\alpha_s$. While a variety of techniques exist (see, e.g., \Refs{Bethke:2000ai,Bethke:2009jm,Bethke:2011tr,Bethke:2012jm,Moch:2014tta} for a review), it is important to obtain consistent results for values extracted in a variety of measurements, in particular, those sensitive to all-orders resummation, and those insensitive to all-orders resummation. While the perturbative aspects of event shapes in $e^+e^-$ collisions are under exceptional control, with resummation to NNNLL \cite{Becher:2008cf,Abbate:2010xh,Hoang:2014wka,Hoang:2015hka}, and fixed-order corrections to NNLO, the standard event shapes receive large non-perturbative corrections. These can be treated by also fitting for moments of appropriate shape functions, but these moments are highly correlated with the value of $\alpha_s$. There is currently some tension between values of $\alpha_s$ extracted from event shapes as compared with other approaches, and it is therefore interesting to understand if tools from jet substructure can be used to provide a theoretically improved, complementary measurements of $\alpha_s$ from event shapes.

An interesting approach may be to use groomed observables. We have seen that grooming can be used to reduce the numerical impact of non-perturbative effects, rendering the observable perturbative over a larger range in which the fit to $\alpha_s$ can be performed. In this case there are other issues that need to be addressed, namely power corrections of parameters of the groomer. However, these are perturbative corrections, so it is hoped that they their understanding can be improved through analytic calculations.  This is in contrast to non-perturbative corrections, which are difficult to understand analytically.

To enter into the world average, measurements of $\alpha_s$ must involve NNLO calculations.  In $e^+e^-$ collisions, the required matrix elements have been known for some time \cite{Garland:2002ak,GehrmannDeRidder:2007hr,GehrmannDeRidder:2007jk,Gehrmann-DeRidder:2007nzq,Weinzierl:2008iv,DelDuca:2016ily}, and even groomed event shapes have been calculated to NNLO accuracy \cite{Kardos:2018kth}.   However, the corresponding matrix elements for NNLO accuracy of jets in $pp$ collisions are not available yet, although they should be in the coming years. It would nevertheless be interesting to understand the theoretical and experimental issues inherent in such a measurement to evaluate if it may prove competitive in the future. While groomed event shapes were not measured at LEP it would also be interesting to perform a re-analysis of the LEP data to see if such an approach can provide a competitive measurement of $\alpha_s$.  First steps for such an $\alpha_s$ extraction program have been outlined in the Les Houches study of \Ref{Bendavid:2018nar}, but a complete analysis will require effort from a large number of theorists and experimentalists.

\subsubsection{Impact on the Broader QCD Community}\label{sec:help_others}

While we have already noted a variety of issues, such as resummation of NGLs or logarithms of $R$, where advances have been motivated by jet substructure, here we emphasize several areas where ideas from jet substructure can play an important role in the QCD community more broadly.  We highlight the use of jet substructure observables to formulate infrared subtractions for fixed-order calculations, as well as the use of jet substructure observables to improve parton shower modeling. We hope that there will be significantly more such examples in the future.\\

\noindent {\small{\sf Infrared Subtractions for Fixed Order Calculations}}\\

As emphasized in \Sec{sec:calcs}, the behavior of jet substructure observables is intimately related to the infrared and collinear limits of QCD. The design of jet substructure observables for different purposes is therefore equivalent to understanding how applying physical measurements can control, or modify this infrared structure. This topic is of general interest to the QCD community more generally, particularly as better theoretical control over the observables is achieved.

To obtain a reliable description of more differential kinematic distributions, including those relevant for jet substructure, as well as to match the experimental precision of LHC measurements, calculations are often required at NNLO in perturbative QCD. This represents the current state of the art for predictions involving jets in the final state, and has been achieved only for a few key processes, namely $W/Z/H+$ jet \cite{Boughezal:2015aha,Boughezal:2015dva,Boughezal:2015ded,Boughezal:2015dra,Ridder:2015dxa,Chen:2016zka}, $t \bar t$ \cite{Czakon:2013goa}, and inclusive single jet production \cite{Ridder:2013mf,Currie:2016bfm}, as well as the all gluon channel for dijet production \cite{Ridder:2013mf}.

As discussed in this review, higher order calculations in QCD have infrared singularities which cancel between real and virtual contributions for an IRC safe observable. While this cancellation is guaranteed, it is difficult to isolate and analytically cancel these singularities. At NLO this is well understood and implemented for generic processes \cite{Frixione:1995ms,Catani:1996vz,Catani:1996jh,Frixione:1997np,Catani:2002hc}. At NNLO the structure of singularities is significantly more complicated. Extending the NLO techniques, several working schemes have proven successful with colored particles in the final state \cite{GehrmannDeRidder:2005cm,Czakon:2010td,Boughezal:2011jf,Czakon:2014oma}.

Recently a general subtraction scheme based on the $N$-jettiness observable \cite{Stewart:2010tn}, familiar from jet substructure, was proposed \cite{Boughezal:2015dva,Boughezal:2015aha,Gaunt:2015pea}.   It uses a physical observable, namely $N$-jettiness, to isolate and control infrared divergences, in much the same way that the $N$-subjettiness observable is used in jet substructure, namely to isolate $N$-jet configurations from $N-1$-jet configurations. $N$-jettiness subtractions have been applied to $W/Z/H/\gamma+$ jet~\cite{Boughezal:2015dva, Boughezal:2015aha, Boughezal:2015ded, Boughezal:2016dtm,Campbell:2017dqk} and inclusive photon production \cite{Campbell:2016lzl} at NNLO. They have been implemented in MCFM8 for color-singlet production~\cite{Campbell:2016jau, Campbell:2016yrh, Boughezal:2016wmq, Campbell:2017aul}, and also applied to single-inclusive jet production in $ep$ collisions \cite{Abelof:2016pby}.

This scheme has the advantage that is is related to the singular behavior of IRC safe observables, which can be understood using the techniques discussed in \Sec{sec:calcs}. Furthermore, power corrections can be analytically computed to improve the performance of the technique. This has been demonstrated for color singlet processes both in direct QCD \cite{Boughezal:2016zws}, and in SCET \cite{Moult:2016fqy} using subleading power operator bases \cite{Feige:2017zci,Moult:2017rpl}.

\begin{figure}[t]
\begin{center}
\includegraphics[width=0.95\columnwidth]{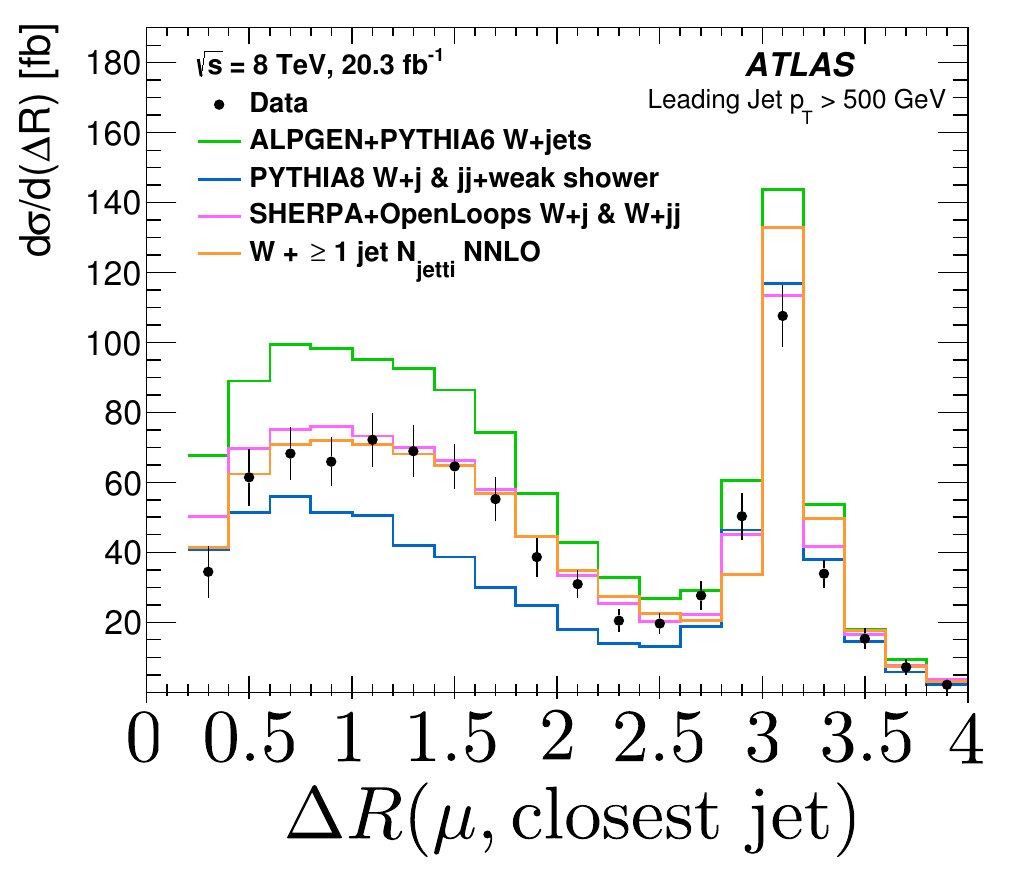} 
\end{center}
\vspace{-0.4cm}
\caption{A plot of the $\Delta R$ between a muon and the closest jet in $pp\to W+$jets events compared with an NNLO calculation using the $N$-jettiness subtraction scheme.  Adapted from Ref.~\cite{Aaboud:2016ylh}.} 
\label{fig:Delta_R_W}
\end{figure}

Implemented as a global subtraction, this technique has been applied to calculate $W/Z/H~+$ jet distributions at NNLO~\cite{Boughezal:2015dva, Boughezal:2015aha, Boughezal:2015ded, Boughezal:2016dtm}, and implemented in MCFM for color-singlet production~\cite{Campbell:2016jau, Campbell:2016yrh, Boughezal:2016wmq}. It has also been applied to single-inclusive jet production in $ep$ collisions \cite{Abelof:2016pby}. \Fig{fig:Delta_R_W} shows an application of a higher-order perturbative QCD calculation using the $N$-jettiness subtraction technique to jet substructure \cite{Aaboud:2016ylh}.  Here, the measurement of the angle between the muon from a decaying $W$ boson to the closest jet is compared to several predictions.  Excellent agreement is observed with the NNLO prediction, though no uncertainties are reported.  Improved subtraction schemes involving jets in the final state will advance the precision of jet substructure calculations at the LHC.  With more comparisons to hadronic jet substructure measurements, this opens up a new precision regime.

The $N$-jettiness subtraction scheme represents an interesting application of ideas from jet substructure to perturbative QCD calculations. It would be of significant interest to understand whether other jet substructure observables can be used to develop further improved subtractions. This is an active area of research to which there is interest from a variety of different communities.\\

\noindent {\small{\sf Improving Parton Shower Modeling}}\\

The wealth of jet substructure measurements, calculations, and observables will also have an important impact on improving parton shower generators. The jet substructure observables highlighted in this review probe the physics of a jet in interesting and new ways. This in turn provides a stringent test of both the perturbative and non-perturbative aspects of parton shower generators. While parton showers have been tuned to event shape observables at LEP, this provides limited access to gluon jets, or initial state radiation. Furthermore, the wide variety of new observables designed for jet substructure exhibit different sensitivities to perturbative and non-perturbative effects, which provides a handle when tuning generators. For a recent application of using newly developed jet substructure observables to tune parton showers using LEP data, see Ref.~\cite{Fischer:2014bja}.

While we have emphasized throughout this review how techniques, such as jet grooming, can be used to minimize sensitivity to non-perturbative or soft physics, it is precisely these contributions which are the least well modeled.  The measurement of a wide range of observables both with and without grooming will be essential, not only for understanding the behavior of these techniques, but also for a comprehensive parton shower tuning program based on jet substructure measurements. Precise measurements of both one- and two-prong observables, before and after grooming, will provide a variety of handles to isolate physics effects and tune event simulation generators. This will in turn feed into reduced uncertainties in the modeling of QCD jets at the LHC.

\begin{figure}[t]
\begin{center}
\includegraphics[width=0.95\columnwidth]{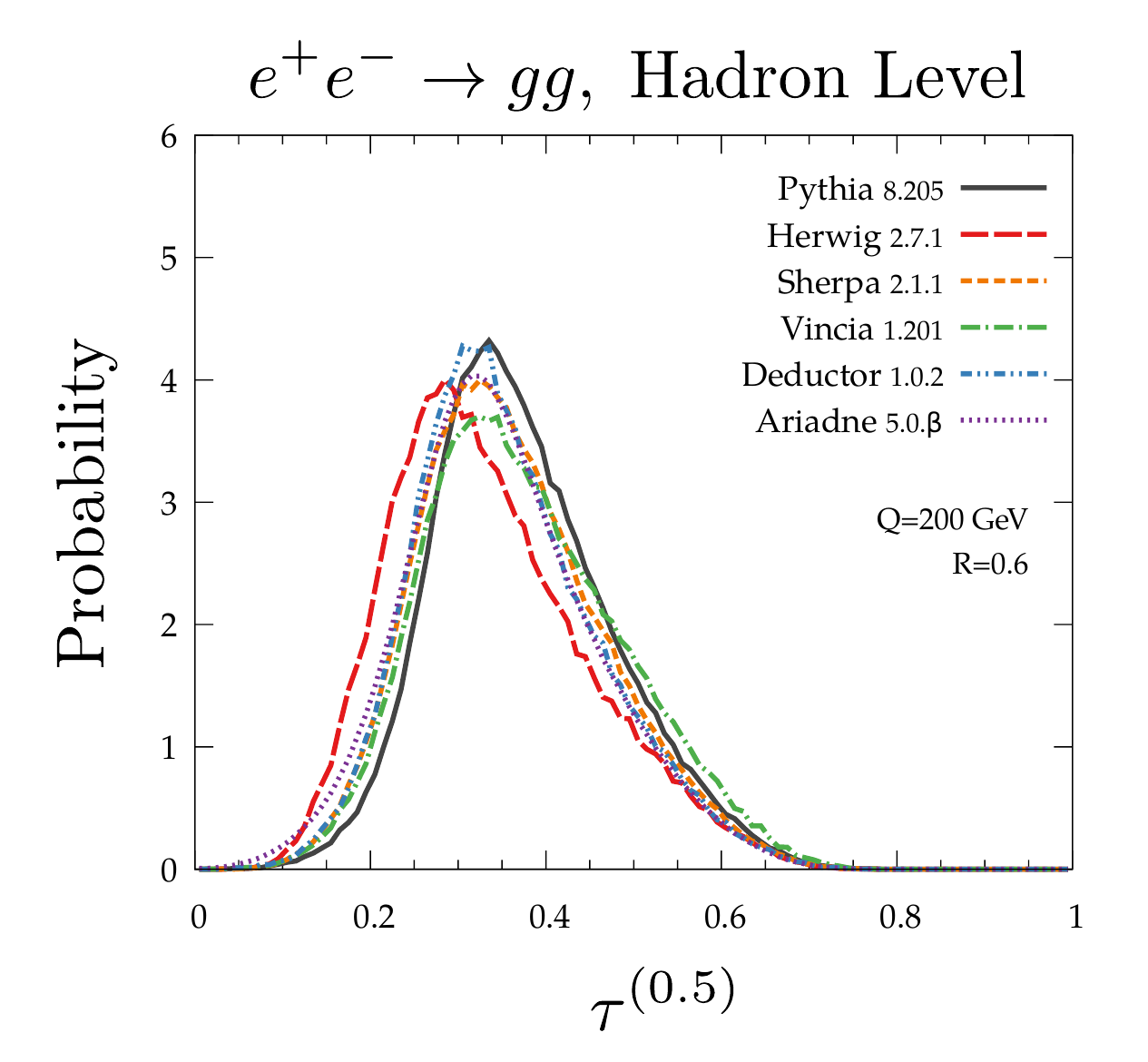} 
\end{center}
\vspace{-0.4cm}
\caption{A comparison of the distribution of the angularity $\tau^{(0.5)}$ as measured on gluon jets in $e^+e^-\to gg$ events for a variety of parton showers. A significant spread is observed, sandwiched by the Pythia and Herwig$++$ results. Adapted from Ref.~\cite{Badger:2016bpw}.  }
\label{fig:sandwich}
\end{figure}

As an important example, we highlight quark vs.~gluon discrimination, and in particular the modeling of gluon jets. The difference between quark and gluon jets is non-parametric, being largely driven by their different color charges.  For detailed studies of a variety of observables for quark vs.~gluon discrimination see \Refs{Gallicchio:2011xc,Gallicchio:2011xq,Gallicchio:2012ez,Larkoski:2013eya,Larkoski:2014pca,Badger:2016bpw,Moult:2016cvt,Komiske:2016rsd}.  Observables sensitive to the difference between quark and gluon jets often receive large non-perturbative contributions, so that an accurate modeling of these non-perturbative effects is important. However, particularly for gluon jets, there is a significant discrepancy between different implementations of the parton shower. An example of this is shown in \Fig{fig:sandwich}, which shows the distribution of the angularity $\tau^{(0.5)}$ measured on gluon jets simulated in $e^+e^-\to H\to gg$ events from a variety of different parton showers.  We note that such differences between parton showers are not restricted to angularities measured on gluon jets, and was also illustrated in the efficiency of the $D_2^{(\alpha)}$ observable as predicted by Herwig and Pythia, as shown in \Fig{fig:D2_shape}. The direct tuning to observables measured on gluon jets should allow for an improved modeling of gluon jets. The ability to perform reliable quark vs.~gluon discrimination will have important implications for a variety of measurements and searches at the LHC (see for example \Ref{Bhattacherjee:2016bpy}), and should be a goal of the jet substructure theory community.

We also emphasize that there has been significant recent work on improving the perturbative description of the parton shower, including a recent implementation of a $2\to 4$ NNLL shower \cite{Li:2016yez}, and triple collinear splitting functions \cite{Hoche:2017hno,Hoche:2017iem}. As the perturbative accuracy improves this will place tighter constraints on perturbative parameters, reducing the flexibility in parton shower tuning. Observables with different sensitivities to perturbative and non-perturbative physics, and an interplay with analytic resummation, will also be essential for this goal.

\subsubsection{Open Data}\label{sec:od}

All of the LHC experiments have released data or full detector simulations in some form for educational purposes.  Recently, CMS took this a step further by publishing research grade data and simulation from the $\sqrt{s}=7$ TeV physics runs in 2010 and 2011 and at $\sqrt{s}=8$ TeV in 2012. The CMS Open Data format contains the full information of the particle-flow candidates, including jet energy correction factors and jet quality criteria information. These data represent a potentially interesting opportunity for jet substructure studies. 

\begin{figure}[t!]
\begin{center}
\includegraphics[width=0.7\textwidth]{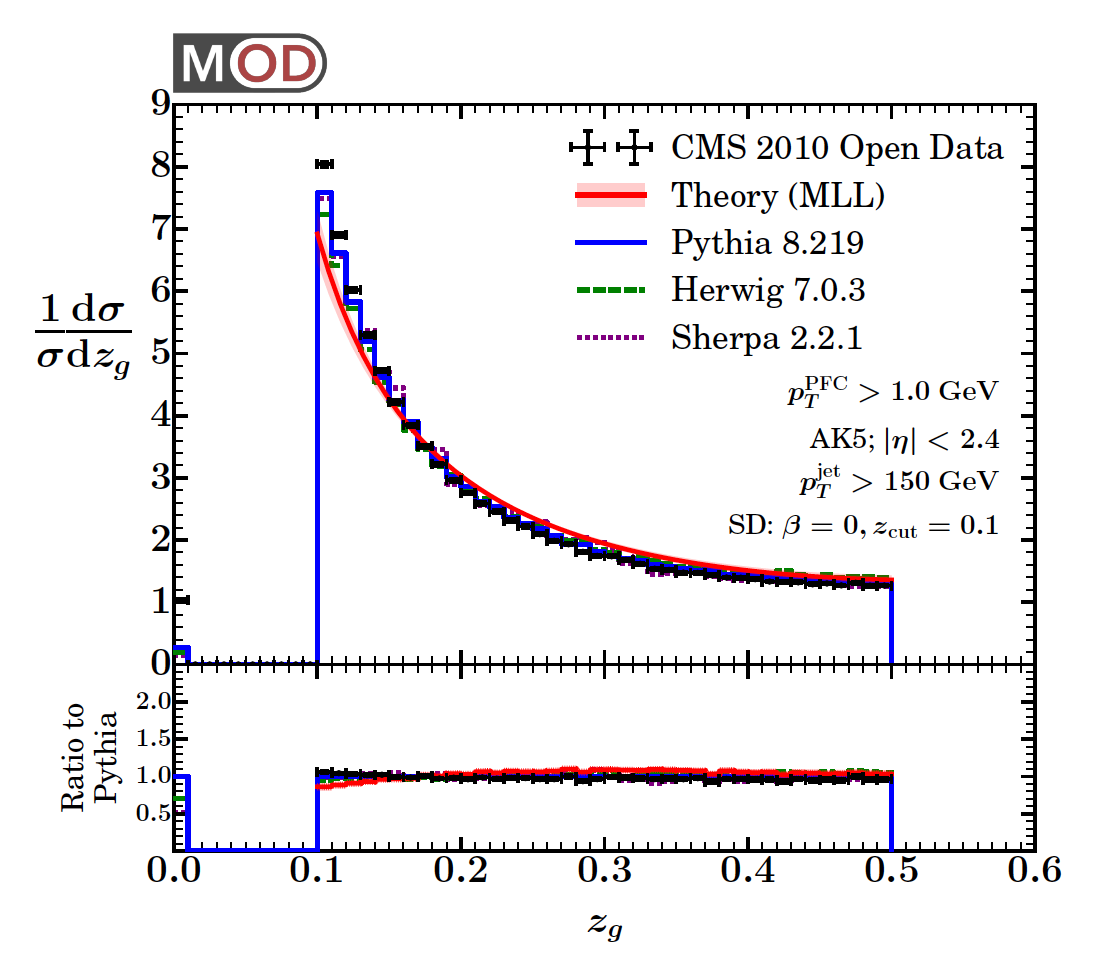}
\caption{Linear distribution of the $z_g$ variable in CMS Open Data compared to three MC predictions and the analytic calculation at leading-logarithmic accuracy. Taken from Ref.~\cite{Larkoski:2017bvj,Tripathee:2017ybi}.}
\label{fig:opendata2}
\end{center}
\end{figure}

The first analysis performed with the CMS Open Data studied jet substructure variables and compared them to Monte Carlo generator predictions as well as to analytic jet substructure calculations performed to leading-logarithmic accuracy~\cite{Larkoski:2017bvj,Tripathee:2017ybi}. In \Fig{fig:opendata2} we show a comparison of the $z_g$ distribution from the on Open Data, with an analytic calculation. The $z_g$ observable was specifically chosen for this analysis due to its robustness. The original data release was without accompanying detector simulations so this first analysis compared detector-level data directly to particle-level MC predictions.  Without correcting for detector effects and without systematic uncertainties, these comparisons must be interpreted carefully.

An important aspect of the discussions surrounding these data is the extent to which calibrations and systematic uncertainties should be provided as well as the amount of simulation to provide alongside the data.  There is currently no consensus in the field for how recent data should be treated, but there is general agreement that it would be ideal to perform legacy analyses on data (long) after the experiments have ended.  This will always be fundamentally limited by the experimental expertise of the practitioners at the time of data taking, but HEP datasets are expensive and are often unique.  Certainly this will be an important topic for many years to come and it seems likely that studies related to jets will continue to push the community in this area.

\subsubsection{Recommendations and Summary}

We conclude this section by emphasizing several important directions where we believe that progress is needed, and summarizing several wish lists. While precision wishlists, for example, Ref.~\cite{Andersen:2014efa}, are common, they have been primarily in the context of fixed order calculations, where the desires of the community are fairly well defined. We believe that the field of jet substructure is now sufficiently mature that several well-defined goals can be proposed on the theory side, and we therefore hope that this can serve as a reference to members of different communities. Since jet substructure calculations rely on developments from the fixed order and resummation communities, we discuss important theory developments that would be highly desirable in both areas. We also hope that there will be significant improvement in jet substructure observables themselves, and we discuss important features in the design of future observables going beyond simply performance. Finally, we conclude with a list of data-theory comparisons which we believe would further illuminate the theoretical and experimental underpinnings of the field of jet substructure.\\

\noindent {\small{\sf Resummation}}\\

As discussed in \Sec{sec:calcs}, due to the hierarchies of scales present in the regions of phase space probed by jet substructure observables an important aspect of the calculation of such observables is the resummation of large logarithms. Here we highlight several areas where we believe that progress can be made. We separate the discussion into two sections, namely the calculation of higher order anomalous dimensions for linear renormalization group equations, and the understanding of higher order corrections to non-linear evolution equations.

Observables which identify a fixed number of subjets are typically described by functions which obey linear renormalization group equations, such as given in \Eq{eq:RG_linear}. Examples of relevance for jet substructure discussed in this review are the jet mass and groomed jet mass, as well as two-prong observables such as $D_2^{(\alpha)}$. The structure of linear renormalization group equations is well understood, and increased perturbative accuracy can be obtained by computing the required anomalous dimensions to higher orders. Since this is a well-defined problem, here we simply list some observables for which higher order resummation is achievable, and would play an important role in improving the precision of jet substructure calculations. We believe that a realistic goal would be groomed masses at NNNLL, ungroomed mass in $pp\to W/Z/H$ + jet at NNLL, and two-prong groomed observables, such as $D_2^{(\alpha)}$, at NNLL. This will provide a solid foundation of theoretical calculations for a variety of different observables at a precision which can be reliably compared with experimental data.

Another class of observables, which are becoming more prevalent in the study of jet substructure, are those whose description involves non-linear evolution equations. A well-known example highlighted in this review are observables involving non-global logarithms, which can be resummed by the non-linear BMS evolution equation of \Eq{eq:BMS}. Other examples of observables relevant for jet substructure which obey non-linear evolution equations are logarithms of the jet radius \cite{Dasgupta:2014yra,Dasgupta:2016bnd}, as well as jet charge, or track based observables \cite{Waalewijn:2012sv,Chang:2013iba,Chang:2013rca}. For such observables the structure of the evolution equation itself can change at higher perturbative orders, greatly increasing the complexity of the problem. Due to the importance of these observables for jet substructure it will be essential to understand the structure of the evolution equations beyond the leading order to ensure that these effects are under theoretical control. From a theory perspective it would also be interesting to have a more comprehensive theory of such observables, as presumably they will become more prevalent in the study of increasingly differential observables in jet substructure. For the particular case of NGLs, higher-order corrections have been considered in \Ref{Caron-Huot:2015bja}, although their numerical impact has not been assessed.\\

\noindent {\small{\sf Fixed Order}}\\

Underlying the calculation of all jet observables at the LHC are perturbative scattering amplitudes. While the calculation of scattering amplitudes is a mature field, amplitudes involving many legs, and their corresponding cross sections, are difficult to calculate. Due to the rapid maturation of the field of jet substructure, we are now confronting issues with the availability of fixed-order amplitudes, and the next level of precision will require the next generation of perturbative calculations. Since these calculations are already on the horizon for the fixed order community the goal of this section will be to describe how advances in the calculation of fixed order amplitudes will have an impact on the field of jet substructure. We highlight two examples, namely improving the description of the internal substructure of jets, and improving the description of hard processes for which jet substructure techniques are of particular relevance.

As discussed in \Sec{sec:calcs} precision calculations of jet substructure observables at the LHC are difficult due to the fact that high multiplicity amplitudes are required. Indeed, substructure observables of interest require at least a single emission within a jet to be non-zero. While all $2 \to 2$ NNLO amplitudes in QCD have been known for a number of years \cite{Glover:2001af,Bern:2002tk,Glover:2003cm,Bern:2003ck,Freitas:2004tk,Glover:2004si}, NNLO predictions for single inclusive jet production \cite{Ridder:2013mf,Currie:2016bfm} and dijet production \cite{Ridder:2013mf} are just becoming available. To achieve NNLO precision for the jet mass will require $2\to 3$ amplitudes at NNLO. While there has been significant recent progress in the computation of such amplitudes \cite{Badger:2013yda,Gehrmann:2015bfy,Dunbar:2016aux,Badger:2017jhb,Abreu:2017hqn,Chawdhry:2018awn,Abreu:2018jgq,Badger:2018enw,Abreu:2018zmy,Abreu:2018aqd}, as well as in the development of subtraction techniques for computing the cross section, which were briefly discussed in this review, there is still significant progress to be made. The completion of the calculation of such amplitudes will be required to truly push jet substructure to the precision regime. It is important to emphasize that even after grooming has been applied such amplitudes are still required to perform matching for an observable such as the jet mass. Nevertheless, it is important to ensure that all aspects of the calculations specific to jet substructure, such as the resummation, are theoretically well understood so that higher loop amplitudes can be incorporated as they become available. 

One of the original uses of jet substructure at the LHC was to discover the Higgs in the hadronic decay channel $H\to \bar b b$ \cite{Butterworth:2008iy}. With the increase in data, the shift will be to the precision measurement of differential spectra of highly boosted Higgs bosons. Accurate theory predictions in this region of phase space will require the full NNLO amplitudes with massive quark dependence, since the loop coupling to the Higgs is resolved and the standard effective theory approach is no longer applicable. This in turn offers a powerful channel to probe any beyond the Standard Model contributions to Higgs production. The two-loop master integrals with full quark mass dependence were recently completed for this process \cite{Bonciani:2016qxi}. Combined with recently developed NNLO subtraction techniques, this should allow for the precision study of boosted hadronically decaying Higgs bosons at the LHC.\\

\noindent {\small{\sf Observable Definitions}}\\

\begin{figure}[t]
\begin{center}
\includegraphics[width=0.95\columnwidth]{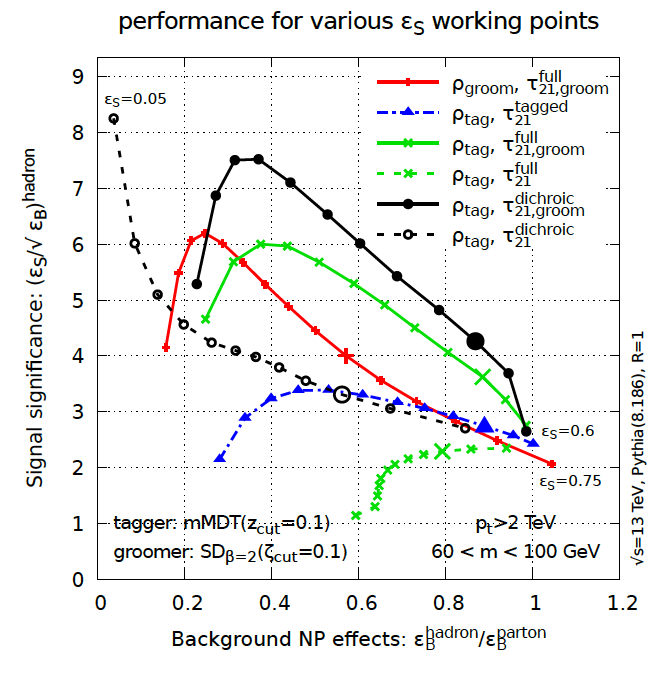} 
\end{center}
\vspace{-0.4cm}
\caption{A comparison of the impact of non-perturbative corrections as a function of the signal significance for a variety of different variants of $\tau_{2,1}$.  Taken from Ref.~\cite{Salam:2016yht}.} 
\label{fig:np_eff}
\end{figure}

One of the cornerstones of the field of jet substructure has been new observables. To continue to progress the field it will be essential to continue to develop new observables which probe jets in novel ways. Since the field of jet substructure is now more mature we have a more sophisticated set of criteria by which to judge observables, beyond simply their tagging performance.

For jet substructure taggers, for which the baseline performance is already quite high, it will be important to push for simplicity and robustness of the observables used in experiment. Since it is difficult to anticipate progress in the design of tagging observables we will instead highlight several features which should be taken into account when designing future taggers:
\begin{itemize}
\item {\bf Stability:} For identifying mass peaks it has recently been emphasized that the stability of the tagging observable as a function of $m_J$ and $p_{TJ}$ is important \cite{Dolen:2016kst}. Dependence on $m_J$ and $p_{TJ}$ can be eliminated using the DDT or CSS approaches \cite{Dolen:2016kst,Moult:2017okx}, or directly built into to observables, such as for the $N_2$ observable \cite{Moult:2016cvt}, and should be an important consideration when designing future observables.
\item {\bf Minimal Sensitivity to Non-Perturbative Effects:} Observables which probe structures within jets are also typically sensitive to non-perturbative effects. These appear as uncertainties in parton shower modeling or analytic calculations. A promising approach is to study the behavior of observables in a ``performance--non-perturbative sensitivity plane" \cite{Salam:2016yht}, although more studies are required to have an understanding of a reasonable ``metric" in this space for comparing observables. An example from \Ref{Salam:2016yht} is shown in \Fig{fig:np_eff}, which motivated the introduction of the dichroic $\tau_{2,1}^{(\alpha)}$ ratio observable.
\item {\bf Theoretical Simplicity:} While theoretical simplicity is harder to define, ideally newly proposed observables will be designed in such a way so as to facilitate theoretical calculations. This includes, for example, simplifying phase space restrictions for automation of fixed order calculations, or ideally, enabling the derivation of factorization theorems. 
\end{itemize}
Beyond this, another direction which requires study from the theory side is in the analytic understanding of correlations between observables, which will allow one to go beyond the single variable paradigm in a theoretically-controlled manner. Ideally this will enable the construction of a set of tagging observables which are both powerful and robust experimentally, and theoretically well understood.

%

\begin{figure}[t]
\begin{center}
\includegraphics[width=0.95\columnwidth]{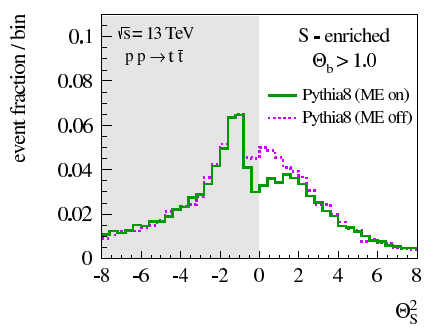} 
\end{center}
\vspace{-0.4cm}
\caption{A simulation of the distribution of $\Theta_s^2=\text{sgn}(X)(X^2+Y^2)$, where $X,Y$ are the coordinates of radiation relative to top quark, as described in the text.  The dead cone effect, which causes a suppression for $\left| \Theta_s^2 \right |<1$ is clearly visible.  Taken from Ref.~\cite{Maltoni:2016ays}.
} 
\label{fig:dead_cone}
\end{figure}

As the amount of data at the LHC increases and experimental precision improves, it will be possible to use techniques from jet substructure to study increasingly differential distributions in increasingly extreme regions of phase space.  This will enable detailed probes of the gauge theories underlying the Standard Model. Some examples already highlighted in this review include the soft-dropped splitting fraction $z_g$ shown in \Fig{fig:zg_dist}, and the angle between a muon and the closest jet shown in \Fig{fig:Delta_R_W}. Other examples, such as the jet pull \cite{Gallicchio:2010sw}, which measures aspects of the color flow within an event, have already been measured in experiment \cite{Aad:2015lxa,Aaboud:2018ibj}, but await theory calculations. As a further example highlighting a measurement of an increasingly subtle aspect of gauge theories, \Ref{Maltoni:2016ays} showed how tools from jet substructure, namely top tagging and soft drop grooming, can be used to observe the dead cone effect at the LHC. The dead cone effect is a universal effect in gauge theories in vacuum, which causes radiation from a particle of mass $m$ and energy $E$ to be suppressed in vacuum within an angle $\theta \lesssim m/E\equiv \theta_D$ \cite{Dokshitzer:1991fd,Dokshitzer:1991fc}. This effect is absent in the presence of a medium, where the dead cone is filled \cite{Armesto:2003jh}. \Fig{fig:dead_cone} shows the expected distribution of radiation around the top quark obtained in Pythia after selection criteria, clearly showing the dead cone effect. To describe the distribution of radiation we have defined the vector relative to the top quark using $\Theta\equiv \theta/\theta_D$, and $X=\Theta \cos \phi$, $Y=\Theta\sin \phi$, so that the top quark direction of flight is $(X,Y)=(0,0)$, and $\Theta_s^2=\text{sgn}(X)(X^2+Y^2)$. The dead cone peak then occurs at $\Theta^2\simeq 1$, and radiation is suppressed below this value.  We hope that a number of other features of gauge theories can be elucidated at the LHC using techniques from jet substructure.\\

\noindent {\small{\sf Goals for Data-Theory Comparison}}\\

We conclude this section of the review with a list of several observables for which we think precision data-theory comparisons would be particularly useful, and will be possible in the timescale of a few years.  Because of this emphasis on precision data-theory comparison we focus on observables for which the theoretical ingredients exist, or are understood, which therefore limits the scope of the list. We do not include, for example, heavy ion measurements, although we expect significant theory and experimental progress in this area. 

We highlight here the jet mass and other one-prong observables, two-prong observables, track-based observables, fragmentation, and the groomed top mass. We discuss in some detail the processes for which data-theory comparison will be simplest, as well as some of the physics goals of these particular measurements.


\begin{itemize}

\item {\bf Jet mass measurements:} The jet mass is a benchmark observable in jet substructure. Measurements of both the groomed and ungroomed jet mass are therefore central as a test of theoretical methods, as well as a probe of grooming procedures. For achieving data-theory comparison, the simplest process in which to measure the jet mass is $pp\to \text{jet}+ L$, where $L$ is a color singlet, namely $W/Z/H/\gamma$. The restriction to the case of a jet recoiling against a color singlet greatly simplifies the theoretical description, while maintaining the relevant physics, namely non-trivial color flow and initial and final state partonic channels. It has therefore been the focus of most theoretical studies \cite{Dasgupta:2012hg,Jouttenus:2013hs,Kolodrubetz:2016dzb,Frye:2016okc,Frye:2016aiz}. We wish to emphasize the importance of performing measurements with different sensitivities to soft radiation; (ideally both groomed and ungroomed measurement).  This is important both for testing analytic resummation, as well as for tuning parton shower generators, and will allow for a detailed study of color flow, hadronization, and the underlying event. Since measurements without grooming are difficult due to the presence of pileup, it may be more feasible to perform measurements with different grooming parameters, or on tracks, as discussed later in this section. More thought is required on how information about soft radiation can be extracted in a manner that is useful for theoretical comparisons.
For measurements of the groomed jet mass, using mMDT or soft drop are strongly theoretically preferred to other groomers because they remove NGLs and enable the most precise calculations.


The extension to $pp \to$ dijets will also be interesting, although more theoretical work is required due to the more complicated color structure. The renormalization group evolution has been worked out in Refs.~\cite{Kidonakis:1998nf,Kelley:2010fn,Moult:2015aoa}, and studied in Ref.~\cite{Hornig:2016ahz}.

\item {\bf ``Big Five'' of Quark vs.~Gluon Discrimination:}

Generalizing from the jet mass, an important set of measurements for improving our understanding of the radiation patterns of single prong jets, which have applications to quark vs.~gluon discrimination, are the ``big five", namely multiplicity, pTD, the Les Houches Angularity (LHA), width, and mass (see \cite{Badger:2016bpw} for detailed definitions). These observables each probe different aspects of soft and collinear physics, as well as non-perturbative fragmentation, and their combined measurement will allow a significant improvement in the parton shower description of jet radiation patterns. As with the case of mass, it will be essential to measure these with and without grooming (or with grooming parameters varied), and in events with different color flow, namely for $W/Z/H/\gamma+$jet and dijet processes.

\item {\bf Two-Prong Observables:}  With more theoretical studies of two-prong observables, it is well-motivated to measure those observables that can be directly compared to calculations.  Again we recommend the process $pp\to \text{jet}+ L$ for data-theory comparison.  The energy correlation function ratio observable $D_2^{(2)}$ and the $N$-subjettiness ratio $\tau_{2,1}^{(2)}$ are the most theoretically studied, and so should be the first measured.  We emphasize the angular exponent of $\alpha = 2$ for both observables, since with this value they are more closely related to the jet mass which greatly simplifies theory calculations.  Existing measurements of these observables use the angular exponent $\alpha = 1$ since this value has performed optimally in tagging studies, which makes quantitative comparison with theory difficult, at least at the current stage.

As with the jet mass, both ungroomed and groomed measurements would provide insight for parton shower tuning.  With recent demonstrations of robustness and first calculations, predictions of groomed two-prong observables will likely be the focus of the theory community in the forseeable future.  As with the jet mass, the preferred jet groomers are mMDT and soft drop, for the same reasons.  Calculations of two-prong observables designed to be robust, like the $N_2$ observable of Ref.~\cite{Moult:2016cvt}, are also likely to be completed in the near future.


\item {\bf Track-Based Mass and Two-Prong Observables:} In addition to the jet mass and two-prong observables, we believe it would be of interest to measure track versions of these observables, again in $pp\to \text{jet}+ L$. From the theory perspective track based observables are interesting since they are not IRC safe, and therefore provide a test of our understanding of non-perturbative physics \cite{Waalewijn:2012sv,Chang:2013iba,Chang:2013rca}. For this same reason, they are also useful for tuning parton showers. Track based observables were measured at LEP and used for precision tests of QCD \cite{Buskulic:1992hq,Abreu:1996na}.  Experimentally, track based observables are interesting since they make use of the excellent angular resolution of the tracker allowing for more precise measurements, and are often less sensitive to pile up. They may therefore play an important role at future colliders, or the high luminosity LHC, and have been used in a variety of studies of boosted object tagging \cite{Schaetzel:2013vka,Larkoski:2015yqa,Spannowsky:2015eba,ATLAS-CONF-2016-035} in such regimes. We therefore believe that it is important to ensure that they are under good theoretical and experimental control.

\item {\bf Fragmentation Functions:}  The non-perturbative fragmentation of QCD partons into hadrons represents the fundamental barrier to any precision jet substructure analysis.  Ultimately, universal non-perturbative parameters that describe fragmentation are extracted from data \cite{Sato:2016wqj,Anderle:2015lqa,deFlorian:2007ekg,deFlorian:2007aj}, and then can be used to describe a broad range of jet processes.  For progress into the future, more of these extractions must be performed, with a broad range of precision data.  Understanding fragmentation in more detail can shed light on open problems in QCD, like understanding the medium in heavy ion collisions, discrimination of quark versus gluon jets, the $J/\psi$ production mechanism, and others.  These are fundamental issues that are currently being probed at all four of the major LHC experiments (ATLAS, CMS, ALICE, and LHCb) and other colliders like the Relativisitic Heavy Ion Collider (RHIC) at Brookhaven.

%
%

\item {\bf Groomed Top Mass:} As discussed in \Sec{sec:newfronts}, a precision measurement of the top mass using groomed observables is an interesting future goal for the jet substructure program. Experimentally, a precision measurement of the top quark mass using a new technique at a hadron collider is an important goal. Theoretically, such a measurement probes both the physics of the top quark and the dynamics of the jet, and most interestingly, the interplay between the two. In this case, more study is needed both to identify the optimal choice of grooming parameters, as well as to understand the effects of a top tagger on the mass distribution.

\end{itemize}

We hope that these goals motivate both the theory and experimental communities, and we believe they will provide considerable insight into the physics relevant for a wide array of measurements using jet substructure.

\section{Machine Learning}\label{sec:machine_learning}

The previous section focused on the development of an analytic, first-principles understanding of jet substructure.  This QCD-driven approach has resulted in many new techniques for measurements and searches as well as insights into the fundamental structure of the strong force.  While we must continue this line of research, there is now a new set of tools for a complimentary line of research.  Machine learning (ML) tools allow for a data-driven approach to optimize jet substructure analyses and to uncover new patterns in nature.  Jet substructure practitioners have been leading the adaptation of deep learning in high energy physics, especially using low-level (high-dimensional) inputs, and establishing connections between data-driven and theory-driven approaches to physics analysis at the LHC and beyond.

{\it Machine learning} is a generic term to describe procedures for identifying and classifying structure within a dataset.   As such, most analysis techniques can be described as a form of machine learning.  However, there is a deeper connection between machine learning and jet physics: the fundamental object of study only exists in the context of machine learning.  A jet is defined by a clustering algorithm, which is an example of an {\it unsupervised machine learning} technique.  Unlike the output of most clustering procedures, jets have a {\it physical meaning} - see Sec.~\ref{sec:theory}.   Even though there is an extensive machine learning literature on clustering techniques, the most commonly used jet algorithms were established within the high energy physics community\footnote{There are attempts to bridge these communities - e.g. Ref.~\cite{Mackey:2015hwa} adapts a classical clustering algorithm to jet finding.}.  This is because the physical meaning of a jet is valid only if the defining algorithm satisfies particular properties such as infrared and collinear safety.   Connecting physical meaning with machine learning algorithms is an important theme for the rest of this section.


The complement to clustering is {\it supervised learning}, which includes nearly all forms of jet tagging (``classification'').  High energy physics is a unique setting for supervised learning because it is possible to generate large high fidelity simulation (and to a limited extent, real collision) datasets that have a known type or origin (``labeled'').  This review implicitly and explicitly describes a plethora of jet substructure observables (``features'') that are useful for separating jets initiated by different partons or particles.  The optimal tagger is one that uses the likelihood ratio based on the full radiation pattern within the jet~\cite{nplemma}.  In practice, training datasets are not large enough, computers are not fast enough, and simulations are not reliable enough to use the full likelihood directly.  Instead, there are many powerful techniques to approximate the full likelihood given a limited number of labeled examples.  These include (boosted) decision trees (BDTs), random forests, and neural networks (NNs).  There is a long history of using these techniques trained on a relatively small number of one-dimensional projections of the full radiation pattern such as the jet mass and the number of subjets.  However, recent advances in machine learning have led to the ``deep learning revolution'' in which it has become relatively fast and reliable to train NNs with many layers and an increasingly complex set of architectures.  With these networks, it is possible to build classifiers trained directly on all of the available information contained in the jet radiation pattern.  The field of deep neural networks (DNNs) is rapidly expanding, but Ref.~\cite{Goodfellow-et-al-2016,1404.7828,DNNnaturereview} are rather comprehensive general introductions/reviews.

\begin{figure}[tb]
\centering
\includegraphics[width=0.5\textwidth]{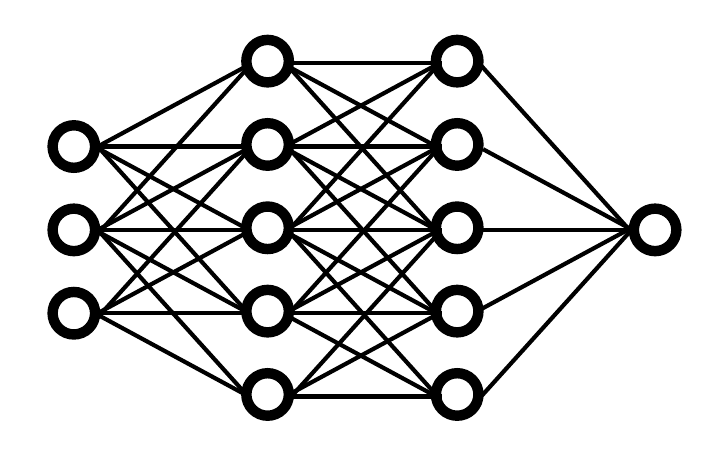}
\caption{A schematic picture of a deep neural network architecture.}
\label{fig:ML:NN}
\end{figure}

While BDTs\footnote{BDTs through TMVA have been used extensively in HEP, but there have been extensive recent innovation as well.  See for instance XGBoost~\cite{Chen:2016:XST:2939672.2939785}, which is now broadly used by the ML community.  In fact, this technique was developed as part of a Kaggle challenge designed to improve the Higgs boson search~\cite{Adam-Bourdarios:2015pye}.  Neural networks will be the focus of the rest of this section due to their scalability to large number of input and output dimensions, but BDTs will continue to play an important role in HEP and jet physics in particular.} are still the most common form of ML in high energy physics, DNNs are gaining in popularity due to their flexibility.  A NN is a composition of linear operations with non-linear \textit{activation} functions: 

\begin{align}
\label{eq:nn}
\text{NN}(x|w,b)=f(wx+b),
\end{align} 

\noindent where $x\in\mathbb{R}^n$ is the input data, $w\in\mathbb{R}^{n\times n}$ is a weight matrix and $b\in\mathbb{R}$ is called the bias.  The non-linear functions $f$ can take many forms such as a sigmoid or hyperbolic tangent.  The rectified linear unit (ReLU), $f(x)=\max\{0,x\}$, is the most popular non-linear function because it is non-linear enough~\cite{Cybenko1989,HORNIK1991251,NIPS2017_7203,hanin} but the gradient of the composition of many ReLUs does not disappear as fast the analogous composition of sigmoids.   While the weights $w$ and biases $b$ are typically optimized using gradient-based methods such as stochastic gradient decent, the functions $f$ are chosen ahead of time and typically varied as part of a hyper-parameter scan.  Equation~\ref{eq:nn} can be extended to have many dimensional outputs with a $w,b$ pair per dimension and can be iterated to form a deep neural network.  A schematic diagram of such a network is shown in Fig.~\ref{fig:ML:NN}.  Each line represents a function like Eq.~\ref{eq:nn} and there are three inputs, two \textit{hidden layers} with 5 nodes each, and one output.

In addition to providing a set of powerful tools, the bridging of DNNs and jet substructure has introduced a new mindset for approaching jet analysis.  In particular, approaching a problem with ML in mind requires an explicit notion of optimality.  Much of the observable development in Sec.~\ref{sec:theory} focused on heuristic notions of optimality, where parametrically separated regions of phase space are combined to form analytically-tractable observables.  In an ML approach, the notion of optimality is encoded in a \textit{loss function}.  For jet tagging with two types of jets (``binary classification''), the goal of ML is to make the output of the NN as close as possible to 0 for one class (often a background process) and 1 for the other class (often a signal process).   Optimality of this procedure is quantified by what \textit{close} means.   The most common loss function is binary cross-entropy, $L(x,y)=-y\log(x)+(1-y)\log(1-x)$, where $x$ is the estimate and $y\in\{0,1\}$ is the true label.

\begin{figure}[tb]
\centering
\includegraphics[width=0.5\textwidth]{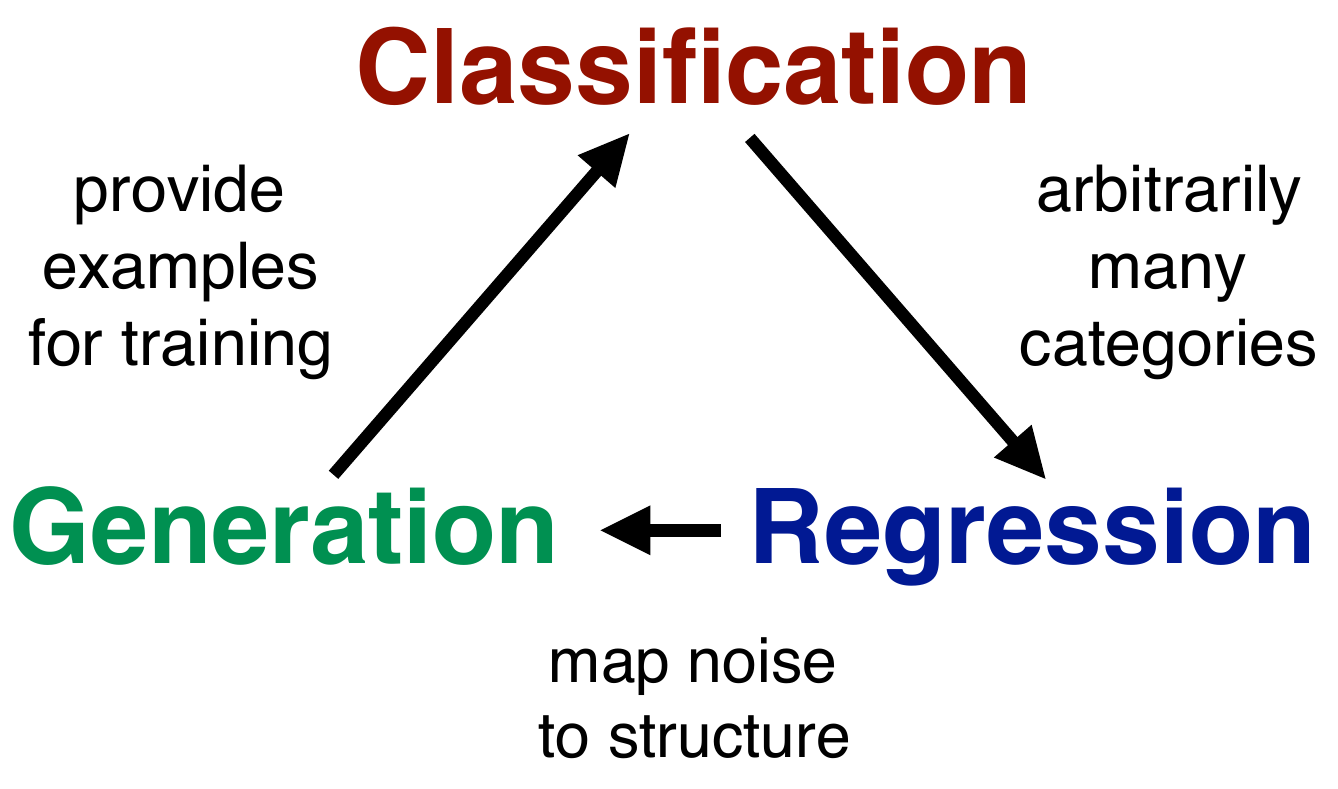}
\caption{A schematic illustration of the connection between NN applications for classification, regression, and generation.  Specific examples for jet physics are provided in later sections.}
\label{fig:ML:triangle}
\end{figure}

Discriminating between two classes of jets is not the only application of DNNs.  One of the benefits of using neural networks for ML is that it is straight-forward to have multiple outputs.  While the most common application of NNs in high energy physics use binary classification with a single output between 0 (more background-like) and 1 (more signal-like), one can have multi-class classification with one output per class.  Individual binary classifiers can be created using ratios of outputs.  Classification networks are trained by minimizing the distance between the NN output and a integral encoding of the classes.  Beyond classification, NNs can also be used for regression when there are arbitrarily many categories.  In a regression task, the goal is to predict a (multi-dimensional) continuous output from a set of inputs.  The most common loss function for regression is the mean squared error loss, $L(x,y)=(x-y)^2$.  Yet another application of DNNs beyond classification is generation.  A generator (like a Parton Shower Monte Carlo program) is a map from random numbers (noise) to structured data.  One can learn this map as a NN and so a generative NN is a regression model that learns to map random numbers to structure (by approximating the Jacobian).  Figure~\ref{fig:ML:triangle} schematically shows the connection between classification, regression, and generation.

This section is organized as follows.  We begin with a discussion of how to represent and preprocess jets as input to NNs in Sec.~\ref{sec:ML:reps}.  Then, applications for classification, regression, and generation are described in Sec.~\ref{sec:ML:classification},~\ref{sec:ML:regression}, and~\ref{sec:ML:simulation}, respectively.   Before closing, Sec.~\ref{sec:ML:anomaly} discusses new proposals to use ML and jet substructure to find new, unexpected structures in nature (anomaly detection).   The section ends with an outlook to the future in Sec.~\ref{sec:ML:future}.

\subsection{Jet Representations and Preprocessing}
\label{sec:ML:reps}

\subsubsection{Images, Sequences, and Sets}

The first step in any machine learning problem is to decide how to represent the data.  Jets have a rich structure and there is no unique way for encoding information about the radiation pattern into a particular data structure.  The earliest uses of machine learning for jet substructure used a set of one-dimensional physically-motivated observables -  see Ref.~\cite{Gallicchio:2011xq} for an example of this in the context of quark-versus-gluon jet tagging.  The downside of such an approach is that there is no guarantee that all of the available information is available to the NN.  The first applications of DNNs using all of the available information~\cite{Almeida:2015jua,deOliveira:2015xxd,Baldi:2016fql} represented jets as images\footnote{Jet images were actually first used in the early 90's in Ref.~\cite{PhysRevD.44.2025}, but machine learning techniques were not applied to jets-as-images until the LHC-era.}~\cite{Cogan:2014oua}.  A jet image is a pixelated grayscale image, where the pixel intensity represents the energy (or transverse momentum) of all particles that deposited energy in a particular location.  Grayscale images were expanded to include additional layers (`colors') to encode more information such as charge-energy versus neutral-energy~\cite{Komiske:2016rsd}.  The natural machine learning tool for analyzing images is the convolution neural network (CNN).  Unlike the generic dense network shown in Fig.~\ref{fig:ML:NN}, a CNN has \textit{weight sharing} so that the number of trainable parameters is much smaller.  For an $n\times n$ image, a filter of size $m\times m$, $m\ll n$ is constructed and convolved across the input image to form hidden layer features.  Since $m$ is small compared to $n$, the number of parameters that need to be optimized can be small and does not necessarily scale with $n$.  These networks also have the benefit that they are translationally invariant, though this is not necessarily useful for jet images that have been preprocessed (more on that below).   A CNN has a maximal amount of weight sharing (same filter applied to the entire image) - there have also been proposals to do classification with a smaller amount of weight sharing~\cite{deOliveira:2017pjk} in which the image is first broken into pieces and then a CNN is used on each piece.  While most jets-as-images applications have used standard $\eta-\phi$ coordinates to construct the images, Ref.~\cite{Dreyer:2018nbf} described an innovative alternative scheme where the jet clustering history is used to build an image that mimics the QCD splitting function.  As such, background jets tend to be uniformly distributed and signal jets have a non-trivial structure. 

Jets constructed with a calorimeter are naturally represented as images due to the finite cell granularity; however, images may not be the most efficient representation when finer granularity is available.  Machine learning techniques applied to individual particles may loose relevant information if the particles are first pixelated.  A useful method for representing jet substructure in this case is to consider the constituents as a sequence.  Natural language processing has developed powerful tools for analyzing sequences (like sentences), which can be adapted for jet tagging.  The analog to the CNN for sequences is the recurrent neural network (RNN).  Like the CNN, RNN's exploit the structure of the input to use weight sharing for a reduced number of trainable parameters.  The set of particles inside the jet are ordered (often by the jet $p_T$, but there is no unique choice) and then processed one at a time by the NN.  Each time an input is given to the NN, it combines the input with an internal state.  This process can be repeated any number of times and the final internal state can be used for classification or any other task.  RNNs were first applied to jet substructure in the context of flavor tagging~\cite{Guest:2016iqz,ATL-PHYS-PUB-2017-003}, where charged-particle tracks and displaced vertices are used to identify long-lived $b$-hadrons inside $b$-quark jets.  A natural extension of jets-as-sequences is to represent jets as binary trees.  As jets are constructed from $2\rightarrow 1$ clustering algorithms, the clustering history is a binary tree and can be used as input to a neural network.  Recurrent neural networks are the natural choice for operating on trees, where the neural network now takes multiple inputs to add to the internal state.  The first application of recurrent NNs to jet physics was in the context of boosted boson tagging~\cite{Louppe:2017ipp} and has also been applied to quark-versus-gluon tagging~\cite{Cheng:2017rdo}.  A similar recurrent structure is part of the JUNIPR model~\cite{Andreassen:2018apy} that will be described more in Sec.~\ref{sec:ML:simulation}.  One last generalization of sequences and trees is to consider a jet as a generic graph, which is a set of nodes and edges.  An adjacency matrix represents the connection strength between the various nodes of the graph and the natural NN structure is the graph convolutional neural network.  One of the standard jet clustering algorithms can be used to derive an adjacency matrix, but more exotic options are also possible.  The first application of graph networks to jet physics was in the context of boosted boson tagging~\cite{henrion} and has since been studied for pileup mitigation as well~\cite{Martinez:2018fwc}.

One downside of both the image and sequence/tree/graph approaches is that a spatial or temporal structure has to be imposed on the jet constituents.  Due to quantum mechanics, there is no unique ordering or history that can be associated with the final state hadrons inside a jet and so the most `natural' representation of a jet is simply as an unordered set of 4-vectors.  The earliest applications of jets-as-sets still imposed an order (and truncated the size)~\cite{Pearkes:2017hku} as methods for variable-length unordered sets were not yet available.  Since that time, there have been advances in machine learning on point clouds (unordered variable-length sets) which have been applied to jet substructure~\cite{Komiske:2018cqr,Qu:2019gqs}.  Like CNNs and recurrent/recursive/graph networks, techniques like deep sets~\cite{NIPS2017_6931} applied to point clouds exploit weight sharing to reduce the number of tunable parameters. These algorithms can naturally encode additional information about particles beyond their kinematic properties, leading to `Particleflow Networks' and `ParticleNet with PID'.

While adapting tools from the machine learning community to fit new data structures has been a fruitful program, there is another possibility that has been pursued.  In particular, there have been a variety of studies to pick a physics-inspired representation such that the subsequent machine learning is simplified.  One such approach is the energy flow polynomials~\cite{Komiske:2017aww}, which are a basis of functions that provably span IRC-safe jet observables\footnote{For machine learning methods that have not been engineered to be IRC safe, see Ref.~\cite{Choi:2018dag} for a study of the approximate IRC-safety of deep networks-with-images for top tagging.}.   As a bonafide basis, machine learning can proceed by simple linear regression to learn the optimal combination of basis functions.  Another approach has been to use the $N$-subjettiness observables $\tau_k^{(\beta)}$~\cite{Thaler:2010tr} to `span' $m$-body phase space for sufficiently many $\tau_k^{(\beta)}$'s~\cite{Datta:2017rhs}.  The authors of Ref.~\cite{Datta:2017lxt,Datta:2019ndh} proposed to extract analytically tractable observables from the $\tau_k^{(\beta)}$ set by learning product observables which approximate the full NNs.  Another way to automate physically-inspired learning was presented with the Lorentz layer~\cite{Butter:2017cot} and Lorentz boost~\cite{Erdmann:2018shi} networks, which encode Lorentz invariant feature extraction as the first layers to a deep network acting directly on four-vectors.

\begin{figure}[tb]
\centering
\includegraphics[width=0.95\textwidth]{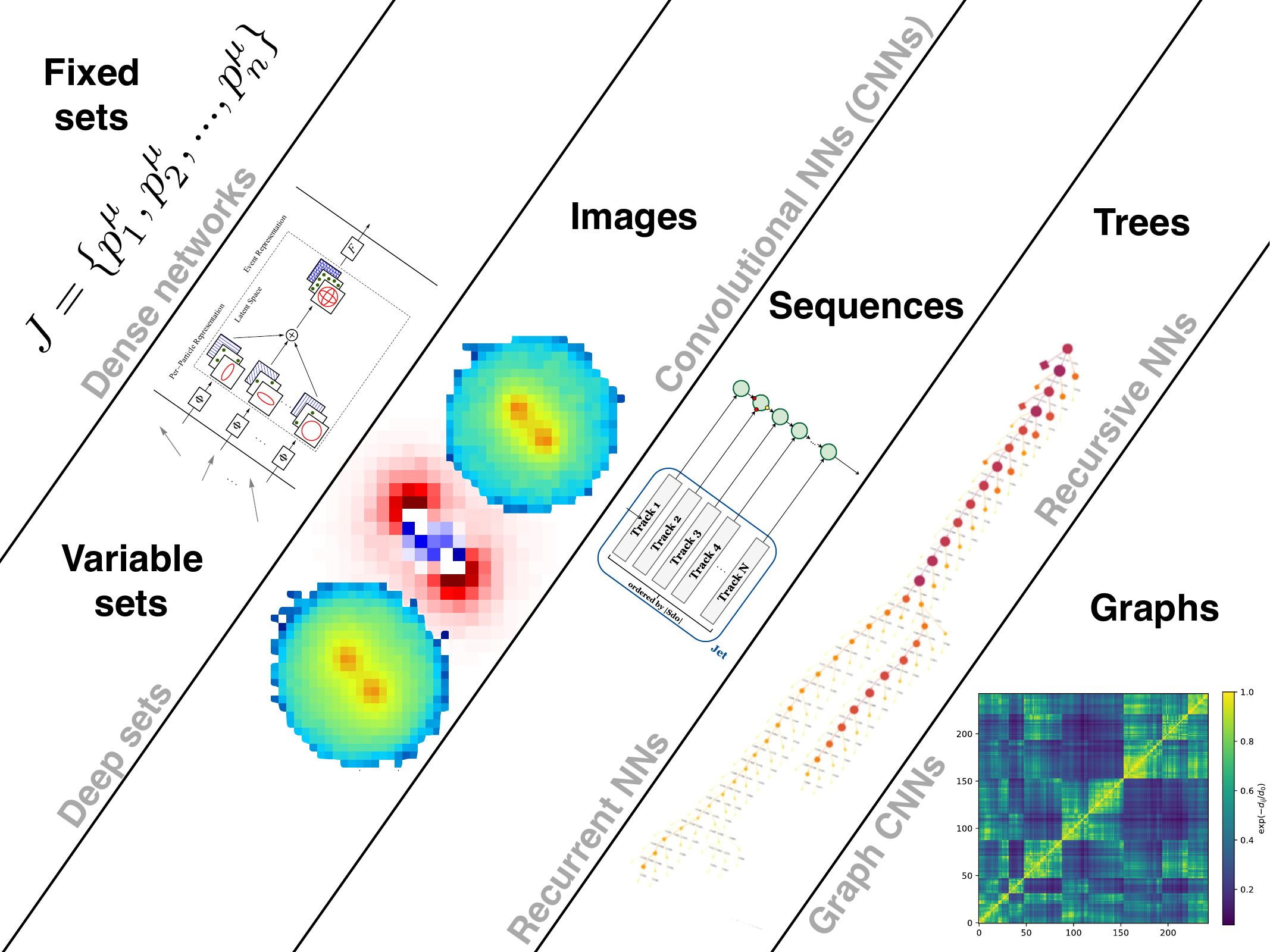}
\caption{A schematic diagram of the ways to represent jets and the natural NN architectures that go with each representation.  The deep sets image is from Ref.~\cite{Komiske:2018cqr}, the recurrent NN image is from Ref.~\cite{ATL-PHYS-PUB-2017-003}, the tree image is from Ref.~\cite{Cheng:2017rdo}, and the graph image is from Ref.~\cite{henrion}.}
\label{fig:ML:representations}
\end{figure}

Figure~\ref{fig:ML:representations} provides an overview of the representations that have been used to analyze jet substructure in the context of machine learning, along with the NN architectures used for each representation.

\subsubsection{Preprocessing and the Symmetries of Spacetime}

\begin{figure}[tb]
\centering
\includegraphics[width=0.95\textwidth]{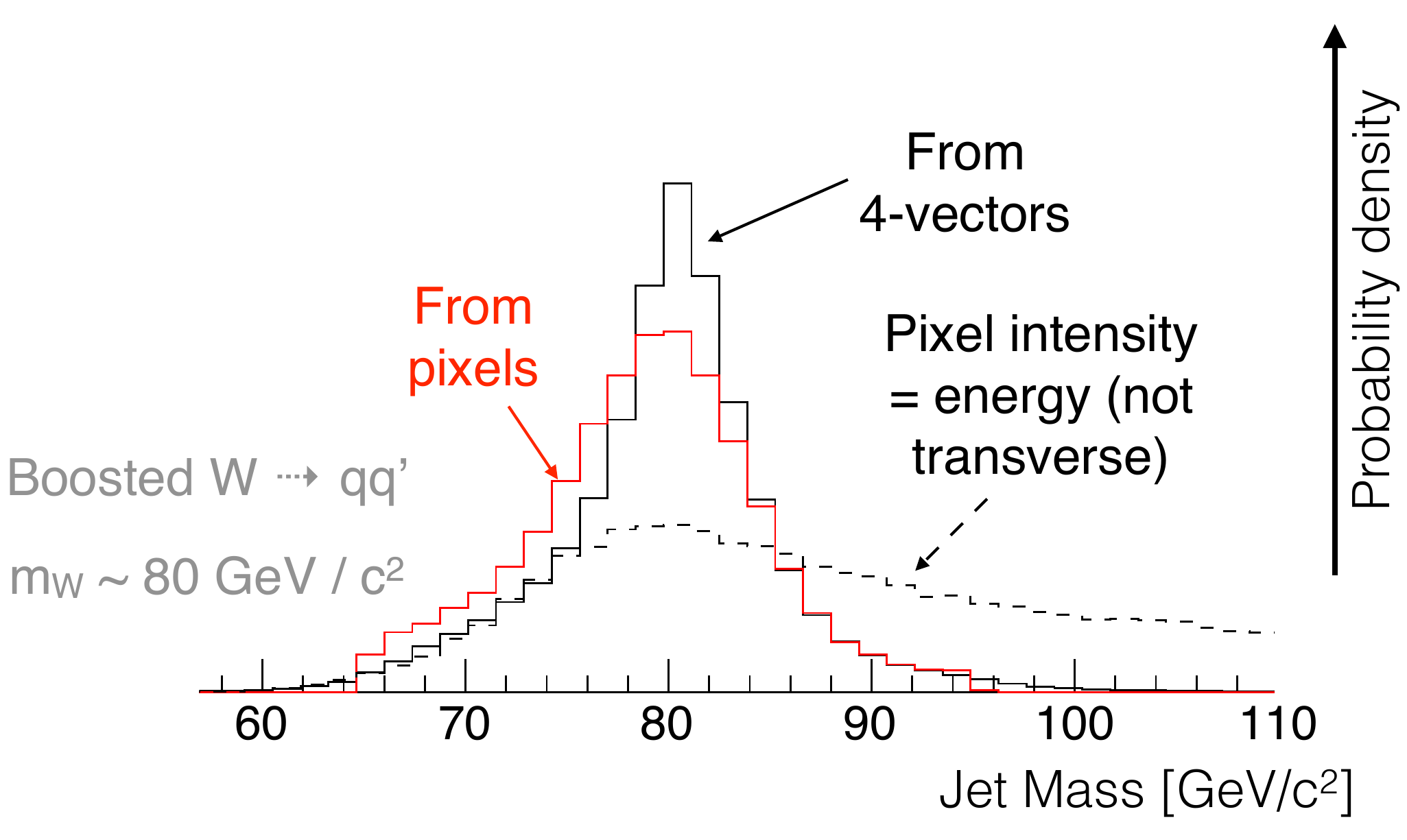}
\caption{The probably distribution of the jet (image) mass after various preprocessing steps for boosted hadronically decaying $W$ bosons.  The black distribution is the jet mass prior to any preprocessing while the red distribution is the result of coarse pixelation.  The dashed line is the mass distribution following an $L^2$-based normalization.}
\label{fig:ML:preprocess}
\end{figure}

The first step to process a full jet's substructure is to preprocess them into the proper format based on the machine learning architecture.  Preprocessing steps such rotating the jet along a pre-defined axis can significantly reduce the amount of training data required, but such steps can also distort the physical information contained in the resulting representation.   The impact of preprocessing in the case of images was extensively studied in Ref.~\cite{deOliveira:2015xxd,deOliveira:2017pjk,Pearkes:2017hku}.  Steps as benign as centering an image can have important consequences for the information content of a jet image.  For example, translations in $\eta$ correspond to boosts along $z$.  If the pixel intensity of an image is the deposited energy, then the jet image mass will not be the same before and after translating if the pixel intensities are untouched.  For this reason, it is important to use $p_T$ instead of energy for the pixel intensity.  Another natural preprocessing step with images is rotation.  The radiation pattern inside a jet is approximately symmetric about the jet axis in the $\eta$ and $\phi$ plane.  Therefore, many analyses using jet images rotate the jets to remove this symmetry.  Jet rotations leave observables like $n$-subjettiness invariant, but do not preserve the jet mass.   Alternatively, one could perform a proper rotation that would preserve the mass, but distort $n$-subjettiness~\cite{deOliveira:2017pjk,Pearkes:2017hku}.  As a final example, consider image normalization.  In industrial image processing, it is common to normalize the pixel intensities so that the sum of the squares of all intensities is unity ($L^2$ norm).  The problem with this normalization is that the $L^2$ norm of a jet image is correlated with the jet mass so performing the normalization procedure removes information about the jet mass from the image.  This is illustrated in Fig.~\ref{fig:ML:preprocess}.  After applying the $L^2$ normalization, the jet image mass is significantly distorted, resulting in a loss of usual information for jet classification.

Similar considerations apply for each representation discussed in the previous section.  Some representations require less preprocessing and therefore are less susceptible to information loss.   The first layers of a neural network can be thought of as automated preprocessing steps (e.g. conventional convolutional layers, $1\times 1$ convolutional layers~\cite{CMS-DP-2017-013}, or physically-inspired Lorentz layers).  Pushing more of the preprocessing to automated steps of the training is another strategy for preserving information for later parts of deep networks.

\subsubsection{Event-level Classification}

While the information inside jets can be an effective tool for identifying a particular process, the context around the jet can provide additionally useful information for increasing classification performance.   Various image-~\cite{Bhimji:2017qvb,Andrews:2019faz,Lin:2018cin} and tree-based~\cite{Louppe:2017ipp} approaches have been studied for event-level classification.  In the image case, one has to be cautious about edge effects due to the cylindrical geometry, which can be solved by appending the part of the image near $\phi=\pi$ to the image at $\phi=-\pi$ for every convolutional layer~\cite{Lin:2018cin}.  For maximally exploiting the data and aiding interpretation (see the end of Sec.~\ref{sec:ML:classification}), it can be useful to decompose images into jet representations that then are combined with other jets and event-level information~\cite{Lin:2018cin,Louppe:2017ipp}. Additionally, decomposing the structure of an image can be useful for calibrating the tagger in data, where the piece of the network corresponding to an individual jet can be testing in a control sample with low expected BSM acceptance. 

\subsection{Jet Tagging with Machine Learning}
\label{sec:ML:classification}

\subsubsection{Tagging examples}

Deep learning has now been applied to jets of every origin: electroweak bosons, gluons, light quarks, heavy quarks, and even BSM particles~\cite{Datta:2019ndh,Shimmin:2017mfk}.   It would not be practical to go through each application; instead, the goal of this short subsection is to provide references for further reading.  Table~\ref{tab:bigtagle} provides a table with example applications of deep learning for tagging, sorting by jet representation and target jet origin.

\begin{table}[h!]
\begin{adjustbox}{width=\columnwidth,center}

    \begin{tabular}{|c|c|c|c|c|c|c|c|}
    \hline
    & quark/gluon  & $W/Z$ & $H$ & $b/c$ & $t$ & BSM \\
    \hline\hline
   Image &\cite{Komiske:2016rsd,ATL-PHYS-PUB-2017-017,CMS-DP-2017-027}  &\cite{deOliveira:2015xxd,Baldi:2016fql} &\cite{Lin:2018cin} & &\cite{Kasieczka:2017nvn,Kasieczka:2019dbj} &\\
   Sequences &\cite{CMS-DP-2017-027}  & & & \cite{ATL-PHYS-PUB-2017-003}& &\\
   Tree &\cite{Cheng:2017rdo}  & \cite{Louppe:2017ipp,Andreassen:2018apy}& & &\cite{Kasieczka:2019dbj} &\\
   Graph & & & \cite{henrion} & & &\\
   Ordered Set & & &  &\cite{CMS-DP-2017-013,ATL-PHYS-PUB-2017-013} &\cite{Pearkes:2017hku,Kasieczka:2019dbj} &\cite{Shimmin:2017mfk}\\
   Point Cloud &\cite{Komiske:2018cqr,Qu:2019gqs}  & & & &\cite{Komiske:2018cqr,Qu:2019gqs,Kasieczka:2019dbj} &\\
   Physics-Inspired  & \cite{Komiske:2017aww}& \cite{Datta:2017rhs} & \cite{Datta:2019ndh,Datta:2017lxt}& & \cite{Butter:2017cot,Kasieczka:2019dbj}&\cite{Datta:2019ndh}\\
    \hline           
    \end{tabular}
  \end{adjustbox}
  \label{tab:bigtagle}
  \caption{A table summarizing applications of deep learning for classifying jets.  Each row corresponds to a different method for representing jets while each column stands for the target jet origin.  Only cases with supervised learning are given here - there are more cases for BSM tagging using semi-supervised and unsupervised approaches in Sec.~\ref{sec:ML:anomaly}.}
\end{table}
 
One of the assumptions of Table~\ref{tab:bigtagle} is that the various categories defined by the column headers are well-defined.  For boosted electroweak bosons, the radiation pattern is formally isolated from the rest of the event.  However, there can be ambiguities related to `containment', i.e. how much of the boson decay is captured by the jet.  Typical containment definitions use unphysical partons in leading order simulations.  This is even more complicated for jets originating from particles with a net strong force charge - in that case, it is not formally possible to treat the jet in isolation from its environment (though in some cases, these effects may be small~\cite{Bright-Thonney:2018mxq}).  For these reasons and more, there has been a significant effort within the deep-learning-for-jets community to develop methods that do not rely on labeling schemes.  

\subsubsection{Learning directly from (unlabeled) data}
\label{sec:weaksupervision}

Even without ambiguities in labeling, learning directly from data is not possible due to quantum mechanics: a jet's origin is not knowable on a jet-by-jet basis in data.  This is in stark contrast to industry and many other physical and biological sciences where experts can provide high fidelity labels to be used for machine learning.  One innovative possibility is to use machine learning tools to \textit{define} the labels~\cite{Metodiev:2018ftz}, where relatively pure regions of phase space are used as anchor bins to isolate two classes from mixed samples.

With traditional labeling schemes, it is still possible to apply machine learning to datasets where the per-jet labels are not known.  Suppose that there are two classes of jets $A$ and $B$ and there are two mixed samples $M_1$ and $M_2$ that are a super-position of $A$ and $B$: $M_1 = g_1A+(1-g_1)B$, $M_2=g_2A+(1-g_2)B$.  When $g_1=1$ and $g_2=0$, then $M_1$ and $M_2$ are pure samples where every jet has a known label.  Such a setting is known as \textit{fully supervised} learning.  The usual procedure of supervised learning for classification is as follows:

\begin{align}
\label{eq:supervisedlearning}
f_\text{full}=\text{argmin}_{f':\mathbb{R}^n\rightarrow[0,1]}\sum_{i=1}^N\ell(f'(x_i)-t_i),
\end{align}

\noindent where $n$ is the dimensionality of the feature space (number of jet attributes used for training), $f$ is a classifier (could be a neural network), $\ell(\cdot)$ is the loss function, $N$ is the number of jets available for training, $x_i\in\mathbb{R}^n$ is the information available about a single jet, and $t_i\in\{0,1\}$ is the per-jet label.  When the samples $M_1$ and $M_2$ are not pure, than the labels $t_i$ are not known and so the usual procedure summarized in Eq.~\ref{eq:supervisedlearning} is not applicable.  One possibility is that the label proportions $g_1$ and $g_2$ are known.  It is often the case that these fractions, usually set by PDF and hard-scatter matrix elements, are much better known than the phase space of the full radiation pattern inside jets (the space used for machine learning).  One can modify Eq.~\label{eq:supervisedlearning} for this case to learn on average, in a procedure called learning from label proportions (LLP)~\cite{Dery:2017fap}:

\begin{align}
\label{eq:weaklearning}
f_\text{weak}=\text{argmin}_{f':\mathbb{R}^n\rightarrow[0,1]}\sum_j\ell\left(\sum_{i=1}^N\frac{f'(x_i)}{N}-g_j\right),
\end{align}

\noindent where $g_j$ is the fraction of class $A$ in sample $j$.  The new loss function in Eq.~\ref{eq:weaklearning} acts on batches of events where the fractions $g$ are known.  In the limit of infinite training statistics and an arbitrarily flexible function $f$, Eq.~\ref{eq:weaklearning} converges to the optimal classifier derived with per-instance labels.  In fact, there is a neighborhood around the $g_i$ where Eq.~\ref{eq:weaklearning} results in exactly the same classifier and so there is some tolerance for uncertainty in the fractions~\cite{Cohen:2017exh}. 

Two disadvantages of the LLP approach is that the sample fractions have to be known (though not precisely), and the loss function has to be modified from the normal supervised approach.  An alternative method that does not suffer from either of these considerations is the Classification without Labels (CWoLa) approach~\cite{Metodiev:2017vrx}.  In the CWoLa method, one simply assigned labels to the mixed samples (say $i$ to sample $M_i$) and then uses any supervised technique to distinguish the two mixed samples from each other.  The resulting classifier is then used to distinguish $A$ and $B$.  As with LLP, in the limit of infinite training statistics and an arbitrarily flexible functions, this classifier amazingly converges to the optimal classifier derived with per-instance labels.  A critical assumption of CWoLa is that the statistics of $A$ in $M_1$ are the same as the $A$ in $M_2$ (and the same for $B$).  In other words, there is no way to distinguish $M_1$ $A$ jets from $M_2$ $A$ jets - the only difference between the mixed samples is that the proportions of $A$ is different.

Both LLP and CWoLa are examples of \textit{weak supervision}, which is a class of learning procedures where less than per-instance labels are known.

\subsubsection{Decorrelation procedures}
\label{sec:decorrelation}

Given a classifier trained directly on data or on simulation, it is often desirable to estimate the background directly with data (especially for multijet backgrounds).  One of the most common background estimation techniques is the sideband approach, where the signal is expected to be localized (usually in jet or dijet mass) and regions away from the signal are used to fit a smooth function.  This function is then interpolated into the signal region.  In order for this procedure to work, jet tagging cannot sculpt non-smooth features in the distribution.  Unfortunately, jet features used for tagging are often correlated with the resonant feature.  Physics-inspired methods such as designed decorrelated taggers~\cite{Dolen:2016kst} or convolved substructure~\cite{Moult:2017okx} manually decorrelate prong-tagging features (such as $D_2$) with the jet mass so that using the former to tag boosted bosons and the latter to localize potential signals is amenable to side-band approaches.  However, these methods to not generalize to machine learning classifiers that use many features.  One proposal is to use adversarial techniques to simultaneously optimize tagging performance and independence from the jet mass~\cite{Shimmin:2017mfk}.  A hyper-parameter trades off these two competing goals.  This method was compared with the physics-inspired approaches and many other related automated techniques in Ref.~\cite{ATL-PHYS-PUB-2018-014}.

\subsubsection{Jet Metacognition}

\begin{figure}[tb]
\centering
\includegraphics[width=0.45\textwidth]{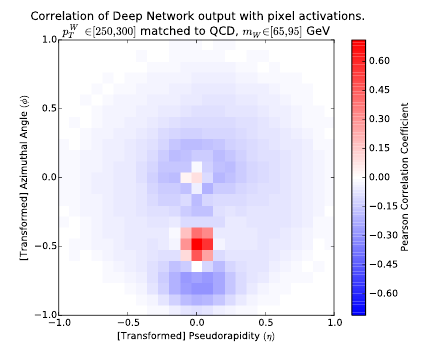}\includegraphics[width=0.45\textwidth]{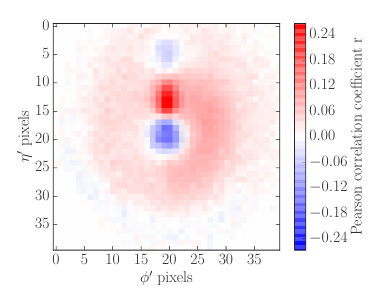}
\caption{Correlation images between the deep neural network for $W$ tagging (left) and top tagging (right).  Higher values correspond to more signal ($W$ or top) like.  For the $W$ tagging case, the energy and location of the sub-leading subjet (the leading subjet is at the origin by construction) is strongly correlated with the network output.  In the top tagging case, the situation is more complicated because of the additional prong in the decay.  The left plot is reproduce from Ref.~\cite{deOliveira:2015xxd} and the right is from Ref.~\cite{Kasieczka:2017nvn}.}
\label{fig:ML:correlations}
\end{figure}

Validating classifier performance in data is critical to build confidence in the power of deep learning techniques.  Understanding what the networks are learning is also important for robustness (and perhaps teaching the physicist something new!), because it is usually only possible to calibrate/validate classifiers in a limited kinematic range.  For example, Ref.~\cite{deOliveira:2015xxd} provides a list of diagnostic tests that one can run to help visualize what information is learned by convolutional neural networks (though some are applicable more generally).  That list\footnote{Some of these items were also covered in Ref.~\cite{Almeida:2015jua}.} is summarized here:

\begin{itemize}
\item {\bf Low-level correlations}: Correlations between the network inputs and outputs can show which areas of the input space are most useful for discrimination.   For a jet image $J$, this results in another image $C$ where the pixel intensity is the correlation between the network output $N$ and the pixel intensity, $C_{ij}=\rho(J_{ij},N(J))$.   This only identifies linear information about the network output but can illustrate how this is distributed non-linearly in space.  Examples are shown in Fig.~\ref{fig:ML:correlations} for $W$ and top tagging.  Extensions to non-linear generalizations of the correlation coefficient are also possible.
\item {\bf High-level correlations}: The joint distribution of standard physically-inspired features (e.g. jet mass) and the network output (or intermediate node activations) illustrate if and how the network is learning about known physical effects.  
\item {\bf High-level input}: Building a new classifier that combines the network output and a standard physically-inspired feature can demonstrate to what extent the information about that feature is learned by the network.
\item {\bf Redacted phase space}: Studying the distribution of inputs and the network performance after conditioning on standard physically-inspired features can help to visualize what new information the network is using from the jet.  Training the network on inputs that have been conditioned on specific values of known features can also be useful for this purpose.
\item {\bf Re-weighted phase space}: A complementary approach to redacting is to re-weight phase space so that the marginal likelihood ratio for standard physically-inspired features is unity, $p_s(m) / p_b(m)=1$, where $m$ is a feature of the full image $J$ and $p_{s,b}(m)=\int p_{s,b}(J)\delta(m(J)=m)$ is the marginal probability distribution.  With this weighting, the known feature $m$ is not useful for classification.  Reference~\cite{Chang:2017kvc} named this `planning'.  
\item {\bf Weights}: The activations for the various layers can sometimes be useful in identifying what the network is learning.  This is particularly true for convolutional layers where the filters encode activated features.  An interesting further step is to convolve the filters with the average image from the two classes and then visualize their difference.
\item {\bf Most activating images}: A complementary approach to visualizing the network weights is to find which sets of inputs most activate a particular node or the entire network.  In the case of jet images, one can plot the average of the $n$ most activating images.
\end{itemize}

\subsection{Regression Techniques}
\label{sec:ML:regression}

The remainder of Sec.~\ref{sec:machine_learning} will describe extensions of machine learning to jet physics beyond classification.  This section begins with regression.  In contrast to classification, the goal of regression is to learn a continuous target.  The most important regression task in jet physics is calibration: taking the observable features of a jet after detector distortions and predicting true properties of the jet as they would have been prior to reaching the detector.  One common example is the energy calibration, where tracking detectors and calorimeters are used to measure momenta and energies and they are combined and calibrated to predict the true jet energy.  A typical machine-learning based regression procedure would be setup to learn the true energy $y$ given some detector-level observables $x$, which could be the `reconstructed' jet energy from tracks and calorimeter-energy deposits.  Symbolically, a neural network $f$ trained in this way would likely be chosen as follows:

\begin{align}
f=\text{argmin}_g\sum_i(g(x_i)-y_i)^2.
\end{align}

\noindent One can show that in the asymptotic limit (enough training data, sufficiently flexible neural network), $f$ will approach $\langle y|x\rangle$, that is the average value of the true energy given the reconstructed energy\footnote{If instead, one uses the mean absolute error, $|g(x_i)-y_i|$, then the neural network will approach the median of $y$ given $x$.  If the loss is $\text{lim}_{\alpha\rightarrow 0}|g(x_i)-y_i|^\alpha$, the the neural network will learn the mode.}.  This is not ideal, because then the classifier $f$ depends on the spectrum of $y$ used during the training.  One can compute the method bias:

\begin{align}
\label{eq:closure}
\langle f(x)|y\rangle=\int dx dy' y' p_\text{train}(y'|x)p_\text{test}(x|y),
\end{align}

\noindent where $p_\text{train}$ is the distribution uses for training $p_\text{test}$ is the distribution used for testing.  Note that Eq.~\ref{eq:closure} is not equal to $y$ even when $p_\text{train}=p_\text{test}$.  For the current non-neural network-based calibrations of both the ATLAS and CMS experiments, a method called numerical inversion is used instead of the above procedure.  Instead of learning $y$ given $x$, one learns $x$ given $y$ and then inverts the function.  Symbolically, if $g:y\rightarrow x$, then the calibration is $g^{-1}(x)$.  This procedure is independent of $p_\text{train}$ by construction and results in a nearly unbiased procedure under mild assumptions~\cite{Cukierman:2016dkb}.  One can generalize this procedure with deep learning, to take advantage of multi-dimensional features of the measured jet radiation pattern~\cite{ATL-PHYS-PUB-2018-013}.  This \textit{generalized numerical inversion} starts by learning a function $g:(y,\theta)\rightarrow x$, where $\theta$ are the auxiliary (detector-level) features that may be useful for calibration.  Ideally, one would then invert $g$ for every $\theta$ and then the calibration is simply $g^{-1}(x|\theta)$.  Since $\theta\in\mathbb{R}^n$, it is not practical to invert $g$ for each $\theta$.  Instead, one can simply learn another neural network $h$ which approximates the inverse.  This new function is then used to perform the calibration. 

A critical task of jet calibration is to mitigate the impact of pileup.  While one could in principle include local pileup corrections as a part of generalized numerical inversion, another possibility is to first correct for pileup and then apply a standard or machine learning-based energy calibration.  Prior to the LHC Run 2, the main pileup mitigation technique was the jet areas correction~\cite{Cacciari:2007fd}, $p_\text{T,J}\mapsto p_\text{T,J}-\rho A$, where $\rho$ is an estimate of the pileup density in the event and $A$ is the jet area~\cite{Cacciari:2008gn}.  Since that time, ATLAS, CMS, and the theory community have studied a large number of \textit{constituent-based} pileup mitigation techniques (see e.g. Ref.~\cite{Soyez:2018opl}).  These proposals correct the inputs to jets and therefore can mitigate the pileup dependence of all jet substructure observables at once.  Two natural questions are if machine learning can improve upon these ideas or at least automate their optimization given the existence of tunable parameters in each method.

Both of these questions have been addressed with various studies.  The authors of Ref.~\cite{Komiske:2017ubm} used convolutional neural networks to take `detector-level' jet image information and predict a pileup-removed jet image.  This end-to-end approach can use all of the available information with the potential to improve upon other methods that use physics-inspired constructions.   Note that this is a regression model where the input is many-dimensional \textbf{and} the output is also many-dimensional (an image).  Another study described in Ref.~\cite{Martinez:2018fwc} uses graph neural networks to automate the constituent-based pileup mitigation proposed by the pileup per particle identification (PUPPI) method~\cite{Bertolini:2014bba}.  In the PUPPI approach, the pileup removal follows a particular form, where a kernel is used to synthesize local information around a particle to determine the probability that it is from pileup.  The regression task is then performed by weighting particles based on the probability.  This kernel is essentially replaced with a neural network and the classification task is set up as a machine learning problem.  While there has been no direct comparison between the more structured PUPPI-like approaches (classification for regression) and the image-based approaches (direct multi-dimensional regression), it is likely that the optimal procedure will take advantage of the strengths of both methods.

\subsection{Machine Learning for Jet Simulations}
\label{sec:ML:simulation}

Another promising application of machine learning beyond classification is generation.  This is a natural extension of regression, where the goal is to learn a probability distribution, or at least learn to sample from a probability distribution (`likelihood free').  There have been a variety of proposals for using deep generative models for jet physics, ranging from replacing / augmenting parton shower models to accelerating the detector response to hadrons.  

\subsubsection{Matrix element calculations}

One of the challenges with performing complete differential cross section calculations of jet substructure is the slow numerical calculations that are required to match with fixed-order matrix elements.  A significant challenge with these calculations is the construction phase space discretization.  Many of the common matrix element tools are built on variations of the VEGAS~\cite{PETERLEPAGE1978192} algorithm.  Recently, multiple proposals have suggested that deep learning can be used to optimize the grid construction~\cite{Bendavid:2017zhk,Klimek:2018mza}.   One exciting prospect of these approaches is that phase space regions can have non-trivial shapes and therefore hold great promise for significantly reducing the time required to compute complex matrix elements.

At hadron colliders, another important input to a complete calculation is the PDF.  Neural networks for PDFs have been studied by the NNPDF collaboration for many years~\cite{Ball:2008by,Rojo:2018qdd}.

\subsubsection{Parton shower models}

Following matrix element calculations, the next step in a typical simulation for LHC physics is the parton shower.  Current parton shower methods use Markov Chain Monte Carlo approaches that are very efficient for populating the large phase space needed to describe realistic hadronic final states.  However, faster or more flexible methods may be useful to augment existing methods for increased speed and capturing higher-order effects (when fit to data).  The simplest approach would be an end-to-end generator that does not encode any physical structure in the generative process.  This is case for Ref.~\cite{deOliveira:2017pjk}, where a generative adversarial network (GAN)~\cite{NIPS2014_5423} is used to sample jet images encoding the radiation pattern inside a jet.  GAN approaches to simulation are discussed in more detail in Sec.~\ref{sec:fastsim}.

Two other generative models that are built on the structure of QCD were proposed in Ref~\cite{Monk:2018zsb} and Ref.~\cite{Andreassen:2018apy}.  The (approximately) scale invariant structure of QCD is combined with convolutional neural networks in Ref.~\cite{Monk:2018zsb} to construct a shower program where the convolutional filters encode the effective splitting function.  In contrast, Ref.~\cite{Andreassen:2018apy} uses a recurrent structure, where a parton-shower like structure is imposed to generate histories.  Each component of the shower evolution (such as the splitting functions) is replaced with a neural network.  By using a physically-inspired structure, both the image- and tree-based approaches use far fewer parameters than typical deep neural networks.  Therefore, these are promising methods for generation as well as related task such as phase space re-weighting.

The three parton shower approaches mentioned so far strive to generate an entire parton shower history.  An alternative possibility is to re-weight an existing history produced from a standard parton shower program.  Analytic methods may be applicable in some cases~\cite{Mrenna:2016sih,Bellm:2016voq,Badger:2016bpw}, but Ref.~\cite{Bothmann:2018trh} has started to explore the possibility of using neural networks as well.  Combinations of re-weighting and ab initio methods may provide the most flexible solution to jet phase space reweighting in the future.

\subsubsection{Fast simulation}
\label{sec:fastsim}

Typically the slowest part of any full simulation of an LHC collision is the calorimeter response of hadrons.  State-of-the-art physics-based simulations like Geant4~\cite{Agostinelli:2002hh} provide an excellent description of hadronic interactions with detector material over an wide range in energy and particle type.  However, these models are slow (can take minutes for one LHC event) and make some approximations.  Therefore, there has been a growing interest in using generative neural networks to speed up simulations and maybe one day be tuned directly on collision or testbeam data.

One method with growing popularity is the GAN~\cite{NIPS2014_5423}.  This method uses two neural networks: one maps noise to structure (the generator) and one called the discriminator that classifies batches of events as either `real' (from the physics-based generator / data) or from the generator network.  These two networks `compete' and when the discriminator is confused, the generator will be a good simulator.  This method has been studied in Ref.~\cite{Paganini:2017hrr,Paganini:2017dwg,Erdmann:2018jxd,ATL-SOFT-PUB-2018-001,Carminati:2018khv} for calorimeter simulations of individual hadrons.  An alternative to the GAN is the variational autoencoder (VAE)~\cite{Rezende:2014:SBA:3044805.3045035,DBLP:journals/corr/KingmaW13}, which was studied in Ref.~\cite{ATL-SOFT-PUB-2018-001}.  The goal of an autoencoder is to learn a model that compresses an image and than uncompresses it to reconstruct the same image.  The goal of a VAE is to learn an autoencoder where the distribution of the compressed image (the `latent space') is as close as possible to a multidimensional Gaussian as possible.  Both GANs and VAEs are promising methods for accurate generation in high dimensional feature spaces, but significant effort is still required to reach the high level of fidelity needed for accurate detector simulations.  The biggest challenge for these methods is to quantify their performance - unlike classification or regression, the performance is not well-summarized by the value of the loss function.  Therefore, new visualization and other methods will be required to know when a deep generative model is good enough to replace or augment existing fast / full simulation techniques.

In addition to individual hadron-level detector simulation, there have been proposals for jet-level fast simulations based on GANs~\cite{Erdmann:2018kuh,DiSipio:2019imz}.  One advantage of individual hadron-level generators is that the energy deposited in a given location of a calorimeter factorizes: the energy from many particles is simply the sum of the energies.  As a result, the extreme tails of event-level distributions are mostly due to less-extreme tails of individual hadron distributions.  In contrast, when entire jet or events are generated, the distribution tails are limited to what the GAN has seen during training.  In other words, generative models are excellent at interpolation but are likely unreliable for extrapolation.  Despite this limitation, interpolation models may still be useful for a variety of applications. 

\subsubsection{Beyond simulation}

While generative models have gained significant attention for accelerating or improving jet simulations, they may be useful for other purposes as well.  For example, Ref.~\cite{Lin:2019htn} proposed GANs as an un-binned approach to the template procedure from Ref.~\cite{Cohen:2014epa} for exploiting factorization to estimate multijet backgrounds in high-multiplicity jet searches.  In addition to removing binning effects, this GAN-based method can naturally be conditioned on other jet observables such as those that indicate if the jet was produced from a quark or a gluon.

\subsection{Anomaly detection}
\label{sec:ML:anomaly}

\begin{figure}[tb]
\centering
\includegraphics[width=0.65\textwidth]{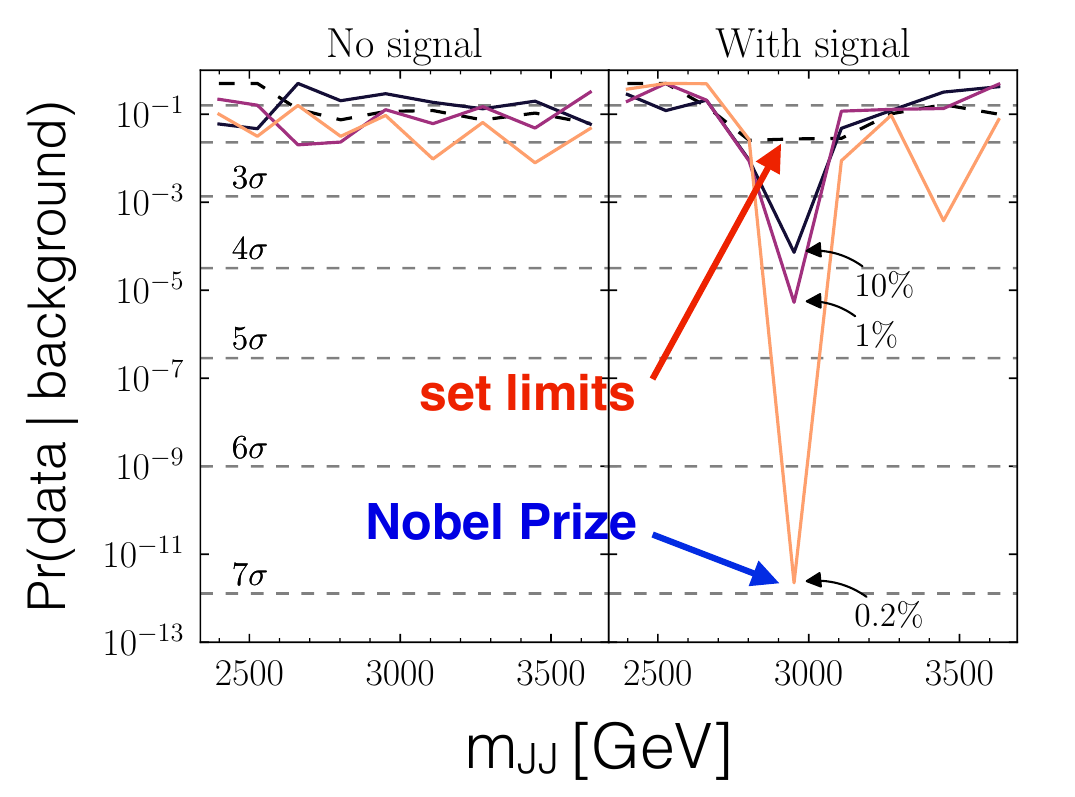}
\caption{The local $p$-value for a dijet resonance search using the CWoLa method proposed in Ref.~\cite{Collins:2018epr,Collins:2019jip}.  The left plot shows the case where there is no injected signal and the right panel shows the case where a signal is injected near 3 TeV.  The different lines correspond to different classifier working points (lower efficiency means more signal-like).  Figure adapted from Ref.~\cite{Collins:2018epr,Collins:2019jip}.}
\label{fig:ML:cwolahunting}
\end{figure}

One last application of modern machine learning to jet physics is the identification of new structure: anomaly detection.  For all the methods described so far, there is a particular target distribution in mind: either a particular signal model for classification, a certain `true' value for regression, or a given probability distribution for generation.  Given that there is no significant evidence for new particles for forces at the LHC so far, it is critical that the existing supervised methods be complemented with more model-agnostic approaches.  The weak supervision approaches from Sec.~\ref{sec:weaksupervision} already showed that one can learn directly from unlabeled data.  The authors of Ref.~\cite{Collins:2018epr,Collins:2019jip} took this idea one step further to apply it for anomaly detection.  In the CWoLa weak supervision approach, one can train on mixed samples of two classes to learn classifiers that are asymptotically optimal for distinguishing examples from sets of pure samples.  The anomaly detection extension of this method uses some known feature where a signal is expected to be resonant (if it exists) to created potentially mixed samples.  Classifiers are trained to distinguish signal region from sideband regions and then applied to enhance low-purity signals from jets with a nonstandard jet substructure.  One must take great care in the training to not sculpt artificial resonance structures, but the method may be able to identify many signals that are currently uncovered by conventional searches.  Figure~\ref{fig:ML:cwolahunting} illustrates how this method would work for a signal $W'\rightarrow WX$, where the $X\rightarrow WW$.  The signature of this signal is one two-prong jet with a mass near $m_W$ and another 4-prong jet with a mass near $m_X$.  

One drawback of the CWoLa approach is that there must be sufficient signal present in order for the classifier to learn anything - in the language of Sec.~\ref{sec:weaksupervision}, the two mixed samples must have different class proportions.  An alternative anomaly detection technique\footnote{There have been other proposals for anomaly detection such as Ref.~\cite{DAgnolo:2018cun,DeSimone:2018efk}, but these rely heavily on a background model and thus are likely not well-suited for complex hadronic final states where data-driven background estimates are critical.} that has been proposed to find new resonances in hadronic final states is the use of autoencoders~\cite{Farina:2018fyg,Heimel:2018mkt,Roy:2019jae,Hajer:2018kqm,Cerri:2018anq}.   As was introduced in Sec.~\ref{sec:fastsim}, the goal of an autoencoder is to learn a model that can compress and then decompress.  For anomaly detection, the idea is that jets which have not been used during training will not be reconstructed well from this compression and decompression.  The autoencoder proposals can be trained entirely on background and then applied to a region where there may or may not be signal.  The difference between the original jet representation (e.g. a jet image) and the one passed through the autoencoder compression and decompression is used for model-agnostic classification.  References~\cite{Farina:2018fyg,Heimel:2018mkt,Roy:2019jae} have shown the potential to be sensitive to jets with non-SM substructure and have additionally investigated the correlations with the jet mass - the most important feature for background estimation in the single jet mass search (see Sec.~\ref{sec:decorrelation}).   Both Ref.~\cite{Heimel:2018mkt,Roy:2019jae} demonstrated how automatic or manual methods can be used to reduce the mass dependence.  Future comparisons between autoencoders and the CWoLa method (and its variations) will likely show that both are needed to broadly cover BSM scenarios with complementary strengths and weaknesses. 

\subsection{The future}
\label{sec:ML:future}

Deep learning for jet physics is a rapidly expanding field that continues to adapt state-of-the-art algorithms from industry and produce innovative methods for characterizing jets.  This work spans the entire spectrum from nearly physics-agnostic to strongly physics-inspired.  A dedicated workshop series~\cite{workshop1,workshop2} has grown from the growing interest in this work and there is a bright future for combining deep learning with QCD in order to maximally exploit the data at the LHC and beyond.

\section{Conclusions}
\label{sec:conclusions}

Jet substructure is now playing a central role at the LHC. Over the past several years a wealth of experimental and theoretical techniques have been developed to exploit the substructure of jets. In this review we have attempted to provide a comprehensive, yet pedagogical overview of this rapidly developing field, focusing on advances in both theory and machine learning. While substructure techniques will continue to evolve in parallel with our improving understanding of QCD (and quantum field theory more generally) as well as machine learning, we believe that by focusing on the underlying physical principles guiding jet substructure observables, calculations, and applications of machine learning to jet substructure, this review will remain a useful document for the foreseeable future. We look forward to the future developments and applications of jet substructure, and we hope that this review serves as a useful starting point for the interested reader to study these topics in greater detail and contribute to this exciting field.

\section*{Acknowledgements}

We thank Jon Butterworth for suggesting this jet substructure review article, as well as the participants and organizers of the BOOST 2016 conference for discussions and input.  We greatly appreciated detailed comments, criticisms, and helpful suggestions from Yang-Ting Chien, Jan Kieseler, Eric Metodiev, Duff Neill, Stefan Prestel, La\'is Sarem Schunk, Ariel Schwartzman, and Jesse Thaler. I.M. also thanks Frank Tackmann for help with illustrations.  This work is supported by the Office of High Energy Physics of the U.S. DOE under Contract No. DE-AC02-05CH11231, and the LDRD Program of LBNL.

\bibliography{report-bibdesk}

\end{document}